\documentclass[11pt,twoside]{article}
\usepackage{amsmath,amsthm,amssymb,amsbsy,bm,calc,comment,fancyhdr,fixltx2e,
            float,glossaries,ifthen,mathrsfs,textgreek,titlesec,tocloft,xy}
\input epsf
\PassOptionsToPackage{ctagsplt,Righttag}{amstex}
\usepackage[title]{appendix}
\usepackage{cite}
\usepackage[dvips,dvipsnames,usenames]{color} 
\usepackage[colorlinks=true,citecolor=MidnightBlue,linkcolor=BrickRed,breaklinks=true]{hyperref} 
\usepackage[usenames,dvipsnames]{xcolor}
\usepackage[super]{nth}
\usepackage{graphicx}
\usepackage{mathtools}
\usepackage{wrapfig}
\usepackage[font={footnotesize}]{caption}
\usepackage{tikz}

\definecolor{darkgreen}{rgb}{0,0.65,0}
\definecolor{darkred}{rgb}{0.65,0,0}
\include{Commands}

\newcommand{\fvect}[1]{\overset{\rightarrow}{#1}}
\newcommand{\bvect}[1]{\overset{\leftarrow}{#1}}
\newcommand{\scp}{\mathcal{P}}
\newcommand{\bscp}{\partial \scp}
\newcommand{\lo}{{L\textsubscript{o}}}
\newcommand{\ld}{{L\textsubscript{d}}}
\newcommand{\textspace}{\smallskip}
\newcommand{\eqnspace}{\smallskip}
\newcommand{\kb}{k_{\mathrm{b}}}
\newcommand{\kg}{k_{\mathrm{g}}}
\newcommand{\kc}{k_{\mathrm{c}}}

\newcommand{\bmxb}{\bm{x}_{\mathrm{b}}}
\newcommand{\scmp}{\mathscr{M}_{\mathrm{p}}}
\newcommand{\npb}{n_{\mathrm{p}}^{\mathrm{b}}}
\newcommand{\mupb}{\mu_{\mathrm{p}}^{\mathrm{b}}}
\newcommand{\bmjq}{\bm{J}_{\mathrm{q}}}
\newcommand{\bmjs}{\bm{J}_{\mathrm{s}}}
\newcommand{\jq}{J_{\mathrm{q}}}

\advance\textheight by \topskip
\begin{document}


\begin{center}
	{\textbf{\Large{The irreversible thermodynamics of curved lipid membranes}}
	} \\
	\vspace{0.21in}

	Amaresh Sahu$^{1, \ddag}$, Roger A. Sauer$^{2, \S}$, and Kranthi K. Mandadapu$^{1, \dag}$ \\
	\vspace{0.25in}

	\footnotesize{
		{
			$^1$ Department of Chemical \& Biomolecular Engineering, University of California at Berkeley,\\
			Berkeley, CA, 94720, USA \\[3pt]
			$^2$ Aachen Institute for Advanced Study in Computational Engineering Science (AICES),\\
			RWTH Aachen University, Templergraben 55, 52056 Aachen, Germany \\
		}
	}
\end{center}

\vspace{13pt}
%
%

\begin{abstract}
	The theory of irreversible thermodynamics for arbitrarily curved lipid membranes is presented here.
	The coupling between elastic bending and irreversible processes such as intra-membrane lipid flow, intra-membrane phase transitions, and protein binding and diffusion is studied.
	The forms of the entropy production for the irreversible processes are obtained, and the corresponding thermodynamic forces and fluxes are identified.
	Employing the linear irreversible thermodynamic framework, the governing equations of motion along with appropriate boundary conditions are provided.
\end{abstract}
\vspace{15pt}


\noindent\rule{4.6cm}{0.4pt}

\noindent \small{\ddag \, amaresh.sahu@berkeley.edu \\
	\S \, sauer@aices.rwth-aachen.de \\
	\dag \, kranthi@berkeley.edu
}

\vspace{25pt}

%
%

\small \tableofcontents
\vspace{08pt}

%
%

\section*{List of important symbols}

\begin{tabbing}
	$\bm{1}$ \hspace{28pt}
	\= identity tensor in $\mathbb{R}^3$ \\
	$a$ 
	\> current area of the membrane \\
	$a_k$
	\> chemical activity of component $k$ \\
	$A$
	\> area of the membrane in the reference configuration \\
	$\mathscr{A}$
	\> chemical affinity of a chemical reaction \\
	$\bm{a}_\alpha$ 
	\> in-plane covariant basis vectors \\ 
	$\bm{a}_{\alpha , \beta}$
	\> partial derivative of $\bm{a}_\alpha$ in the $\theta^\beta$ direction \\
	$\bm{a}_{\alpha ; \beta}$
	\> covariant derivative of $\bm{a}_\alpha$ in the $\theta^\beta$ direction \\
	$a_{\alpha \beta}$
	\> covariant metric tensor \\ 
	$a^{\alpha \beta}$
	\> contravariant metric tensor \\
	$\alpha_k$
	\> stoichiometric coefficient of the $k^{\textrm{th}}$ species in a chemical reaction \\
	$\bm{b}$
	\> body force per unit mass \\
	$\bm{b}_k$
	\> body force per unit mass on the $k^{\textrm{th}}$ membrane component \\
	$b_{\alpha \beta}$
	\> covariant curvature tensor \\ 
	$b^{\alpha \beta}$
	\> contravariant curvature tensor \\
	$\bar{b}^{\alpha \beta}$
	\> cofactor of curvature \\
	$c_k$
	\> mass fraction of the $k^{\textrm{th}}$ membrane component \\
	$C$
	\> spontaneous curvature \\
	$\bm{d}$
	\> director field \\
	$d^{\alpha \beta}$
	\> contravariant, symmetric components of the in-plane velocity gradients \\
	$D_k$
	\> in-plane diffusion constant of the $k^{\textrm{th}}$ membrane component \\ 
	$\Delta$
	\> surface Laplacian operator \\
	$e$
	\> energy per mass, including kinetic and internal energy \\
	$\epsilon^{\alpha \beta}$
	\> permutation tensor \\
	$\eta_\mathrm{e}$
	\> external entropy supply per unit mass \\ 
	$\eta_\mathrm{i}$
	\> internal entropy production per unit mass \\
	$\bm{f}$
	\> total force at the membrane boundary \\
	$\bm{f}_i$
	\> total force on the $i^{\textrm{th}}$ corner of the membrane boundary \\
	$\gamma$
	\> coefficient of the energetic penalty of phase boundaries \\
	$\Gamma^\alpha_{\mu \lambda}$
	\> Christoffel symbols of the second kind \\
	$H$
	\> mean curvature \\ 
	$\bm{i}$ 
	\> surface identity tensor \\
	$\bm{j}_k$
	\> diffusive flux of species $k$ relative to the mass-averaged velocity \\
	$J$
	\> areal membrane expansion relative to original configuration \\ 
	$J^k$
	\> thermodynamic flux conjugate to $X_k$ \\
	$\bmjq$
	\> in-plane heat flux \\
	$\bmjs$
	\> in-plane  entropy flux \\ 
	$\kb$
	\> mean bending modulus \\
	$\kc$
	\> compression modulus \\
	$\kg$ 
	\> Gaussian bending modulus \\
	$k_2, \, k_3$
	\> energy penalty parameters for unbound and bound PI(4,5)P\textsubscript{2} lipids, respectively \\
	$\fvect{k}, \bvect{k}$
	\> forward and reverse reaction rate constants, respectively \\
	$K$
	\> Gaussian curvature \\ 
	$K_\mathrm{eq} (T)$
	\> equilibrium constant for the protein binding and unbinding reaction \\
	$\kappa$
	\> scalar thermal conductivity \\
	$\kappa^{\alpha \beta}$
	\> thermal conductivity tensor \\
	$\kappa_\nu$, $\kappa_\tau$
	\> normal curvatures in the $\bm{\nu}$ and $\bm{\tau}$ directions \\
	$\ell$
	\> arc length parametrization of a curve \\
	$L^{ik}$
	\> phenomenological coefficient between $J^i$ and $X_k$ \\
	$\lambda$
	\> bulk viscosity coefficient for in-plane flow \\
	${\Lambda^\nu}_\sigma$
	\> unimodular transformation tensor \\
	$\bm{m}$
	\> bending moment per unit length at the membrane boundary \\
	$m_\nu$, $m_\tau$
	\> bending moment components in the $\bm{\nu}$ and $\bm{\tau}$ directions, respectively \\
	$M$
	\> scalar moment acting on the membrane boundary in the $\bm{\nu}$ direction \\
	$\bm{M}$
	\> director traction at the membrane boundary \\
	$\mathscr{M}_k$
	\> molar mass of component $k$ \\
	$\scmp$
	\> molar mass of an unbound epsin-1 protein \\
	$M^{\alpha \beta}$
	\> contravariant bending moment tensor \\
	$\bm{\mu}$
	\> couple-stress tensor \\
	$\mu_k$
	\> chemical potential of the $k^{\textrm{th}}$ membrane component \\
	$\mupb$
	\> chemical potential of proteins in the fluid surrounding the membrane \\
	$\mu_k^\circ (T)$
	\> chemical potential of the $k^{\textrm{th}}$ membrane component at standard thermodynamic conditions \\
	$n_k$
	\> concentration of the $k^{\textrm{th}}$ species in the chemical reaction \\
	$\npb$
	\> concentration of proteins in bulk phase (surrounding fluid) \\
	$\bm{n}$
	\> normal vector to the membrane surface \\ 
	$N^{\alpha \beta}$
	\> in-plane contravariant components of the membrane stress \\
	$\bm{\nu}$
	\> in-plane unit normal on the membrane boundary \\
	$\omega^{\alpha \beta}$
	\> contravariant bending dissipation tensor \\
	$p$
	\> pressure normal to the membrane \\
	$\scp$
	\> membrane patch under consideration \\
	$\bscp$
	\> boundary of the membrane patch $\scp$ \\
	$\pi^{\alpha \beta}$
	\> contravariant viscous dissipation tensor \\
	$\phi$
	\> concentration parameter used for notational simplicity \\
	$\psi$
	\> Helmholtz energy density per unit mass, with fundamental variables $a_{\alpha \beta}$ and $b_{\alpha \beta}$ \\
	$\bar{\psi}$
	\> Helmholtz energy density per unit mass, with fundamental variables $\rho$, $H$, and $K$ \\
	$r$
	\> heat source or sink per unit mass \\
	$R$
	\> ideal gas constant \\
	$\mathcal{R}$
	\> reaction rate per unit area \\
	$\fvect{\mathcal{R}}, \bvect{\mathcal{R}}$
	\> forward and reverse reaction rates, respectively \\
	$\fvect{\mathcal{R}}_{\mathrm{e}}, \bvect{\mathcal{R}}_{\mathrm{e}}$
	\> forward and reverse reaction rates at equilibrium, respectively \\
	$\rho$
	\> areal mass density \\
	$\rho_k$
	\> areal mass density of the $k^{\textrm{th}}$ membrane component \\
	$s$
	\> entropy per unit mass \\
	$S^\alpha$
	\> out-of-plane contravariant components of the membrane stress \\
	$\bm{\sigma}$
	\> Cauchy stress tensor \\
	$\hat{\sigma}$
	\> in-plane membrane tension \\
	$\sigma^{\alpha \beta}$
	\> contravariant in-plane stress components due to stretching and viscous flow \\
	$T$
	\> local membrane temperature \\
	$\bm{T}$
	\> traction at the membrane boundary \\
	$\bm{T}^\alpha$
	\> stress vector along a curve of constant $\theta^\alpha$ \\
	$\bm{\tau}$
	\> in-plane unit tangent at the membrane boundary \\
	$\theta^\alpha$
	\> fixed surface parametrization of the membrane surface \\
	$u$
	\> internal energy per unit mass \\
	$\bm{v}$
	\> barycentric velocity \\ 
	$\bm{v}_k$
	\> velocity of the $k^{\textrm{th}}$ membrane component \\
	$w$
	\> total areal membrane energy density \\
	$w_{\mathrm{c}}$
	\> areal membrane energy density of compression and expansion \\
	$w_{\mathrm{dw}}$
	\> double-well areal membrane energy density \\
	$w_{\mathrm{g}}$
	\> areal membrane energy density of maintaining concentration gradients \\
	$w_{\mathrm{h}}$
	\> areal Helfrich energy density \\
	$w_{\mathrm{sw}}$
	\> single-well areal membrane energy density \\
	$W_c$
	\> total membrane compression energy \\
	$w_\alpha$
	\> normal component of $\dot{\bm{a}}_\alpha$ \\
	$w_{\alpha \beta}$
	\> in-plane component of $\dot{\bm{a}}_\alpha$ in the $\theta^\beta$ direction \\
	$\bm{x}$ 
	\> position of the membrane surface, in $\mathbb{R}^3$ \\
	$\bmxb$
	\> position of a point on the membrane boundary \\
	$X_k$
	\> thermodynamic force \\
	$\xi$
	\> twist at the membrane boundary \\
	$\xi^\alpha$
	\> convected coordinate parametrization of the membrane surface \\
	$\zeta$
	\> shear viscosity coefficient for in-plane flow \\
	$\otimes$
	\> dyadic or outer product between two vectors 
\end{tabbing}

\newpage


%
%

\section{Introduction} \label{sec:intro}

In this paper we develop an irreversible thermodynamic framework for arbitrarily curved lipid membranes to determine their dynamical equations of motion.
Using this framework, we find relevant constitutive relations and use them to understand how bending and intra-membrane flows are coupled.
We then extend the model to include multiple transmembrane species which diffuse within the membrane, and learn how phase transitions are coupled to bending and flow.
Finally, we model the binding and unbinding of surface proteins and their diffusion along the membrane surface.
\textspace

Biological membranes comprised of lipids and proteins make up the boundary of the cell, as well as the boundaries of internal organelles such as the nucleus, endoplasmic reticulum, and Golgi complex.
Lipid membranes and their interactions with proteins play an important role in many cellular processes, including endocytosis \cite{kishimoto-pnas-2011, baumgart-ncomm-2015, rao-cmls-2011, karotki-jcb-2011, liu-plosbio-2009, walther-nature-2006, schmid-nature-2007, kukulski-cell-2012, mcmahon-nrmcb-2011}, exocytosis \cite{zhang-bpj-2010, chernomordik-cosb-1995}, vesicle formation \cite{gruenberg-nat-rev-cell-bio-2004}, intra-cellular trafficking \cite{barlowe-cell-1994}, membrane fusion \cite{harrison-nsmb-2008, yang-tamm-nat-comm-2016, chernomordik-cosb-1995}, and cell-cell signaling and detection \cite{su-vale-science-2016, carlson-mahadevan-plos-comp-bio-2015, qi-pnas-2001}. \\[-8pt]

Protein complexes that have a preferred membrane curvature can interact with the membrane surface and induce bending \cite{bacia-nature-2011}, important in processes where coat proteins initiate endocytosis \cite{liu-plosbio-2009, schmid-nature-2007, mcmahon-nrmcb-2011, walther-nature-2006, karotki-jcb-2011} and BAR proteins sense and regulate membrane curvature \cite{rao-cmls-2011, baumgart-bpj-2015, peter-science-2004, baumgart-ncomm-2015}.
In all of these processes, lipid membranes undergo morphological changes in which phospholipids flow to accommodate the shape changes resulting from protein-induced curvature.
These phenomena include both the elastic process of bending and irreversible processes such as lipid flow.
\textspace

Another important phenomena in many biological membrane processes is the diffusion of intra-membrane species such as proteins and lipids to form heterogeneous domains.
For example, T cell receptors are known to form specific patterns in the immunological synapse when detecting antigens \cite{carlson-mahadevan-plos-comp-bio-2015, qi-pnas-2001}.
In artificial giant unilamellar vesicles, a phase transition between liquid-ordered (\lo) and liquid-disordered (\ld) membrane phases has been well-characterized \cite{veatch-keller-bpj-2003, veatch-keller-prl-2002, veatch-keller-bba-2005}.
Such phase transitions have also been observed on plasma membrane vesicles \cite{veatch-acscb-2008}.
Furthermore, morphological shape changes in which either the \lo\ or the \ld\ phase domains bulge out to reduce the line tension between the two phases have been observed \cite{veatch-keller-bpj-2003, baumgart-hess-nature-2003}.
These phenomena clearly indicate the coupling between elastic membrane bending and irreversible processes such as diffusion and flow, which must be understood to explain the formation of tubes, buds, and invaginations observed in various biological processes \cite{baumgart-hess-nature-2003, stachowiak-natcellbio-2012, zhao-cell-reports-2013, hu-kozlov-science-2008}.
\textspace

The final phenomena of interest is the binding and unbinding of proteins to and from the membrane surface, and the diffusion of proteins once they are bound.
Protein binding and unbinding reactions are irreversible processes and are ubiquitous across membrane-mediated phenomena \cite{liu-plosbio-2009, walther-nature-2006, kukulski-cell-2012, schmid-nature-2007, gruenberg-nat-rev-cell-bio-2004, harrison-nsmb-2008, yang-tamm-nat-comm-2016, bacia-nature-2011}.
As an example, epsin-1 proteins can bind to specific membrane lipids during the early stages of endocytosis and induce bending \cite{stachowiak-natcellbio-2012, schmid-nature-2007}.
Moreover, antigen detection by T cells can be sensitive to the kinetic rates of T cell receptor binding and unbinding \cite{davis-chien-ari-1998, matsui-davis-pnas-1994, sykulev-eisen-pnas-1994}.
The kinetic binding of proteins also plays a crucial role in viral membrane fusion \cite{harrison-nsmb-2008}, where proteins and membranes are known to undergo kinetically restricted conformational changes in the fusion of influenza \cite{wilson-skehel-wiley-nature-1981, bullough-nature-1994, chen-cell-1998} and HIV \cite{yang-tamm-nat-comm-2016}.
The case of HIV is particularly interesting, as fusion proteins primarily reside at the interface between \lo \,and \ld \,regimes and fusion is believed to be favorable because it reduces the total energy due to line tension between the two phases \cite{yang-tamm-nat-comm-2016}.
\textspace

All of the above phenomena involve elastic bending being fully coupled with the irreversible processes of lipid flow, the diffusion of lipids and proteins, and the surface binding of proteins.
Comprehensive membrane models which include these effects are needed to fully understand the complex physical behavior of biological membranes.
Our work entails developing a non-equilibrium thermodynamic framework that incorporates these processes.
\textspace

Previous theoretical developments have modeled a range of lipid bilayer phenomena.
The simplest models apply the theory of elastic shells \cite{naghdi-1973-theory} and model membranes with an elastic bending energy given by Canham \cite{canham-jtb-1970} and Helfrich \cite{helfrich-1973}.
Many studies focus on solving for the equations of motion for simple membrane geometries such as the deviations from flat planes \cite{seifert-epl-1993, fournier-prl-1996, safran-prl-1995, seifert-long, leitenberger-langmuir-2008, ramaswamy-prl-2014} and cylindrical or spherical shells \cite{seifert-long, prost-prl-2002, rahimi-soft-matter-2013, seifert-pra-1991}.
Some of these works also model the coupling between elastic effects and either inclusions \cite{fournier-prl-1996, dan-langmuir-1993, faris-prl-2009}, surrounding protein structures \cite{ramaswamy-prl-2014}, or fluid flow \cite{seifert-long, leitenberger-langmuir-2008, voth-bpj-2004a} on simple geometries.
More general geometric frameworks based on theories of elastic shells were established to model lipid membranes of arbitrary geometry.
Such models used variational methods to determine the constitutive form of the stress components \cite{steigmann-mms-1998, rahimi-arroyo-pre-2012, arroyo-pre-2009, guven-jpa-2004, capovilla-guven-jpa-2002, powers-rmp-2010}.
Models developed from variational methods have been built upon to include protein-induced curvature \cite{steigmann-agrawal-bmmb-2009}, viscosity \cite{arroyo-pre-2009}, and edge effects \cite{agrawal-cmt-2009}, and are able to describe various membrane processes.
\textspace

While the formulation of models from variational methods is theoretically sound, the techniques involved are not easily extendable to model the aforementioned coupling between fluid flow and protein-induced spontaneous curvature, phase transitions, or the binding and unbinding of proteins on arbitrary geometries.
Recently, membrane models have been developed by using fundamental balance laws and associated constitutive equations \cite{steigmann-fluid-film-arma-1999, kranthi-bmm-2012, steigmann-agrawal-zamp-2011, sauer-liquid-shell-corr-2016, scriven-1960, steigmann-mms-1998, edwards-brenner}.
In addition to reproducing the results of variational methods, they have had great success in understanding the specific effects of protein-induced curvature on membrane tension \cite{kranthi-biophys-2014, walani-pnas-2015} and simulating non-trivial membrane shapes \cite{sauer-liquid-shell-corr-2016}.
However, comprehensive models including all of the irreversible phenomena mentioned thus far and their coupling to bending have not been developed for arbitrary geometries.
\textspace

In this work we develop the general theory of irreversible thermodynamics for lipid membranes, inspired by the classical developments of irreversible thermodynamics by Prigogine \cite{prigogine} and de Groot \& Mazur \cite{degroot-mazur}.
While these classical works \cite{prigogine, degroot-mazur} are for systems modeled using Cartesian coordinates, developing this procedure for two-dimensional lipid membranes is difficult because lipid membranes bend elastically out-of-plane and behave as a fluid in-plane.
As a consequence, a major complexity arises because the surface on which we apply continuum and thermodynamic balance laws is itself curved and deforming over time, thereby requiring the setting of differential geometry.
We address these issues systematically.
\textspace

The following aspects are new in this work:
\begin{enumerate}
	\item A general irreversible thermodynamic framework is developed for arbitrarily curved, evolving lipid membrane surfaces through the fundamental balance laws of mass, linear momentum, angular momentum, energy, and entropy, as well as the second law of thermodynamics.
	\item The contributions to the total entropy production are found and the viscous contribution agrees with earlier variational approaches \cite{arroyo-pre-2009} as well as a balance law formulation using an interfacial flow-based result \cite{kranthi-bmm-2012}.
	\item The thermodynamic framework is extended to model membranes with multiple species, the phase transitions between \lo\ and \ld\ domains, and their coupling to fluid flow and bending.
	\item The model is expanded to incorporate the coupling between protein binding, diffusion, and flow, and the thermodynamic driving force governing protein binding and unbinding is determined.
\end{enumerate}
Our paper is organized as follows: We develop our model by using the fundamental balance laws of mass, momentum, energy, and entropy to determine the equations of motion governing membrane behavior.
We then apply irreversible thermodynamics to determine appropriate constitutive relations, which describe how membrane energetics affect dynamics.
Section \ref{sec:kinematics} reviews concepts from differential geometry which are necessary to describe membranes of arbitrary shape and presents general kinematic results.
Section \ref{sec:1c} models a single-component lipid membrane with viscous in-plane flow, elastic out-of-plane bending, inertia, and protein-induced or lipid-induced spontaneous curvature.
In Section \ref{sec:intramembrane} we extend the model to include multiple lipid components and determine the equations of motion when phase transitions are possible.
In Section \ref{sec:binding}, we model the binding and unbinding of proteins to and from the membrane surface.
Throughout all of these sections, we find membrane phenomena are highly coupled with one another.
For each of Sections \ref{sec:1c}, \ref{sec:intramembrane}, and \ref{sec:binding}, we end by giving the expressions for the stresses and moments, writing the equations of motion, and providing possible boundary conditions to solve associated initial-boundary value problems.
We conclude in Section \ref{sec:conclusion} by including avenues for future work, both in advancing the theory and in developing computational methods.


\section{Kinematics} \label{sec:kinematics}

We begin by reviewing concepts from differential geometry, as presented in \cite{carroll}, which are essential in describing the shape of the membrane and its evolution over time.
We model the phospholipid bilayer as a single differentiable manifold about the membrane mid-plane, implicitly making a no-slip assumption between the two sheets of the bilayer.
We will follow a similar notation to that presented in \cite{kranthi-bmm-2012, sauer-liquid-shell-corr-2016, sauer-shell-2015}.
\textspace

Consider a two dimensional membrane surface $\scp$ embedded in Euclidean 3-space $\mathbb{R}^3$. 
The membrane position $\bm{x}$ is a function of the surface parametrization of the patch $\theta^\alpha$ and time $t$, and is written as
\begin{equation} \label{eq:position-euler}
	\bm{x} = \bm{x} \big( \theta^\alpha, t \big) ~.
\end{equation}
Greek indices in equation \eqref{eq:position-euler} and from now on span the set $\{1, 2\}$.
At every location on the membrane surface, the parametrization $\theta^\alpha$ defines a natural in-plane basis $\bm{a}_\alpha$ given by
\begin{equation} \label{eq:in-plane-basis}
	\bm{a}_\alpha := \bm{x}_{, \alpha} ~.
\end{equation}
The notation $( \, \cdot \, )_{, \alpha}$ denotes the partial derivative with respect to $\theta^\alpha$.
At every point $\bm{x}$ on the patch $\scp$, the set $\{ \bm{a}_1, \bm{a}_2 \}$ forms a basis for the plane tangent to the surface at that point.
The unit vector $\bm{n}$ is normal to the membrane as well as the tangent plane, and is given by \\[-10pt]
\begin{wrapfigure}{r}{0.5\textwidth}
	\vspace{-13pt}
	\begin{center}
		\includegraphics[width=0.45\textwidth]{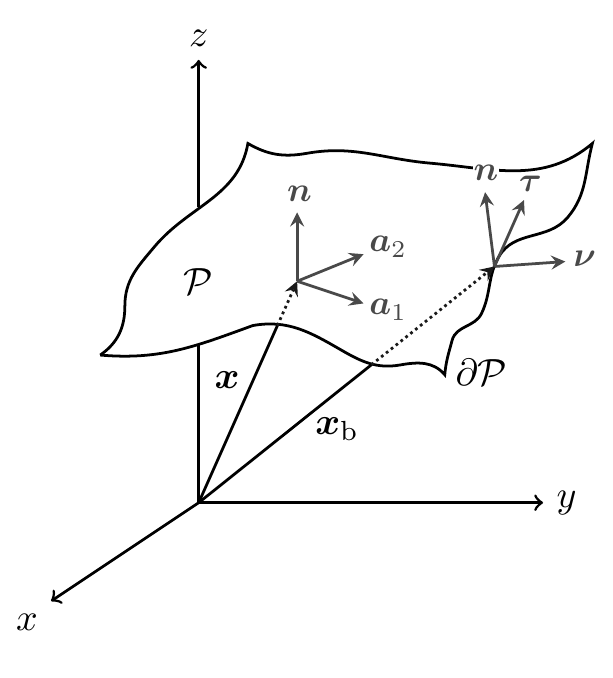}
	\end{center}

	\vspace{-17pt}
	\captionsetup{width=0.45\textwidth}
	\caption{
		A schematic of a membrane patch $\scp$.
		At each point $\bm{x}$ on the membrane patch $\scp$, we define the in-plane tangent vectors $\bm{a}_1$ and $\bm{a}_2$ as well as the normal vector to the plane $\bm{n}$.
		The set $\{ \bm{a}_1, \bm{a}_2 \}$ constitutes a basis for the tangent plane at any location, while the set $\{\bm{a}_1, \bm{a}_2, \bm{n}\}$ forms a basis of $\mathbb{R}^3$.
		At every point $\bmxb$ on the patch boundary $\bscp$, we define the in-plane unit tangent $\bm{\tau}$ and in-plane unit normal $\bm{\nu}$ which also form a basis of the tangent plane.
	}
	\label{fig:geometry}
	\vspace{12pt}
\end{wrapfigure}
\begin{equation} \label{eq:normal}
	\bm{n} := \dfrac{\bm{a}_1 \times \bm{a}_2}{\lvert \bm{a}_1 \times \bm{a}_2 \rvert} ~.
\end{equation}
The set $\{ \bm{a}_1, \bm{a}_2, \bm{n} \}$ forms a basis of $\mathbb{R}^3$, and is depicted in Figure \ref{fig:geometry}.
\textspace

At every point $\bm{x}$, we define the dual basis to the tangent plane, $\{ \bm{a}^1, \bm{a}^2 \}$, such that
\begin{equation} \label{eq:basis-orthogonality}
	\bm{a}^\alpha \cdot \bm{a}_\beta = \delta^\alpha_\beta ~,
\end{equation}
where $\delta^\alpha_\beta$ is the Kronecker delta given by $\delta^1_1 = \delta^2_2 = 1$ and $\delta^1_2 = \delta^2_1 = 0$.
The covariant basis vectors $\bm{a}_\alpha$ and contravariant basis vectors $\bm{a}^\alpha$ are related through the metric tensor $a_{\alpha \beta}$ and contravariant metric tensor $a^{\alpha \beta}$, which are defined as
\begin{equation} \label{eq:metric}
	a_{\alpha\beta}
	:= \bm{a}_\alpha \cdot \bm{a}_\beta
\end{equation}
and
\begin{equation} \label{eq:dual-metric}
	a^{\alpha \beta}
	:= \bm{a}^\alpha \cdot \bm{a}^\beta
	= \big( a_{\alpha\beta} \big) ^ {-1}
	~.
\end{equation}
The metric tensor and contravariant metric tensor describe distances between points on the membrane surface.
The covariant and contravariant basis vectors are related by
\begin{align} 
	\bm{a}_\alpha
	&= a_{\alpha \beta} \, \bm{a}^\beta
	\label{eq:lower-contravariant}
\shortintertext{and}
	\bm{a}^\alpha
	&= a^{\alpha \beta} \bm{a}_\beta
	~,
	\label{eq:raise-covariant}
\end{align}
where in equations \eqref{eq:lower-contravariant}--\eqref{eq:raise-covariant} and from now on, indices repeated in a subscript and superscript are summed over as per the Einstein summation convention.
\textspace

Any general vector $\bm{h}$ can be decomposed in the $\{\bm{a}_1, \bm{a}_2, \bm{n} \}$ and $\{ \bm{a}^1, \bm{a}^2, \bm{n} \}$ bases as
\begin{equation} \label{eq:general-vector-decomposition}
		\bm{h}
		= h^\alpha \, \bm{a}_\alpha
		+ h \bm{n} 
		= h_\alpha \, \bm{a}^\alpha
		+ h \bm{n}
		~,
\end{equation}
where $h^\alpha$ and $h_\alpha$ are contravariant and covariant components, respectively, and are related by 
\begin{align} 
	h^\alpha
	&= a^{\alpha \beta} h_\beta
	\label{eq:raise-vector}
\shortintertext{and}
	h_\alpha
	&= a_{\alpha \beta} \, h^\beta
	~.
	\label{eq:lower-vector}
\end{align}
In general, the metric tensor $a_{\alpha \beta}$ and contravariant metric tensor $a^{\alpha \beta}$ may be used to raise and lower the indices of vector and tensor components.
For a general tensor $\sigma^{\alpha \beta}$, indices are raised and lowered according to 
\begin{align}
	{\sigma^\alpha}_\beta
	&= \sigma^{\alpha \lambda} a_{\beta \lambda}
	\\
	\shortintertext{and}
	\sigma^{\alpha \beta}
	&= {\sigma^\alpha}_\lambda a^{\beta \lambda} ~.
\end{align}
For a symmetric tensor ${s^\alpha}_\beta$ with one raised and one lowered index, the order of the indices is not important and the tensor may be written as ${s^\alpha}_\beta$, ${s_\beta}^\alpha$, or $s^\alpha_\beta$, as all forms are equivalent.
\textspace

When characterizing a membrane patch, it is useful to define a new basis at the membrane patch boundary $\bscp$.
Consider the tangent plane at a point $\bmxb$ on the membrane boundary, with the membrane normal vector $\bm{n}$.
We define in-plane orthonormal basis vectors $\bm{\tau}$ and $\bm{\nu}$, where $\bm{\tau}$ is tangent to the boundary while $\bm{\nu}$ is orthogonal to $\bm{\tau}$.
If the membrane boundary $\bscp$ is parameterized by its arc length $\ell$, then the in-plane unit tangent $\bm{\tau}$ and in-plane unit normal $\bm{\nu}$ are defined as
\begin{align} 
	\bm{\tau}
	:= \bm{a}_\alpha \dfrac{\mathrm{d}\theta^{\alpha}}{\mathrm{d}\ell}
	\label{eq:curve-unit-tangent}
\end{align}
and
\begin{align}
	\bm{\nu}
	:= \bm{\tau} \times \bm{n} ~.
	\label{eq:curve-unit-normal}
\end{align}
The basis vectors $\bm{\nu}$ and $\bm{\tau}$ may be expressed in the covariant and contravariant bases as 
\begin{align}
	\bm{\nu}
	= \nu^\alpha \bm{a}_\alpha
	= \nu_\alpha \bm{a}^\alpha 
	\label{eq:nu-decomposed}
\end{align}
and
\begin{align}
	\bm{\tau}
	= \tau^\alpha \bm{a}_\alpha
	= \tau_\alpha \bm{a}^\alpha ~.
	\label{eq:tau-decomposed}
\end{align}
Similarly, the basis vectors $\bm{a}_\alpha$ may be expressed in terms of the basis vectors $\bm{\nu}$ and $\bm{\tau}$ as
\begin{equation} \label{eq:a-alpha-decomposed-boundary}
	\bm{a}_\alpha
	= \nu_\alpha \bm{\nu} 
	+ \tau_\alpha \bm{\tau} ~.
\end{equation}
The orthonormal basis $\{ \bm{\nu}, \bm{\tau}, \bm{n} \}$ at a position $\bmxb$ on the membrane boundary $\bscp$ is depicted in Figure \ref{fig:geometry}.
\textspace

The surface identity tensor $\bm{i}$ in the tangent plane and the identity tensor $\bm{1}$ in $\mathbb{R}^3$ are given by 
\begin{align}
	\bm{i} &:= \bm{a}^\alpha \otimes \bm{a}_\alpha \label{eq:surface-identity} \\
	\shortintertext{and}
	\bm{1} &:= \bm{i} + \bm{n} \otimes \bm{n} ~, \label{eq:identity}
\end{align}
where $\otimes$ denotes the dyadic or outer product between any two vectors.
The curvature tensor $b_{\alpha \beta}$ is given by 
\begin{equation} \label{eq:curvature-tensor}
	b_{\alpha\beta}
	:= \bm{n} \cdot \bm{x}_{, \alpha \beta} 
	~,
\end{equation}
and describes the shape of the membrane due to its embedding in $\mathbb{R}^3$.
Given the contravariant metric and curvature tensors $a^{\alpha \beta}$ and $b_{\alpha \beta}$, the mean curvature $H$ and the Gaussian curvature $K$ can be calculated as
\begin{equation} \label{eq:mean-curvature}
	H := \frac{1}{2}a^{\alpha\beta}b_{\alpha\beta} 
\end{equation}
and
\begin{equation} \label{eq:gaussian-curvature}
	K := \frac{1}{2}\varepsilon^{\alpha\beta}\varepsilon^{\lambda\mu} b_{\alpha\lambda} b_{\beta\mu} ~,
\end{equation}
where the permutation tensor $\varepsilon^{\alpha \beta}$ is given by $\varepsilon^{12} = -\varepsilon^{21} = 1/\sqrt{\det (a_{\alpha \beta})}$ and $\varepsilon^{11} = \varepsilon^{22} = 0$.
The Gaussian curvature may also be written as $K = \det (b_{\alpha \beta}) / \det (a_{\alpha \beta})$.
The cofactor of curvature $\bar{b}^{\alpha \beta}$ is defined as
\begin{equation} \label{eq:cofactor-curvature}
	\bar{b}^{\alpha \beta}
	:= 2 H a^{\alpha \beta}
	- b^{\alpha \beta} ~,
\end{equation}
where $b^{\alpha \beta} = a^{\alpha \lambda} a^{\beta \mu} b_{\lambda \mu}$ is the contravariant form of the curvature tensor.
\textspace

In general, the partial derivative of the covariant or contravariant components of a vector are not guaranteed to be invariant quantities.
The covariant derivative, denoted $( \, \cdot \,)_{; \alpha}$, produces an invariant quantity when acting on vector components \cite{carroll}.
To define the covariant derivative, we introduce the Christoffel symbols of the second kind, denoted $\Gamma^\alpha_{\lambda \mu}$ and given by 
\begin{equation}
	 \label{eq:christoffel}
		\Gamma^\alpha_{\lambda \mu}
		:= \dfrac{1}{2} a^{\alpha \delta} \left(
			a_{\delta \lambda, \mu}
			+ a_{\delta \mu, \lambda}
			- a_{\lambda \mu, \delta}
		\right) ~.
\end{equation}
The covariant derivatives of the contravariant vector components $v^\lambda$ and covariant vector components $v_\lambda$ are defined as
\begin{align}
	v^\lambda_{; \alpha} 
	&= v^\lambda_{, \alpha} 
	+ \Gamma^\lambda_{\mu \alpha} v^\mu 
	\label{eq:covariant-derivative}
	\\
	\shortintertext{and}
	v_{\lambda; \alpha} 
	&= v_{\lambda, \alpha} 
	- \Gamma^\mu_{\lambda \alpha} v_\mu 
	~,
	\label{eq:covariant-derivative-low}
\end{align}
where $v^\lambda_{; \alpha}$ and $v_{\lambda; \alpha}$ both transform as tensors.
To take the covariant derivative of second or higher order tensors, a more complicated formula is required and may be found in \cite{carroll}.
The covariant derivative of the metric tensor as well as the cofactor of curvature is zero.
The covariant derivative of a scalar quantity is equal to its partial derivative.
It is also useful to note $\bm{v}_{; \alpha} = \bm{v}_{, \alpha}$.
The Gauss and Weingarten equations are 
\begin{align}
	\bm{a}_{\beta; \alpha} &= b_{\beta\alpha} \bm{n} \label{eq:gauss} \\
	\shortintertext{and}
	\bm{n}_{, \alpha} &= -b_\alpha^\mu \bm{a}_\mu  \label{eq:weingarten} ~,
\end{align}
respectively, and provide the covariant derivatives of the basis vectors $\bm{a}_\alpha$ and $\bm{n}$.
\textspace

To model the kinematics of a membrane patch $\scp$, we track the patch over time.
At a reference time $t_0$, we define a reference patch $\scp_0$.
The area of the reference patch $A$ may then be compared to the area of the current patch $a$ at a later time $t$.
For an infinitesimal patch $\mathrm{d}a$, the Jacobian determinant $J$ describes the areal dilation or contraction of the membrane and is defined by
\begin{equation} \label{eq:jacobian-determinant}
	J := \dfrac{\textrm{d}a}{\textrm{d}A} ~.
\end{equation}
The Jacobian determinant may be used to convect integrals over the current membrane patch $\scp$ to integrals over the reference patch $\scp_0$, as for a scalar function $f$ we can write
\begin{equation} \label{eq:convected-integral}
	\int_\scp f ~\mathrm{d}a
	= \int_{\scp_0} \!\! f \, J ~\mathrm{d}A ~,
\end{equation}
and the same can be written for vector- or tensor-valued functions.
The details of the mapping between current and reference membrane configurations, and different coordinate parametrizations, are provided in Appendix~\ref{sec:appendix-convected-coordinates} and a detailed description can also be found in \cite{powers-rmp-2010}.
\textspace

To track how quantities change over time, we define the material derivative $\mathrm{d}/\mathrm{d}t$ according to
\begin{equation} \label{eq:material-derivative}
	\dfrac{\textrm{d}}{\textrm{d}t} ( \, \cdot \, ) := ( \, \cdot \, )_{, t} + v^\alpha ( \, \cdot \, )_{, \alpha} ~.
\end{equation}
Here $( \, \cdot \, )_{, t}$ denotes the partial derivative with respect to time where $\theta^\alpha$ are fixed and $v^\alpha$ are the in-plane components of the velocity vector $\bm{v}$, which may be written as 
\begin{align} 
	\bm{v} 
	:= \dfrac{\mathrm{d} \bm{x}}{\mathrm{d}t}
	= \dot{\bm{x}} 
	= v \bm{n} 
	+ v^\alpha \bm{a}_\alpha ~.
	\label{eq:velocity-euler}
\end{align}
In equation \eqref{eq:velocity-euler}, we have used the shorthand notation $\dot{\bm{x}}$ to express $\mathrm{d}\bm{x} / \mathrm{d} t$, and this notation will be used throughout.
The scalar $v$ is the normal component of the membrane velocity given by
\begin{equation} \label{eq:velocity-normal}
	v = \dot{\bm{x}} \cdot \bm{n} ~.
\end{equation}
Applying the material derivative \eqref{eq:material-derivative} to the basis vectors $\bm{a}_\alpha$ is nontrivial, and requires convecting quantities to the reference patch $\scp_0$.
The material derivatives of the in-plane covariant basis vectors are calculated in Appendix~\ref{sec:appendix-material-derivative}, as well as in \cite{kranthi-bmm-2012}, to be 
\begin{equation} \label{eq:a-alpha-dot}
	\begin{split}
		\dot{\bm{a}}_\alpha 
		= \bm{v}_{, \alpha} 
		&= {w_\alpha}^\beta \bm{a}_\beta + w_\alpha \bm{n} \\
		&= w_{\alpha \beta} \bm{a}^\beta + w_\alpha \bm{n} ~,
	\end{split}
\end{equation}
where the quantities $w_\alpha$, $w^\alpha$, ${w_\alpha}^\beta$, and $w_{\alpha \beta}$ are defined for notational simplicity and are given by
\begin{align}
	& {w_\alpha}^\beta := v^\beta_{; \alpha} - v \, b_\alpha^\beta \label{eq:w-alpha-beta-def} ~, \\
	& w_\alpha := v^\lambda b_{\lambda\alpha} + v_{, \alpha} \label{eq:w-alpha-def} ~, \\
	& w_{\alpha \beta} = {w_\alpha}^\mu a_{\mu \beta} \label{eq:w-sub-alpha-beta-def} ~, \\
	\shortintertext{and}
	& w^\alpha = w_\mu a^{\mu \alpha} \label{eq:w-super-alpha-def} ~.
\end{align}
By applying the material derivative \eqref{eq:material-derivative} to the unit normal $\bm{n}$ \eqref{eq:normal} and using the relation $\bm{n} \cdot \dot{\bm{n}} = 0$, we obtain 
\begin{equation} \label{eq:n-dot}
	\begin{split}
		\dot{\bm{n}} 
		&= - \big(
			v^\lambda b_\lambda^\alpha
			+ v^{, \alpha}
		\big) \bm{a}_\alpha \\
		&= - w^\alpha \bm{a}_\alpha \\
		&= - w_\alpha \bm{a}^\alpha ~.
	\end{split}
\end{equation}
The acceleration $\dot{\bm{v}}$ is the material derivative of the velocity and is calculated as  
\begin{equation} \label{eq:acceleration}
	\dot{\bm{v}}
	= \big(
		v_{, t}
		+ v^\alpha w_\alpha
	\big) \bm{n}
	+ \big(
		v^\alpha_{, t} - v w^\alpha + v^\lambda {w_\lambda}^\alpha
	\big) \bm{a}_\alpha ~.
\end{equation}
The material derivatives of the metric tensor $a_{\alpha \beta}$ and curvature tensor $b_{\alpha \beta}$ are found to be
\begin{align} 
	\begin{split}
		\dot{a}_{\alpha \beta}
		&= v_{\beta; \alpha}
		+ v_{\alpha; \beta}
		- 2 v b_{\alpha \beta} \\
		&= w_{\alpha \beta}
		+ w_{\beta \alpha}
		\label{eq:a-alpha-beta-dot}
	\end{split} \\
	\shortintertext{and}
	\begin{split}
		\dot{b}_{\alpha \beta}
		&= \big(
			v^\lambda_{; \alpha}
			- v b^\lambda_\alpha
		\big) b_{\lambda \beta}
		+ \big(
			v^\lambda b_{\lambda \alpha}
			+ v_{, \alpha}
		\big)_{; \beta} \\
		&= {w_\beta}^\lambda b_{\lambda \alpha}
		+ w_{\beta; \alpha} ~,
		\label{eq:b-alpha-beta-dot}
	\end{split}
\end{align}
where ${w_\beta}^\lambda$, $w_{\alpha}$, and $w_{\alpha \beta}$ are given by equations \eqref{eq:w-alpha-beta-def}--\eqref{eq:w-sub-alpha-beta-def}.
Finally, the time derivative of the Jacobian determinant is found in \cite{kranthi-bmm-2012} as 
\begin{equation} \label{eq:jacobian-determinant-derivative}
	\dfrac{\dot{J}}{J}
	= \dfrac{1}{2} a^{\alpha \beta} \dot{a}_{\alpha \beta}
	= v^\alpha_{; \alpha} - 2 v H ~.
\end{equation}
In three-dimensional Cartesian systems, $\dot{J}/J = \mathrm{div} \, \bm{v}$.
Comparing the Cartesian result with the right hand side of equation \eqref{eq:jacobian-determinant-derivative}, we see that the two-dimensional analog for $\mathrm{div} \, \bm{v}$ is $v^\alpha_{; \alpha} - 2 v H$.


\section{Intra-membrane Flow and Bending} \label{sec:1c}

In this section, we develop a comprehensive model of a single-component lipid membrane which behaves like a viscous fluid in-plane and an elastic shell in response to out-of-plane bending.
We use the balance law framework for single-component membranes previously proposed in several works \cite{steigmann-mms-1998, kranthi-bmm-2012, steigmann-fluid-film-arma-1999, kranthi-biophys-2014, sauer-shell-2015} and in later sections extend it to model multi-component membranes, phase transitions, and the binding of proteins to the membrane surface.
\textspace

We begin by determining local forms of the balances of mass, linear momentum, and angular momentum.
We then go on to determine the form of the membrane stresses through a systematic thermodynamic treatment.
To this end, we develop local forms of the first and second laws of thermodynamics and a local entropy balance.
By postulating the dependence of membrane energetics on the appropriate fundamental thermodynamic variables, we determine constitutive equations for the in-plane and out-of-plane stresses.
We then use these stresses to provide the equations of motion as well as possible boundary and initial conditions
for the membrane, and conclude by briefly discussing how membrane dynamics can be coupled to the surrounding bulk fluid.


\subsection{Balance Laws} \label{sec:balance-laws-1c}

Our general procedure is to start with a global form of the balance law for an arbitrary membrane patch $\scp$, convert each term to an integral over the membrane surface, and invoke the arbitrariness of $\scp$ to determine the local form of the balance law.
To convert terms in the global balance laws to integrals over the membrane patch, we will need tools to bring total time derivatives inside the integral and convert integrals over the patch boundary to integrals over the membrane surface.
\textspace

For a scalar-, vector-, or tensor-valued function $f$ defined on the membrane patch $\scp$, the Reynolds transport theorem describes how time derivatives commute with integrals over the membrane surface.
As described in \cite{kranthi-bmm-2012}, the Reynolds transport theorem is given by 
\begin{equation} \label{eq:rtt}
	\dfrac{\textrm{d}}{\textrm{d}t} \bigg(
		\int_{\scp} f(\theta^\alpha, t) \, \textrm{d}a
	\bigg)
	= \int_{\scp} \dot{f}(\theta^\alpha, t) + \big( v^\alpha_{; \alpha} - 2vH \big) \, f(\theta^\alpha, t) \, \textrm{d}a ~.
\end{equation}
Now consider a vector- or tensor-valued function $\bm{f}$, which may be expressed as $\bm{f} = f^\alpha \bm{a}_\alpha + f \bm{n}$.
The surface divergence theorem describes how an integral of $\bm{f} \cdot \bm{\nu} = f^\alpha \nu_\alpha$ over the membrane boundary $\bscp$, where $\bm{\nu}$ is the boundary normal in the tangent plane defined in equation \eqref{eq:curve-unit-normal}, may be converted to a surface integral over the membrane patch $\scp$.
To this end, the surface divergence theorem states
\begin{equation} \label{eq:divergence-theorem}
	\int_{\partial \scp} \!\! f^\alpha \nu_\alpha \, \textrm{d}s
	= \int_\scp f^\alpha_{; \alpha} \, \textrm{d}a ~,
\end{equation}
where $\mathrm{d}s$ is an infinitesimal line element on the membrane boundary.


\subsubsection{Mass Balance}

Consider a membrane patch $\scp$ with a mass per unit area denoted as $\rho (\theta^\alpha, t)$.
The total mass of the membrane patch is conserved, and the global form of the conservation of mass can be written as
\begin{equation} \label{eq:global-mass-balance}
	\dfrac{\mathrm{d}}{\mathrm{d}t} \left(
		\int_\scp \rho
		~\mathrm{d}a
	\right)
	= 0 ~.
\end{equation}
Applying the Reynolds transport theorem \eqref{eq:rtt} to the global mass balance \eqref{eq:global-mass-balance} brings the time derivative inside the integral, and we obtain
\begin{equation} \label{eq:global-mass-balance-intermediate}
	\int_\scp \dot{\rho}
	+ \big(
		v^\alpha_{; \alpha}
		- 2 v H
	\big) \rho
	~\mathrm{d}a
	= 0 ~.
\end{equation}
Since the membrane patch $\scp$ is arbitrary, the local form of the conservation of mass is given by
\begin{equation} \label{eq:local-mass-balance}
	\dot{\rho}
	+ \big(
		v^\alpha_{; \alpha}
		- 2 v H
	\big) \rho
	= 0 ~.
\end{equation}

As the total mass of the membrane patch is conserved, the mass at any time $t$ is equal to the mass at time $t_0$, \emph{i.e.},
\begin{equation} \label{eq:total-mass-equivalance}
	\int_\scp \rho ~\mathrm{d}a
	= \int_{\scp_0} \rho_0 ~\mathrm{d}A
	~,
\end{equation}
where $\rho_0 = \rho (\theta^\alpha, t_0)$ is the areal mass density of the reference patch.
Using equation \eqref{eq:convected-integral} yields
\begin{equation} \label{eq:total-mass-equivalance-convected}
	\int_{\scp_0} \rho \, J ~\mathrm{d}A
	= \int_{\scp_0} \rho_0 ~\mathrm{d}A
	~.
\end{equation}
As the reference patch $\scp_0$ is arbitrary, the Jacobian determinant $J$ is given by
\begin{equation} \label{eq:jacobian-rho}
	J = \dfrac{\rho_0}{\rho}
	~,
\end{equation}
in addition to the form provided in equation \eqref{eq:jacobian-determinant}.
\textspace

Substituting $f = \rho u$ into the Reynolds transport theorem \eqref{eq:rtt}, where $u$ is an arbitrary quantity per unit mass, and using equation \eqref{eq:local-mass-balance}, we obtain
\begin{equation} \label{eq:rtt-density}
	\dfrac{\textrm{d}}{\textrm{d}t} \bigg(
		\int_{\scp} \rho u ~\textrm{d}a
	\bigg)
	= \int_{\scp} \rho \dot{u} ~\textrm{d}a ~.
\end{equation}
Equation \eqref{eq:rtt-density} is a modified Reynolds transport theorem and is useful in simplifying balance laws where quantities are defined per unit mass.


\subsubsection{Linear Momentum Balance} \label{sec:linear-momentum-balance-1c}

It is well-known from Newtonian and continuum mechanics that the rate of change of momentum of a body is equal to the sum of the external forces acting on it.
Lipid membranes may be acted on by two types of forces: body forces on the membrane patch $\scp$ and tractions on the membrane boundary $\bscp$.
On the membrane patch $\scp$, the body force per unit mass is denoted by $\bm{b}(\theta^\alpha, t)$.
At a point $\bmxb$ on the membrane boundary $\partial \scp$ with in-plane unit normal $\bm{\nu}$, the boundary traction is the force per unit length acting on the membrane boundary and is denoted by $\bm{T}(\bmxb, t; \bm{\nu})$. 
The global form of the balance of linear momentum for any membrane patch $\scp$ is given by
\begin{equation} \label{eq:euler}
	\dfrac{\textrm{d}}{\textrm{d}t} \bigg(
		\int_\scp \rho \bm{v} ~\textrm{d}a
	\bigg) 
	= \int_\scp \rho \bm{b} ~\textrm{d}a
	+ \int_{\partial \scp} \!\! \bm{T} ~\textrm{d}s
	~,
\end{equation}
where the left hand side is the time derivative of the total linear momentum of the membrane patch and the right hand side is the sum of the external forces. 
\textspace

For three-dimensional systems in Cartesian coordinates, one may use Cauchy's tetrahedron arguments to decompose the boundary tractions and define the Cauchy stress tensor, which specifies the total state of stress at any location \cite{chadwick}.
Naghdi \cite{naghdi-1973-theory} performed an analogous procedure on a curvilinear triangle on an arbitrary surface to show boundary tractions may be expressed as a linear combination of the stress vectors $\bm{T}^\alpha$ according to
\begin{equation} \label{eq:traction-decomposition}
	\bm{T} (\bmxb, t; \bm{\nu})
	= \bm{T}^\alpha(\bmxb, t) \, \nu_\alpha ~.
\end{equation}
The stress vectors $\bm{T}^\alpha$ describe the tractions along curves of constant $\theta^\alpha$ and are independent of the in-plane boundary unit normal $\bm{\nu}$.
Substituting the traction decomposition \eqref{eq:traction-decomposition} into the global linear momentum balance \eqref{eq:euler}, applying the surface divergence theorem \eqref{eq:divergence-theorem} on the traction term, and applying the Reynolds transport theorem \eqref{eq:rtt-density} on the left hand side, we obtain
\begin{equation} \label{eq:global-momentum-balance}
	\int_\scp \rho\dot{\bm{v}}~\mathrm{d}a
	= \int_\scp \Big(
		\rho \bm{b} 
		+ \bm{T}^\alpha_{; \alpha}
	\Big) ~\mathrm{d}a
	~.
\end{equation}
Since $\scp$ is arbitrary, equation \eqref{eq:global-momentum-balance} yields the local form of the linear momentum balance as
\begin{equation} \label{eq:local-momentum-balance}
	\rho \dot{\bm{v}} 
	= \rho \bm{b} 
	+ \bm{T}^\alpha_{; \alpha} 
	~.
\end{equation}
\eqnspace

To recast the traction decomposition \eqref{eq:traction-decomposition} into a more familiar form involving the Cauchy stress tensor, we express the stress vectors $\bm{T}^\alpha$ in the $\{ \bm{a}_\alpha, \bm{n} \}$ basis without loss of generality as
\begin{equation} \label{eq:stress-vector-general}
	\bm{T}^\alpha = N^{\alpha \beta}\bm{a}_\beta + S^\alpha\bm{n} ~,
\end{equation}
where $N^{\alpha \beta}$ and $S^\alpha$ are the components of the stress vector $\bm{T}^\alpha$ in the $\{ \bm{a}_\alpha, \bm{n} \}$ basis \cite{kranthi-bmm-2012, steigmann-fluid-film-arma-1999}.
Substituting the form of the stress vectors $\bm{T}^\alpha$ \eqref{eq:stress-vector-general} into the traction decomposition \eqref{eq:traction-decomposition} allows us to write
\begin{equation} \label{eq:traction-stress-normal}
	\bm{T} = \bm{\sigma}^{\mathrm{T}} \bm{\nu} ~,
\end{equation}
where $\bm{\sigma}$ is the Cauchy stress tensor given by
\begin{equation} \label{eq:cauchy-stress-tensor}
	\bm{\sigma}
	= N^{\alpha \beta} \bm{a}_\alpha \otimes \bm{a}_\beta
	+ S^\alpha \bm{a}_\alpha \otimes \bm{n} ~.
\end{equation}
Consequently, $N^{\alpha \beta}$ and $S^\alpha$ can also be interpreted as the in-plane and out-of-plane components of the stress tensor $\bm{\sigma}$.
In specifying $N^{\alpha \beta}$ and $S^\alpha$, we will have completely determined the total state of stress at any location on the membrane.
The in-plane tension $\hat{\sigma}$ is one-half the trace of the stress tensor \eqref{eq:cauchy-stress-tensor}, and as found in \cite{sauer-liquid-shell-corr-2016} is given by
\begin{equation} \label{eq:surface-tension}
	\hat{\sigma}
	= \dfrac{1}{2} \bm{\sigma} : \bm{i}
	= \dfrac{1}{2} N^{\alpha}_{\alpha} ~.
\end{equation}
The equation for the in-plane tension $\hat{\sigma}$ \eqref{eq:surface-tension} reinforces the notion that $N^{\alpha \beta}$ describes in-plane stresses and $S^\alpha$ describes out-of-plane stresses, as only $N^{\alpha \beta}$ enters equation \eqref{eq:surface-tension}.
\textspace

When solving for the strong forms of the dynamical equations of motion, we will need to consider the linear momentum balance \eqref{eq:local-momentum-balance} in component form.
In what follows, we decompose the equations of motion in the directions normal and tangential to the surface.
To this end, the body force $\rho \bm{b}$ may be expressed as
\begin{equation} \label{eq:body-force}
	\rho \bm{b}
	= p \bm{n}
	+ b^\alpha \bm{a}_\alpha ~,
\end{equation}
where $p$ is the pressure normal to the membrane and $b^\alpha$ are the in-plane contravariant components of the body force per unit mass.
To express $\bm{T}^\alpha_{; \alpha}$ in component form, we apply the Gauss \eqref{eq:gauss} and Weingarten \eqref{eq:weingarten} equations to the stress vector decomposition \eqref{eq:stress-vector-general} and obtain
\begin{equation} \label{eq:stress-vector-expanded-general}
	\bm{T}^\alpha_{; \alpha}
	= \Big(
		N^{\lambda \alpha}_{; \lambda}
		- S^\lambda b^\alpha_\lambda
	\Big) \bm{a}_\alpha
	+ \Big(
		N^{\alpha \beta} b_{\alpha \beta}
		+ S^\alpha_{; \alpha}
	\Big) \bm{n} ~.
\end{equation}
Substituting the body force decomposition \eqref{eq:body-force}, divergence of the stress vectors \eqref{eq:stress-vector-expanded-general}, and acceleration \eqref{eq:acceleration} into the local form of the linear momentum balance \eqref{eq:local-momentum-balance}, we find the tangential and normal momentum equations are given, respectively, by
\begin{align}
	\rho \big(
		v^\alpha_{, t}
		- v w^\alpha
		+ v^\lambda {w_\lambda}^\alpha
	\big) 
	&= \rho b^\alpha
	+ N^{\lambda \alpha}_{; \lambda}
	- S^\lambda b^\alpha_\lambda
	\label{eq:tangential-general-eoms-elastic-compressible} \\
	\shortintertext{and}
	\rho \big(
		v_{, t}
		+ v^\alpha w_\alpha
	\big) 
	&= p + N^{\alpha \beta} b_{\alpha \beta}
	+ S^\alpha_{; \alpha}
	~.
	\label{eq:normal-general-eom-elastic-compressible}
\end{align}
The normal component of the linear momentum balance \eqref{eq:normal-general-eom-elastic-compressible} is usually referred to as the shape equation \cite{ou-yang-1999, prost-prl-2002}.
\textspace

Although we do not yet know the form of the stresses $N^{\alpha \beta}$ and $S^\alpha$, from equations \eqref{eq:tangential-general-eoms-elastic-compressible} and \eqref{eq:normal-general-eom-elastic-compressible} we already see coupling between in-plane and out-of-plane membrane behavior.
The in-plane stresses $N^{\alpha \beta}$ and the out-of-plane shear $S^\alpha$ appear in the in-plane equations \eqref{eq:tangential-general-eoms-elastic-compressible} and the shape equation \eqref{eq:normal-general-eom-elastic-compressible}.
In general, we expect in-plane flow to influence out-of-plane bending and vice versa.
\\[-7pt]

The three components of the linear momentum balance \eqref{eq:tangential-general-eoms-elastic-compressible}--\eqref{eq:normal-general-eom-elastic-compressible} and the mass balance \eqref{eq:local-mass-balance} allow us to solve for the four fundamental unknowns: the density $\rho$ and the velocity components $v$ and $v^\alpha$.
To solve the equations of motion, however, we must first determine the forms of $N^{\alpha \beta}$ and $S^\alpha$.
We will now systematically determine the form of the in-plane and shear stresses before returning to the equations of motion.


\subsubsection{Angular Momentum Balance} \label{sec:angular-momentum-balance-1c}

\begin{figure}[!t]
	\centering
		\includegraphics[width=0.80\textwidth]{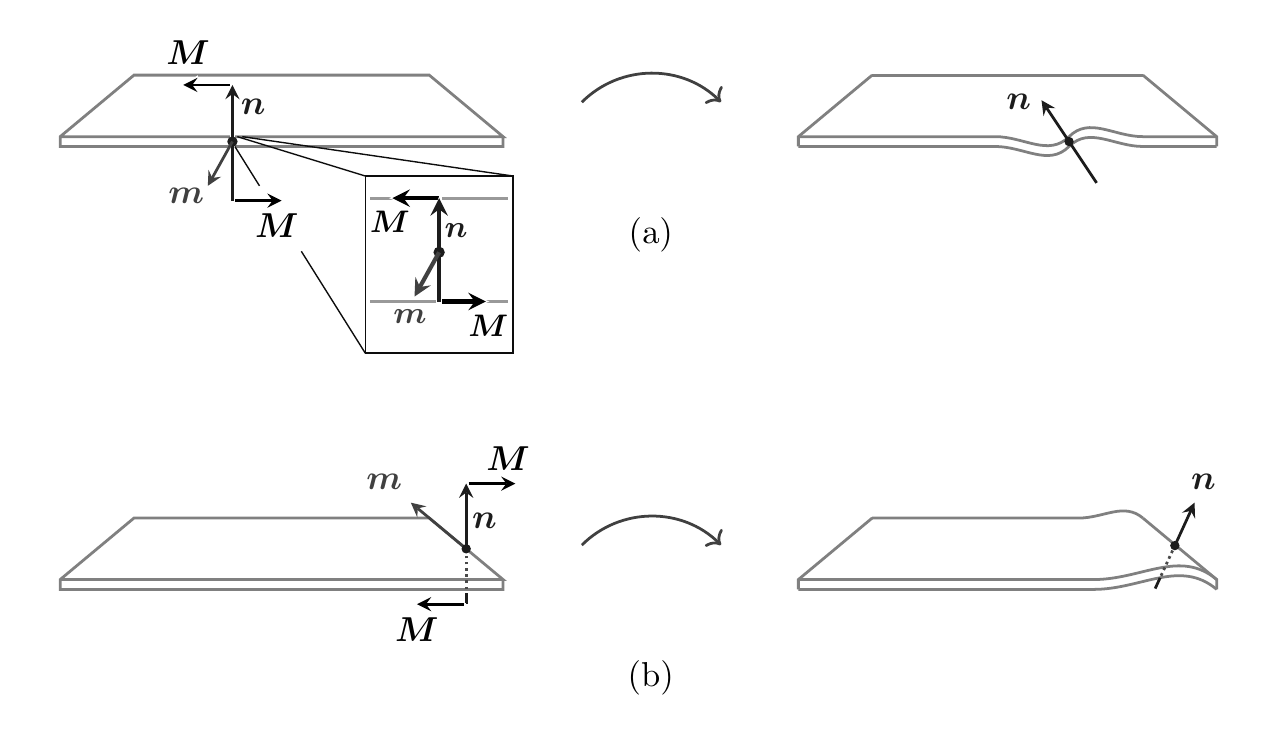}
	\captionsetup{width=0.80\textwidth}
	\caption{
		A director traction $\bm{M}$ acting on the unit normal $\bm{n}$ at the membrane boundary results in edge twisting (a) and bending (b).
		The director traction, which has units of couple per length, acts in the manner of a force on the dimensionless normal vector to produce a moment per length $\bm{m} = \bm{n} \times \bm{M}$.
		The inset in (a) shows a physical representation of the director tractions acting on the membrane.
		In general, the couple per length $\bm{m}$ acting on the membrane is a superposition of that shown in (a) and (b), and lies in the tangent plane.
	}
	\label{fig:moment}
\end{figure}

In this section, we analyze the balance of angular momentum of the membrane.
The rate of change of the total angular momentum of the membrane is equal to the sum of the external torques acting on the membrane patch.
In addition to the torques arising from body forces and boundary tractions, the membrane is able to sustain director tractions on its boundary.
These director tractions give rise to additional external moments on the membrane boundary which will twist the edges of the membrane patch, as depicted in Figure \ref{fig:moment}.
We will show such moments are necessary to sustain the shear stresses $S^\alpha$ introduced in the linear momentum balance.
Moreover, in the absence of director tractions the in-plane stresses are shown to be symmetric.
\textspace

In general, we specify a director field $\bm{d}$ on the membrane patch $\scp$ to account for the finite thickness of the lipid membrane \cite{steigmann-mms-1998, naghdi-1973-theory}.
The director $\bm{d} (\theta^\alpha, t)$ is a unit vector describing the orientation of the phospholipids---when the director $\bm{d}$ does not coincide with the normal to the surface $\bm{n}$, the phospholipids are tilted relative to the normal.
At a point $\bmxb$ on the patch boundary $\bscp$, the director traction $\bm{M} (\bmxb, t)$ describes the equal and opposite forces acting on the director $\bm{d} (\bmxb, t)$.
As the director $\bm{d}$ is dimensionless, the moment per length $\bm{m}$ at the patch boundary is given by the cross product $\bm{d} \times \bm{M}$.
\textspace

To properly account for the director field, it is necessary to include the director velocity $\dot{\bm{d}}$ in an additional balance law for the director momentum as described by Naghdi and Green \cite{naghdi-1973-theory, naghdi-iutam-1982, naghdi-prsla-1976}.
Including the director field would enable us to examine smaller length scale phenomena, for example transmembrane proteins causing phospholipids to tilt and inducing local inhomogeneities in the directors.
In this work, however, we choose not to study such phenomena and treat the membrane as a sheet of zero thickness.
In doing so, the director is forced to be equal to the normal at every point $\bm{x}$ and is prescribed to be 
\begin{equation} \label{eq:d-n}
	\bm{d} (\theta^\alpha, t)
	= \bm{n} (\theta^\alpha, t) ~.
\end{equation}
Note that equation \eqref{eq:d-n} is equivalent to the Kirchhoff-Love assumption \cite{steigmann-mms-1998}.
With this simplification, the moment per unit length of the patch boundary, $\bm{m}$, is given by
\begin{equation} \label{eq:boundary-moment-distributed-moment}
	\bm{m} = \bm{n} \times \bm{M} ~.
\end{equation}
Given equation \eqref{eq:boundary-moment-distributed-moment}, the global form of the angular momentum balance can be written as 
\begin{equation} \label{eq:angular-momentum-balance}
	\dfrac{\mathrm{d}}{\mathrm{d}t} \bigg(
		\int_\scp \rho \bm{x} \times \bm{v} ~\mathrm{d}a
	\bigg) 
	= \int_\scp \rho \bm{x} \times \bm{b} ~\mathrm{d}a
	+ \int_{\partial\scp} \!\! \Big(
		\bm{x} \times \bm{T} 
		+ \bm{n} \times \bm{M}
	\Big) ~\mathrm{d}s
	~,
\end{equation}
where $\rho \bm{x} \times \bm{v}$ denotes the angular momentum density at the point $\bm{x}$, and $\rho \bm{x} \times \bm{b}$ and $\bm{x} \times \bm{T}$ denote the torque densities due to body forces and tractions, respectively.
\textspace

While the director traction $\bm{M}$ may in general have normal and tangential components, the component in the normal direction has no effect on the resulting couple $\bm{m}$ due to equation \eqref{eq:boundary-moment-distributed-moment}.
Thus we restrict $\bm{M}$ to be in the plane of the membrane.
Once again using elementary curvilinear triangle arguments described by Naghdi \cite{naghdi-1973-theory}, the director traction $\bm{M}$ may be written as
\begin{equation} \label{eq:distributed-moment-decomposition}
	\bm{M} (\bmxb, t; \bm{\nu})
	= \bm{M}^\alpha (\bmxb, t) \, \nu_\alpha ~.
\end{equation}
The couple-stress vectors $\bm{M}^\alpha$ in equation \eqref{eq:distributed-moment-decomposition} must be in the plane of the membrane due to our imposed restriction, and may be written without loss of generality as
\begin{equation} \label{eq:stress-moment-decomposition}
	\bm{M}^\alpha
	= - M^{\alpha \beta} \bm{a}_\beta ~.
\end{equation}
Substituting the couple-stress decomposition \eqref{eq:stress-moment-decomposition} into the director traction decomposition \eqref{eq:distributed-moment-decomposition} allows us to write
\begin{equation} \label{eq:distributed-moment-tensor}
	\bm{M} = \bm{\mu}^{\textrm{T}} \, \bm{\nu} ~,
\end{equation}
where $\bm{\mu}$ is the couple-stress tensor given by
\begin{equation} \label{eq:couple-stress-tensor}
	\bm{\mu}
	= -M^{\alpha \beta} \bm{a}_\alpha \otimes \bm{a}_\beta ~.
\end{equation}
Because we require director tractions to not lie in the normal direction, the couple-stress tensor $\bm{\mu}$ in equation \eqref{eq:couple-stress-tensor} does not have any $\bm{a}_\alpha \otimes \bm{n}$ component.
\textspace

Returning to the global form of the angular momentum balance \eqref{eq:angular-momentum-balance}, we substitute the director traction decomposition \eqref{eq:distributed-moment-decomposition} and stress vector decomposition \eqref{eq:traction-decomposition} to obtain
\begin{equation} \label{eq:angular-momentum-balance-simple}
	\dfrac{\mathrm{d}}{\mathrm{d}t} \bigg(
		\int_\scp \rho \bm{x} \times \bm{v}~\mathrm{d}a
	\bigg)
	= \int_\scp \rho \bm{x} \times \bm{b}~\mathrm{d}a
	+ \int_{\partial\scp} \!\! \big(
		\bm{x} \times \bm{T}^\alpha
		+ \bm{n} \times \bm{M}^\alpha
	\big) \nu_\alpha ~\mathrm{d}s
	~.
\end{equation}
Using the Reynolds transport theorem \eqref{eq:rtt-density} and the surface divergence theorem \eqref{eq:divergence-theorem}, equation \eqref{eq:angular-momentum-balance-simple} simplifies to 
\begin{equation} \label{eq:angular-momentum-balance-simplest}
	\int_\scp \rho \bm{x} \times \dot{\bm{v}} ~\mathrm{d}a 
	= \int_\scp \Big(
		\rho \bm{x} \times \bm{b}
		+ \big(
			\bm{x} \times \bm{T}^\alpha
		\big)_{; \alpha} 
		+ \big(
			\bm{n} \times \bm{M}^\alpha
		\big)_{; \alpha} 
	\Big) ~\mathrm{d}a
	~.
\end{equation}
Since the membrane patch $\scp$ is arbitrary, the local form of the angular momentum balance can be obtained as
\begin{equation} \label{eq:angular-momentum-balance-local}
	\rho \bm{x} \times \dot{\bm{v}} 
	= \rho \bm{x} \times \bm{b}
	+ \bm{a}_\alpha \times \bm{T}^\alpha
	+ \bm{x} \times \bm{T}^\alpha_{; \alpha}
	- b^\beta_\alpha \bm{a}_\beta \times \bm{M}^\alpha
	+ \bm{n} \times \bm{M}^\alpha_{; \alpha}
	~,
\end{equation}
where we have distributed the covariant derivatives and used the Gauss \eqref{eq:gauss} and Weingarten \eqref{eq:weingarten} equations.
\textspace

It is useful to know what constraints the local form of the angular momentum balance \eqref{eq:angular-momentum-balance-local} imposes in addition to what was known from the linear momentum balance \eqref{eq:local-momentum-balance}.
Taking the cross product of $\bm{x}$ with the local linear momentum balance \eqref{eq:local-momentum-balance} and subtracting it from the local angular momentum balance \eqref{eq:angular-momentum-balance-local} gives 
\begin{equation} \label{eq:angular-linear-momentum-mix}
	\bm{a}_\alpha \times \bm{T}^\alpha
	- b^\beta_\alpha \bm{a}_\beta \times \bm{M}^\alpha
	+ \bm{n} \times \bm{M}^\alpha_{; \alpha}
	= \bm{0} ~.
\end{equation}
Substituting the couple-stress decomposition \eqref{eq:stress-moment-decomposition} and traction decomposition \eqref{eq:traction-decomposition} into equation \eqref{eq:angular-linear-momentum-mix}, we obtain
\begin{equation} \label{eq:angular-linear-momentum-balance-final}
	\bm{a}_\alpha \times
	\bigg[ \big( N^{\alpha \beta} - b^\beta_\mu M^{\mu \alpha} \big) \bm{a}_\beta
	+ \big( S^\alpha + M^{\beta \alpha}_{; \beta} \big) \bm{n} \bigg]
	= \bm{0} ~.
\end{equation}
Equation \eqref{eq:angular-linear-momentum-balance-final} indicates the following conditions must be true in order for both the linear momentum balance and the angular momentum balance to be locally satisfied:
\begin{align}
	& \sigma^{\alpha \beta} := \big( N^{\alpha \beta} - b^\beta_\mu M^{\mu \alpha} \big) \text{ is symmetric} 
	\label{eq:angular-momentum-symmetric}\\
	\shortintertext{and}
	& S^\alpha = - M^{\beta \alpha}_{; \beta} ~. \label{eq:angular-momentum-s}
\end{align}
In equation \eqref{eq:angular-momentum-symmetric}, the tensor $\sigma^{\alpha \beta}$ describes the components of in-plane tractions due to stretching and viscous flow only, i.e., $\sigma^{\alpha \beta}$ does not include contributions from moments.
This is to say the combination of angular and linear momentum balances impose restrictions between the in-plane stress components $N^{\alpha \beta}$, out-of-plane shear stress components $S^\alpha$, and the components of the couple-stress tensor $-M^{\alpha \beta}$.
As the boundary moment per length $\bm{m}$ is related to the components of $M^{\alpha \beta}$, equation \eqref{eq:angular-momentum-s} indicates the relationship between out-of-plane shear stresses and boundary moments.
If boundary moments had not been included, there would consequently be no shear stresses at any point on the membrane surface.
\textspace

Finally, it will be useful to express the boundary moment per length $\bm{m}$ in terms of the in-plane boundary tangent $\bm{\tau}$ and boundary normal $\bm{\nu}$ as
\begin{align}
	\bm{m} = m_\nu \bm{\nu} + m_\tau \bm{\tau} \label{eq:m-expanded} ~.
\end{align}
Using the identity $\bm{a}_\beta \times \bm{n} = \tau_\beta \bm{\nu} - \nu_\beta \bm{\tau}$, which can be derived from the decomposition of the in-plane unit normal $\bm{\nu}$ \eqref{eq:nu-decomposed} and in-plane unit tangent $\bm{\tau}$ \eqref{eq:tau-decomposed}, and substituting the director traction decomposition \eqref{eq:distributed-moment-decomposition} and couple-stress decomposition \eqref{eq:stress-moment-decomposition} into the equation for the moment per length $\bm{m}$ \eqref{eq:boundary-moment-distributed-moment}, we find the components of the boundary moment per length $\bm{m}$ to be
\begin{align}
	m_\nu &= M^{\alpha \beta} \nu_\alpha \tau_\beta \label{eq:m-nu-expanded} \\
	\shortintertext{and}
	m_\tau &= -M^{\alpha \beta} \nu_\alpha \nu_\beta \label{eq:m-tau-expanded} ~.
\end{align}
\eqnspace

At this stage, all previous works using either the balance law formulation \cite{kranthi-bmm-2012, kranthi-biophys-2014} or variational methods \cite{arroyo-pre-2009} propose constitutive forms of the in-plane viscous stresses and in-plane velocity gradients to model the irreversible processes of fluid flow.
These are then used to determine the equations of motion.
In our work, we will naturally find the constitutive form of the in-plane viscous stresses by evaluating the entropy production and proposing relationships between the thermodynamic forces and fluxes in the linear irreversible regime.
This framework based on entropy production is naturally extendable to multi-component systems and systems with chemical reactions.
In what follows, we proceed to develop such a framework.


\subsubsection{Mechanical Power Balance} \label{sec:mechanical-power-balance-1c}

While a mechanical power balance does not impose any new constraints on the membrane patch, it expresses the relationship between the kinetic energy, internal forces, and external forces, which is useful for the entropy production derivations in subsequent sections.
We begin by taking the dot product of the local momentum balance \eqref{eq:local-momentum-balance} with the velocity $\bm{v}$ and integrating over the membrane patch $\scp$ to obtain
\begin{equation} \label{eq:mechanical-power-balance-global}
	\int_\scp \rho \bm{v} \cdot \dot{\bm{v}} ~ \mathrm{d}a
	= \int_\scp \bm{v} \cdot \bm{T}^\alpha_{; \alpha} ~ \mathrm{d}a
	+ \int_\scp \rho \bm{v} \cdot \bm{b} ~ \mathrm{d}a
	~.
\end{equation}
\eqnspace

The left hand side of equation \eqref{eq:mechanical-power-balance-global} is the material derivative of the total kinetic energy, as an application of the Reynolds transport theorem \eqref{eq:rtt-density} shows 
\begin{equation} \label{eq:kinetic-energy-1c}
	\dfrac{\mathrm{d}}{\mathrm{d}t} \bigg(
		\int_\scp \dfrac{1}{2} \, \rho \bm{v} \cdot \bm{v} ~\mathrm{d}a
	\bigg)
	= \int_\scp \rho \bm{v} \cdot \dot{\bm{v}} ~\mathrm{d}a ~.
\end{equation}
The first term on the right hand side of equation \eqref{eq:mechanical-power-balance-global} may be expanded as
\begin{equation} \label{eq:v-dot-t-alpha-expansion}
	\begin{split}
		\int_\scp \bm{v} \cdot \bm{T}^\alpha_{; \alpha} ~\mathrm{d}a
		&= \int_\scp \Big(
			\big(
				\bm{v} \cdot \bm{T}^\alpha
			\big)_{; \alpha}
			- \bm{v}_{, \alpha} \cdot \bm{T}^\alpha
		\Big) ~\mathrm{d}a \\[8pt]
		&= \int_{\partial \scp} \!\! \big(
			\bm{v} \cdot \bm{T}^\alpha
		\big) \nu_\alpha ~\mathrm{d}s
		- \int_\scp \bm{v}_{, \alpha} \cdot \bm{T}^\alpha ~\mathrm{d}a \\[8pt]
		&= \int_{\partial \scp} \!\! \bm{v} \cdot \bm{T} ~\mathrm{d}s
		- \int_\scp \bm{v}_{, \alpha} \cdot \bm{T}^\alpha ~\mathrm{d}a ~,
	\end{split}
\end{equation}	
where the second equality is obtained by invoking the surface divergence theorem \eqref{eq:divergence-theorem} and the third equality from the boundary traction decomposition \eqref{eq:traction-decomposition}.
By expanding the integrand of the last term in equation \eqref{eq:v-dot-t-alpha-expansion}, we find  
\begin{equation} \label{eq:v-alpha-dot-t-alpha-expansion}
	\begin{split}
		\bm{v}_{, \alpha} \cdot \bm{T}^\alpha
		&= \big(
			w_{\alpha \beta} \bm{a}^\beta + w_\alpha \bm{n}
		\big) \cdot \big(
			N^{\alpha \mu} \bm{a}_\mu + S^\alpha \bm{n}
		\big) \\[8pt]
		&= N^{\alpha \beta} w_{\alpha \beta} + S^\alpha w_\alpha \\[8pt]
		&= \sigma^{\alpha \beta} w_{\alpha \beta}
		+ b^\beta_\mu M^{\mu \alpha} w_{\alpha \beta}
		- M^{\beta \alpha}_{; \beta} w_\alpha ~,
	\end{split}
\end{equation}
where the first equality is obtained with the relation for $\bm{v}_{, \alpha}$ \eqref{eq:a-alpha-dot} and $\bm{T}^\alpha$ \eqref{eq:stress-vector-general}, and the third equality by substituting the results of the angular momentum balance \eqref{eq:angular-momentum-symmetric}--\eqref{eq:angular-momentum-s}.
Using the symmetry of $\sigma^{\alpha \beta}$ found in equation \eqref{eq:angular-momentum-symmetric} and the relation for $\dot{a}_{\alpha \beta}$ \eqref{eq:a-alpha-beta-dot}, the first term in the final equality of equation \eqref{eq:v-alpha-dot-t-alpha-expansion} may be written as $\sigma^{\alpha \beta} w_{\alpha \beta} = \tfrac{1}{2} \sigma^{\alpha \beta} (w_{\alpha \beta} + w_{\beta \alpha}) = \tfrac{1}{2} \sigma^{\alpha \beta} \dot{a}_{\alpha \beta}$.
Using the product rule on the last term in the final equality of equation \eqref{eq:v-alpha-dot-t-alpha-expansion} gives
$M^{\beta \alpha}_{; \beta} w_\alpha = (M^{\beta \alpha} w_\alpha)_{; \beta} - M^{\beta \alpha} w_{\alpha; \beta}$.
Using these simplifications, equation \eqref{eq:v-alpha-dot-t-alpha-expansion} may be rewritten as
\begin{equation} \label{eq:v-alpha-dot-t-alpha-step-2}
	\bm{v}_{, \alpha} \cdot \bm{T}^\alpha
	= \dfrac{1}{2} \sigma^{\alpha \beta} \dot{a}_{\alpha \beta}
	+ M^{\mu \alpha} \Big(
		w_{\alpha \beta} b^\beta_\mu
		+ w_{\alpha; \mu}
	\Big) - \big(
		M^{\beta \alpha} w_\alpha
	\big)_{; \beta} ~.
\end{equation}
The second term on the right hand side of equation \eqref{eq:v-alpha-dot-t-alpha-step-2} is $M^{\mu \alpha} \dot{b}_{\mu \alpha}$, given the relation for $\dot{b}_{\alpha \beta}$ in equation \eqref{eq:b-alpha-beta-dot}.
We rewrite the last term in equation \eqref{eq:v-alpha-dot-t-alpha-step-2} as
\begin{equation} \label{eq:m-beta-alpha-w-alpha-manipulation}
	\big( M^{\beta \alpha} w_\alpha \big)_{; \beta}
	= \big( M^{\beta \alpha} w_\lambda \delta^\lambda_\alpha \big)_{; \beta}
	= \big( M^{\beta \alpha} \bm{a}_\alpha \cdot w_\lambda \bm{a}^\lambda \big)_{; \beta}
	= \big( \bm{M}^\beta \cdot \dot{\bm{n}} \big)_{; \beta} ~.
\end{equation}
With the above simplifications, we find equation \eqref{eq:v-alpha-dot-t-alpha-step-2} reduces to
\begin{equation} \label{eq:v-alpha-dot-t-alpha-final}
	\bm{v}_{, \alpha} \cdot \bm{T}^\alpha
	= \dfrac{1}{2} \sigma^{\alpha \beta} \dot{a}_{\alpha \beta}
	+ M^{\alpha \beta} \dot{b}_{\alpha \beta}
	- \big(
		\dot{\bm{n}} \cdot \bm{M}^\alpha
	\big)_{; \alpha} ~.
\end{equation}

Using equation \eqref{eq:v-alpha-dot-t-alpha-final}, equation \eqref{eq:v-dot-t-alpha-expansion} can be written as
\begin{equation} \label{eq:v-dot-t-alpha-identity}
	\begin{split}
		\int_\scp \bm{v} \cdot \bm{T}^\alpha_{; \alpha} ~\mathrm{d}a
		&= \int_{\partial \scp} \!\! \bm{v} \cdot \bm{T} ~\mathrm{d}s
		- \int_\scp \bigg(
			\dfrac{1}{2} \sigma^{\alpha \beta} \dot{a}_{\alpha \beta}
			+ M^{\alpha \beta} \dot{b}_{\alpha \beta}
			- \left( \dot{\bm{n}} \cdot M^\alpha \right)_{; \alpha}
		\bigg) ~\mathrm{d}a \\[8pt]
		&= \int_{\partial \scp} \!\! \big(
			\bm{v} \cdot \bm{T}
			+ \dot{\bm{n}} \cdot \bm{M}
		\big) ~\mathrm{d}s
		- \int_\scp \bigg(
			\dfrac{1}{2} \sigma^{\alpha \beta} \dot{a}_{\alpha \beta}
			+ M^{\alpha \beta} \dot{b}_{\alpha \beta}
		\bigg) ~\mathrm{d}a ~,
	\end{split}	
\end{equation}
where the second equality is obtained by using the surface divergence theorem \eqref{eq:divergence-theorem}.
Substituting equations \eqref{eq:v-dot-t-alpha-identity} and \eqref{eq:kinetic-energy-1c} into equation \eqref{eq:mechanical-power-balance-global}, we find the total mechanical power balance is given by
\begin{equation} \label{eq:global-power-balance-solution}
	\dfrac{\mathrm{d}}{\mathrm{d}t} \bigg(
		\int_\scp \dfrac{1}{2} \rho \bm{v} \cdot \bm{v} ~\mathrm{d}a
	\bigg)
	+ \int_\scp \bigg(
		\dfrac{1}{2} \sigma^{\alpha \beta} \dot{a}_{\alpha \beta} 
		+ M^{\alpha \beta} \dot{b}_{\alpha \beta} 
	\bigg) ~\mathrm{d}a 
	= \int_{\partial \scp} \!\! \Big(
		\bm{v} \cdot \bm{T} 
		+ \dot{\bm{n}} \cdot \bm{M} 
	\Big) ~\mathrm{d}s
	+ \int_\scp \rho\bm{v}\cdot\bm{b}~\mathrm{d}a
	~.
\end{equation}
The left hand side of equation \eqref{eq:global-power-balance-solution} contains the material derivative of the kinetic energy \eqref{eq:kinetic-energy-1c} and a term describing the internal changes involving the shape and stresses of the membrane, which describe the membrane's internal power.
The terms on the right hand side of the mechanical power balance \eqref{eq:global-power-balance-solution} describe the power due to external forces and moments acting on the membrane.


\subsection{Thermodynamics}

In this section, we develop the thermodynamic framework necessary to understand the effects of bending and intra-membrane viscous flow on the membrane patch.
We develop local forms of the first law of thermodynamics and entropy balance, and we introduce the second law of thermodynamics.
We follow the procedure described by de Groot \& Mazur \cite{degroot-mazur} to understand the internal entropy production, albeit with one difference.
While de Groot \& Mazur \cite{degroot-mazur} begin with the local equilibrium assumption and the Gibbs equation, it is technically difficult to write the Gibbs equation for a system which depends on tensorial quantities.
In this work, we follow the approach demonstrated in \cite{kranthi-thesis} and begin by choosing the appropriate form of the Helmholtz free energy.
Following this framework, one can derive an effective Gibbs equation after the analysis is complete.


\subsubsection{First Law---Energy Balance}

According to the first law of thermodynamics, the total energy of the membrane patch changes due to work being done on the membrane or heat flowing into the membrane.
The mechanical power balance \eqref{eq:global-power-balance-solution} describes the rate of work being done on the membrane due to external tractions, moments, and forces.
Furthermore, heat may enter or exit the membrane patch in one of two ways:
by flowing from the surrounding medium into the membrane along the normal direction $\bm{n}$, or by flowing in the plane of the membrane across the membrane patch boundary.
We denote the heat source per unit mass as $r(\theta^\alpha, t)$, which accounts for the heat flow from the bulk, and the in-plane heat flux as $\bmjq = \jq^{\, \alpha} \, \bm{a}_\alpha$.
By convention, the heat flux $\bmjq$ is positive when heat flows out of the system across the patch boundary.
Defining $e (\theta^\alpha, t)$ to be the total energy per unit mass of the membrane, the global form of the first law of thermodynamics can be written as 
\begin{equation} \label{eq:global-energy-balance}
	\dfrac{\mathrm{d}}{\mathrm{d}t} \bigg(
		\int_\scp \rho e ~\mathrm{d}a
	\bigg)
	= \int_\scp \rho r ~\mathrm{d}a 
	- \int_{\partial \scp} \!\! \bmjq \cdot \bm{\nu} ~\mathrm{d}s
	+ \int_\scp \rho \bm{v} \cdot {\bm{b}} ~\mathrm{d}a
	+ \int_{\partial \scp} \!\! \Big(
		\bm{v} \cdot \bm{T} 
		+ {\dot{\bm{n}}} \cdot \bm{M}
	\Big) ~\mathrm{d}s ~.
\end{equation}

The total energy per unit mass $e$ consists of the internal energy per unit mass $u$ and the kinetic energy per unit mass $\tfrac{1}{2} \bm{v} \cdot \bm{v}$, and is given by
\begin{equation} \label{eq:energy-def}
	\rho e := \rho u + \dfrac{1}{2} \rho \bm{v} \cdot \bm{v} ~.
\end{equation}
Using the Reynolds transport theorem \eqref{eq:rtt-density} and substituting the expression for the total energy per mass $e (\theta^\alpha, t)$ \eqref{eq:energy-def} into equation \eqref{eq:global-energy-balance}, we obtain 
\begin{equation} \label{eq:global-energy-balance-step-1}
	\int_\scp \big( \rho \dot{u} + \rho \bm{v} \cdot \dot{\bm{v}} \big) ~\mathrm{d}a
	= \int_\scp \rho r ~\mathrm{d}a 
	- \int_{\partial \scp} \!\! \bmjq \cdot \bm{\nu} ~\mathrm{d}s 
	+ \int_\scp \rho \bm{v} \cdot {\bm{b}} ~\mathrm{d}a
	+ \int_{\partial \scp} \!\! \Big(
		\bm{v} \cdot \bm{T} 
		+ \dot{\bm{n}} \cdot \bm{M}
	\Big) ~\mathrm{d}s ~.
\end{equation}
Equation \eqref{eq:global-energy-balance-step-1} shares several terms with the mechanical power balance \eqref{eq:global-power-balance-solution}, and by subtracting the two equations, the balance of internal energy can be obtained as 
\begin{equation} \label{eq:global-energy-balance-step-2}
	\begin{split}
		\int_\scp \rho \dot{u}  ~\mathrm{d}a
		&= \int_\scp \rho r ~\mathrm{d}a 
		- \int_{\partial \scp} \!\! \bmjq \cdot \bm{\nu} ~\mathrm{d}s 
		+ \int_\scp \bigg(
			\dfrac{1}{2} \sigma^{\alpha \beta} \dot{a}_{\alpha \beta} 
			+ M^{\alpha \beta} \dot{b}_{\alpha \beta} 
		\bigg) ~\mathrm{d}a \\[4pt]
		&= \int_\scp \bigg(
			\rho r
			- J_{\mathrm{q} \, ; \alpha}^{\, \alpha} 
			+ \dfrac{1}{2} \sigma^{\alpha \beta} \dot{a}_{\alpha \beta} 
			+ M^{\alpha \beta} \dot{b}_{\alpha \beta} 
		\bigg) ~\mathrm{d}a ~,
	\end{split}	
\end{equation}
where the second equality is obtained by using the surface divergence theorem \eqref{eq:divergence-theorem}.
Since the membrane patch $\scp$ is arbitrary, the local form of the internal energy balance is given by
\begin{equation} \label{eq:local-energy-balance}
	\rho \dot{u}
	= \rho r
	- J_{\mathrm{q} \, ; \alpha}^{\, \alpha}
	+ \dfrac{1}{2} \sigma^{\alpha \beta} \dot{a}_{\alpha \beta}
	+ M^{\alpha \beta} \dot{b}_{\alpha \beta} ~.
\end{equation}
The first two terms on the right hand side of equation \eqref{eq:local-energy-balance} describe the heat flow into the system, and the last two terms describe the energy change due to work being done on the system.


\subsubsection{Entropy Balance \& Second Law}

The total entropy of a membrane patch $\scp$ may change in three ways: entropy may flow into or out of the patch across the membrane boundary, entropy may be absorbed or emitted from the membrane body as a supply, or entropy may be produced internally within the membrane patch.
The local quantities corresponding to such changes are the in-plane entropy flux $\bmjs = J_{\mathrm{s}}^{\, \alpha} \bm{a}_\alpha$, the rate of external entropy supply per unit mass $\eta_\mathrm{e} (\theta^\alpha, t)$, and the rate of internal entropy production per unit mass $\eta_\mathrm{i} (\theta^\alpha, t)$, respectively.
For the total entropy per unit mass $s (\theta^\alpha, t)$, the global form of the entropy balance is given by
\begin{equation} \label{eq:global-entropy-balance}
	\dfrac{\mathrm{d}}{\mathrm{d}t} \bigg(
		\int_\scp \rho s ~\mathrm{d}a
	\bigg)
	= -\int_{\partial \scp} \bmjs \cdot \bm{\nu} ~\mathrm{d}s
	+ \int_\scp \Big(
		\rho \eta_\mathrm{e} 
		+ \rho \eta_\mathrm{i}
	\Big) ~\mathrm{d}a ~.
\end{equation}

Applying the Reynolds transport theorem \eqref{eq:rtt-density} and the surface divergence theorem \eqref{eq:divergence-theorem} reduces equation \eqref{eq:global-entropy-balance} to 
\begin{equation} \label{eq:global-entropy-step-1}
	\int_\scp \rho \dot{s} ~\mathrm{d}a
	= \int_\scp \Big(
		-J_{\mathrm{s} \, ; \alpha}^{ \, \alpha}
		+ \rho \eta_{\mathrm{e}}
		+ \rho \eta_{\mathrm{i}}
	\Big) ~\mathrm{d}a ~.
\end{equation}
Again, due to the arbitrariness of the membrane patch $\scp$, the local form of the entropy balance is given by
\begin{equation} \label{eq:local-entropy-balance}
	\rho \dot{s}
	= - J_{\mathrm{s} \, ; \alpha}^{\, \alpha}
	+ \rho \eta_{\mathrm{e}}
	+ \rho \eta_{\mathrm{i}} ~.
\end{equation}
\eqnspace

At this point, it is useful to consider the nature of the entropy flux, external entropy supply, and internal entropy production.
We define the in-plane entropy flux $\bmjs$ and the external entropy supply per unit mass $\eta_{\mathrm{e}}$ to describe  the redistribution of entropy that has already been created.
These terms may be positive or negative.
We now introduce the second law of thermodynamics by requiring the internal entropy production to be non-negative at every point in the membrane.
The second law of thermodynamics is given by
\begin{equation} \label{eq:second-law-thermo-local}
	\eta_{\mathrm{i}} \ge 0 ~.
\end{equation}
The internal entropy production \eqref{eq:second-law-thermo-local} is zero only for reversible processes.


\subsubsection{Choice of Thermodynamic Potential} \label{sec:sc-choice-thermo-potential}

The natural thermodynamic potential for the membrane patch is the Helmholtz free energy \cite{steigmann-fluid-film-arma-1999}.
The Helmholtz free energy per unit mass, $\psi$, is given by
\begin{equation} \label{eq:helmholtz}
	\psi = u - T s ~,
\end{equation}
where $T(\theta^\alpha, t)$ is the local temperature of the membrane patch.
Taking the material derivative of equation \eqref{eq:helmholtz}, solving for $\dot{s}$, and substituting into the local entropy balance \eqref{eq:local-entropy-balance}, we obtain
\begin{equation} \label{eq:entropy-balance-reformulated}
	\rho \dot{s}
	= - J_{\mathrm{s} \, ; \alpha}^{\, \alpha}
	+ \rho \eta_{\mathrm{e}}
	+ \rho \eta_{\mathrm{i}}
	= \dfrac{1}{T} \Big(
		\rho \dot{u}
		- \rho \dot{T} s
		- \rho \dot{\psi}
	\Big) ~.
\end{equation}
Substituting the local form of the first law of thermodynamics \eqref{eq:local-energy-balance} into equation \eqref{eq:entropy-balance-reformulated} yields the total rate of change of entropy, given by 
\begin{equation} \label{eq:entropy-energy-balance}
	\rho \dot{s}
	= - J_{\mathrm{s} \, ; \alpha}^{\, \alpha}
	+ \rho \eta_{\mathrm{e}}
	+ \rho \eta_{\mathrm{i}}
	= \dfrac{1}{T} \bigg(
		\rho r
		- J_{\mathrm{q} \, ; \alpha}^{\, \alpha}
		+ \dfrac{1}{2} \sigma^{\alpha \beta} \dot{a}_{\alpha \beta}
		+ M^{\alpha \beta} \dot{b}_{\alpha \beta}
		- \rho \dot{T} s
		- \rho \dot{\psi}
	\bigg) ~.
\end{equation}
Equation \eqref{eq:entropy-energy-balance} will allow us to determine which terms contribute to the internal entropy production, understand fundamental relationships between the stresses, moments, and energetics of the membrane patch, and finally develop constitutive relations between the stresses, moments, and associated kinematic quantities.


\subsection{Constitutive Relations}

In this section, we choose the fundamental thermodynamic variables for our membrane patch.
With this constitutive assumption, we determine the contributions to the entropy flux, external entropy supply, and internal entropy production.
We then apply linear irreversible thermodynamics to relate generalized thermodynamic forces to their corresponding fluxes.
In doing so, we naturally determine the viscous dissipation due to in-plane fluid flow as well as the dependence of the stresses and moments on the Helmholtz free energy density.


\subsubsection{General Thermodynamic Variables} \label{sec:general-thermo-vars}

Lipid bilayers have in-plane dissipative flow and out-of-plane elastic bending.
The Helmholtz free energy per unit mass $\psi$, as a thermodynamic state function, captures the elastic behavior of lipid membranes.
The general thermodynamic variables that the Helmholtz free energy density of a two-dimensional elastic sheet depends on are the metric tensor $a_{\alpha \beta}$, curvature tensor $b_{\alpha \beta}$, and temperature $T$ \cite{naghdi-1973-theory, steigmann-fluid-film-arma-1999}.
The simplest form of the Helmholtz free energy density $\psi$ that captures this behavior is given by
\begin{equation} \label{eq:psi-constitutive-elastic-compressible}
	\psi = \psi(a_{\alpha\beta}, b_{\alpha\beta}, T) ~.
\end{equation}
In this work, we assume the membrane does not thermally expand or chemically swell, so the metric and curvature tensors capture only elastic behavior.
\textspace

Because the metric and curvature tensors are symmetric, the material derivative of the Helmholtz free energy density $\psi$ is given by
\begin{equation} \label{eq:psi-dot}
	\dot{\psi}
	= \dfrac{1}{2} \Big(
		\dfrac{\partial \psi}{\partial a_{\alpha \beta}}
		+ \dfrac{\partial \psi}{\partial a_{\beta \alpha}}
	\Big) \dot{a}_{\alpha \beta}
	+ \dfrac{1}{2} \Big(
		\dfrac{\partial \psi}{\partial b_{\alpha \beta}}
		+ \dfrac{\partial \psi}{\partial b_{\beta \alpha}}
	\Big) \dot{b}_{\alpha \beta}
	+ \dfrac{\partial \psi}{\partial T} \dot{T} ~.
\end{equation}
Substituting equation \eqref{eq:psi-dot} into the local entropy balance \eqref{eq:entropy-energy-balance}, we obtain
\begin{equation} \label{eq:second-law-thermo-local-expanded}
	\begin{split}
		\rho \dot{s}
		&= - J_{\mathrm{s} \, ; \alpha}^{\, \alpha} 
		+ \rho \eta_{\mathrm{e}}
		+ \rho \eta_{\mathrm{i}} \\[3pt]
		&= \dfrac{1}{T} \bigg\{
			\, \rho r
			- J_{\mathrm{q} \, ; \alpha}^{\, \alpha}
			- \rho \dot{T} \Big(
				s + \dfrac{\partial \psi}{\partial T}
			\Big)
			\\[3pt]
			& \hspace{42pt} + \dfrac{1}{2} \bigg[
				\sigma^{\alpha \beta}
				- \rho \Big(
					\dfrac{\partial \psi}{\partial a_{\alpha \beta}}
					+ \dfrac{\partial \psi}{\partial a_{\beta \alpha}}
				\Big)
			\bigg] \dot{a}_{\alpha \beta}
			+ \bigg[
				M^{\alpha \beta}
				- \dfrac{\rho}{2} \Big(
					\dfrac{\partial \psi}{\partial b_{\alpha \beta}}
					+ \dfrac{\partial \psi}{\partial b_{\beta \alpha}}
				\Big)
			\bigg] \dot{b}_{\alpha \beta} \,
		\bigg\} ~.
	\end{split}
\end{equation}
At this stage, we assume the system is locally at equilibrium, and therefore define the entropy as
\begin{equation} \label{eq:local-equilibrium-entropy-sc}
	s
	= - \Big(
		\dfrac{\partial \psi}{\partial T}
	\Big)_{a_{\alpha \beta}, \, b_{\alpha \beta}}
	~,
\end{equation}
where the partial derivative is taken at constant $a_{\alpha \beta}$ and $b_{\alpha \beta}$.
Rewriting the heat flux and using equation \eqref{eq:local-equilibrium-entropy-sc} reduces equation \eqref{eq:second-law-thermo-local-expanded} to
\begin{equation} \label{eq:second-law-thermo-local-expanded-gradients}
	\begin{split}
		\rho \dot{s} 
		&= - J_{\mathrm{s} \, ; \alpha}^{\, \alpha} 
		+ \rho \eta_{\mathrm{e}}
		+ \rho \eta_{\mathrm{i}} \\[3pt]
		&= - \Big(
			\dfrac{\jq^{\, \alpha}}{T}
		\Big)_{; \alpha} 
		+ \dfrac{\rho r}{T}
		- \dfrac{\jq^{ \, \alpha} \, T_{, \alpha}}{T^2} \\[3pt]
		& \hspace{25pt} + \dfrac{1}{T} \bigg\{ \,
			\dfrac{1}{2} \bigg[
				\sigma^{\alpha \beta}
				- \rho \Big(
					\dfrac{\partial \psi}{\partial a_{\alpha \beta}}
					+ \dfrac{\partial \psi}{\partial a_{\beta \alpha}}
				\Big)
			\bigg] \dot{a}_{\alpha \beta}
			+ \bigg[
				M^{\alpha \beta}
				- \dfrac{\rho}{2} \Big(
					\dfrac{\partial \psi}{\partial b_{\alpha \beta}}
					+ \dfrac{\partial \psi}{\partial b_{\beta \alpha}}
				\Big)
			\bigg] \dot{b}_{\alpha \beta} \,
		\bigg\} ~.
	\end{split}
\end{equation}
From dimensional arguments, only gradients on the right hand side may contribute to the in-plane entropy flux components $J_{\mathrm{s}}^{ \, \alpha}$, which are given by
\begin{equation} \label{eq:entropy-flux-1c}
	J_{\mathrm{s}}^{ \, \alpha} = \dfrac{\jq^{ \, \alpha}}{T} ~.
\end{equation}
The external entropy supply per unit area $\rho \eta_{\mathrm{e}}$ captures entropy being absorbed or emitted across the membrane body.
The only term on the right hand side which describes such a change is the heat source $r$.
Therefore, the external entropy per unit area $\rho \eta_{\mathrm{e}}$ is given by
\begin{equation} \label{eq:external-entropy-1c}
	\rho \eta_{\mathrm{e}} = \dfrac{\rho r}{T} ~.
\end{equation}
In equations \eqref{eq:entropy-flux-1c} and \eqref{eq:external-entropy-1c}, we obtain the familiar result that heat flow into or out of the system is associated with an entropy change.
\textspace

As we have determined the terms on the right hand side of equation \eqref{eq:second-law-thermo-local-expanded-gradients} that contribute to the entropy flux and external entropy, the remaining terms contribute to the internal entropy production.
To this end, the rate of internal entropy production per unit area $\rho \eta_{\mathrm{i}}$ is given by
\begin{equation} \label{eq:internal-entropy-complete}
	\begin{split}
		\rho \eta_{\mathrm{i}} 
		= - \dfrac{\jq^{ \, \alpha} \, T_{, \alpha}}{T^2} 
		+ \dfrac{1}{T} \bigg\{
			\dfrac{1}{2} \bigg[
				\sigma^{\alpha \beta}
				- \rho \Big(
					\dfrac{\partial \psi}{\partial a_{\alpha \beta}}
					+ \dfrac{\partial \psi}{\partial a_{\beta \alpha}}
				\Big)
			\bigg] \dot{a}_{\alpha \beta}
			+ \bigg[
				M^{\alpha \beta}
				- \dfrac{\rho}{2} \Big(
					\dfrac{\partial \psi}{\partial b_{\alpha \beta}}
					+ \dfrac{\partial \psi}{\partial b_{\beta \alpha}}
				\Big)
			\bigg] \dot{b}_{\alpha \beta} 
		\bigg\} ~.
	\end{split}
\end{equation}
The terms on the right hand side of equation \eqref{eq:internal-entropy-complete} are a product of a thermodynamic force, which may be imposed on the system, and a thermodynamic flux.
Denoting the thermodynamic force as $X_k$ and the corresponding flux as $J^k$, equation \eqref{eq:internal-entropy-complete} may be generally written as 
\begin{equation} \label{eq:thermo-force-rate}
	\rho \eta_{\mathrm{i}}
	= J^k \, X_k
	\ge 0 ~.
\end{equation}
In equation \eqref{eq:thermo-force-rate}, the indices $k$ are used as a label, as $X_k$ and $J^k$ may be scalars, vectors, or tensors.
As described by Prigogine \cite{prigogine} and de Groot \& Mazur \cite{degroot-mazur}, we assume in the linear irreversible regime, \emph{i.e.}, near equilibrium, there is a linear relationship between $X_k$ and $J^k$ given by 
\begin{equation} \label{eq:phenomenological}
	J^i
	= L^{i k} X_k ~,
\end{equation}
where $L^{i k}$ are the phenomenological coefficients.
\textspace

In the internal entropy production \eqref{eq:internal-entropy-complete}, there are three thermodynamic forces: the in-plane temperature gradient $T_{, \alpha}$ and the material derivatives of the metric and curvature tensor, $\dot{a}_{\alpha \beta}$ and $\dot{b}_{\alpha \beta}$, respectively.
We invoke the Curie principle \cite{curie}, as done by Prigogine \cite{prigogine} and de Groot \& Mazur \cite{degroot-mazur}, and propose that the phenomenological coefficients between quantities with different tensorial order must be zero.
Therefore, the heat flux $\jq^{\, \alpha}$ is independent of the tensorial forces $\dot{a}_{\alpha \beta}$ and $\dot{b}_{\alpha \beta}$.
Similarly, the stresses and moments are independent of the temperature gradients.
In the spirit of equation \eqref{eq:phenomenological}, the phenomenological relation for the heat flux is then given by
\begin{equation} \label{eq:heat-flux-general}
	\jq^{\, \alpha}
	= - \kappa^{\alpha \beta} T_{, \beta} ~,
\end{equation}
where the tensor $\kappa^{\alpha \beta}$ is the thermal conductivity tensor.
As the fluid is thermally isotropic in-plane, $\kappa^{\alpha \beta} = \kappa \, a^{\alpha \beta}$, where the constant $\kappa$ is the scalar thermal conductivity.
In this case, equation \eqref{eq:heat-flux-general} reduces to
\begin{equation} \label{eq:heat-flux-1c}
	\jq^{ \, \alpha} = - \kappa \, T^{, \alpha} ~,
\end{equation}
where we use the shorthand $T^{, \alpha}$ to denote $T_{, \beta} \, a^{\alpha \beta}$.
We note that in the case of lipid bilayers, there are usually no temperature gradients and equation \eqref{eq:heat-flux-1c} does not play a major role in describing the relevant irreversible processes.
\textspace

To obtain the remaining phenomenological coefficients associated with the other irreversible processes in equation \eqref{eq:internal-entropy-complete}, we define the thermodynamic fluxes
\begin{align}
	\pi^{\alpha \beta}
	&= \sigma^{\alpha \beta}
	- \rho \Big(
		\dfrac{\partial \psi}{\partial a_{\alpha \beta}}
		+ \dfrac{\partial \psi}{\partial a_{\beta \alpha}}
	\Big) \label{eq:pi-alpha-beta-def}
	\\
	\shortintertext{and}
	\omega^{\alpha \beta}
	&= M^{\alpha \beta}
	- \dfrac{\rho}{2} \Big(
	\dfrac{\partial \psi}{\partial b_{\alpha \beta}}
		+ \dfrac{\partial \psi}{\partial b_{\beta \alpha}}
	\Big) 
	\label{eq:omega-alpha-beta-def}
\end{align}	
for notational convenience.
In the linear irreversible regime, the phenomenological relations relating $\pi^{\alpha \beta}$ and $\omega^{\alpha \beta}$ to $\dot{a}_{\alpha \beta}$ and $\dot{b}_{\alpha \beta}$ can be generally written as
\begin{align}
	\pi^{\alpha \beta}
	= R^{\alpha \beta \gamma \mu} \, \dot{a}_{\gamma \mu}
	+ S^{\alpha \beta \gamma \mu} \, \dot{b}_{\gamma \mu} 
	\label{eq:pi-phenom-gen}
	\\
	\shortintertext{and}
	\omega^{\alpha \beta}
	= T^{\alpha \beta \gamma \mu} \, \dot{a}_{\gamma \mu}
	+ U^{\alpha \beta \gamma \mu} \, \dot{b}_{\gamma \mu} ~,
	\label{eq:omega-phenom-gen}
\end{align}
where the fourth-order contravariant tensors $R^{\alpha \beta \gamma \mu}$, $S^{\alpha \beta \gamma \mu}$, $T^{\alpha \beta \gamma \mu}$, and $U^{\alpha \beta \gamma \mu}$ are general fourth-order phenomenological viscous coefficients.
The tensors $S^{\alpha \beta \gamma \mu}$ and $T^{\alpha \beta \gamma \mu}$ describe interference between the two irreversible processes driven by $\dot{a}_{\alpha \beta}$ and $\dot{b}_{\alpha \beta}$, and we assume them to be zero for the case of lipid bilayers. 
The phenomenological relations \eqref{eq:pi-phenom-gen}--\eqref{eq:omega-phenom-gen} then reduce to 
\begin{align}
	\pi^{\alpha \beta}
	&= R^{\alpha \beta \gamma \mu} \, \dot{a}_{\gamma \mu} 
	\label{eq:pi-phenom-no-cross}
	\\
	\shortintertext{and}
	\omega^{\alpha \beta}
	&= U^{\alpha \beta \gamma \mu} \, \dot{b}_{\gamma \mu} ~.
	\label{eq:omega-phenom-no-cross}
\end{align}
Given the form of the internal entropy production \eqref{eq:internal-entropy-complete}, equations \eqref{eq:pi-phenom-no-cross} and \eqref{eq:omega-phenom-no-cross} indicate $\pi^{\alpha \beta}$ captures the dissipation due to in-plane flow and $\omega^{\alpha \beta}$ captures the dissipation due to out-of-plane bending.
In general, $U^{\alpha \beta \gamma \mu}$ need not be equal to zero and bending can provide another way by which the membrane dissipates energy.
However, we assume out-of-plane-bending is not a dissipative process and so $U^{\alpha \beta \gamma \mu} = 0$.
Consequently, $\omega^{\alpha \beta} = 0$, which leads to the constitutive relation for the couple-stress tensor $M^{\alpha \beta}$ being given by
\begin{equation} \label{eq:thermo-m-alpha-beta}
	M^{\alpha \beta}
	= \dfrac{\rho}{2} \Big(
		\dfrac{\partial \psi}{\partial b_{\alpha \beta}}
		+ \dfrac{\partial \psi}{\partial b_{\beta \alpha}}
	\Big) ~.
\end{equation}
\textspace

Because of the in-plane viscous nature of the lipid bilayer, $\pi^{\alpha \beta}$ is nonzero.
Lipid membranes are isotropic in-plane, indicating $R^{\alpha \beta \gamma \mu}$ is an isotropic tensor.
A fourth-order tensor in general curvilinear coordinates is isotropic when it is invariant to all unimodular transformations of the coordinate system, represented by the tensor ${\Lambda^\nu}_\sigma$, such that
\begin{equation} \label{eq:t-isotropic-def}
	R^{\alpha \beta \gamma \mu}
	= {\Lambda^\alpha}_\delta {\Lambda^\beta}_\nu {\Lambda^\gamma}_\lambda {\Lambda^\mu}_\sigma R^{\delta \nu \lambda \sigma} ~.
\end{equation}
For ${\Lambda^\nu}_\sigma$ to represent a unimodular coordinate transformation, it must satisfy ${\Lambda^\nu}_\sigma {\Lambda_\lambda}^\sigma = \delta^\nu_\lambda$ and ${\Lambda^\nu}_\sigma {\Lambda_\nu}^\lambda = \delta^\lambda_\sigma$, where $ {(\Lambda^{-1})^\nu}_\sigma = {\Lambda_\nu}^\sigma$ \cite{carroll}.
In what follows, we choose forms of ${\Lambda^\nu}_\sigma$ satisfying these requirements to determine constraints on the form of $R^{\alpha \beta \gamma \mu}$.
\textspace

First, consider a rotation of the coordinate axes by $\pi/2$ radians about the direction of the normal vector $\bm{n}$.
The transformation tensor ${\Lambda^\nu}_\sigma$ corresponding to this rotation is given by
\begin{equation} \label{eq:lambda-90-rotation}
	{\Lambda^1}_1 =  0 ~, \hspace{30pt}
	{\Lambda^1}_2 =  -1 ~, \hspace{30pt}
	{\Lambda^2}_1 =  1 ~, \hspace{30pt}
	{\Lambda^2}_2 =  0 ~. 
\end{equation}
Applying equation \eqref{eq:lambda-90-rotation} to the definition of an isotropic tensor \eqref{eq:t-isotropic-def}, we obtain
\begin{equation} \label{eq:t-90-rotation}
	\begin{split}
		R^{1111} &= R^{2222} ~, \\
		R^{2111} &= -R^{1222} ~, 
	\end{split}
	\hspace{20pt} 
	\begin{split}
		R^{1112} &= -R^{2221} ~, \\
		R^{1122} &= R^{2211} ~, 
	\end{split}
	\hspace{20pt} 
	\begin{split}
		R^{1121} &= -R^{2212} ~, \\
		R^{1212} &= R^{2121} ~, 
	\end{split}
	\hspace{20pt} 
	\begin{split}
		R^{1211} &= -R^{2122} ~, \\
		R^{1221} &= R^{2112} ~, 
	\end{split}
\end{equation}
reducing the initial 16 variables in $R^{\alpha \beta \gamma \mu}$ to eight.
\textspace

Next, consider a transformation where we exchange $\bm{a}_1$ and $\bm{a}_2$.
The transformation tensor ${\Lambda^\nu}_\sigma$ for this operation is given by
\begin{equation} \label{eq:lambda-a1-a2-swap}
	{\Lambda^1}_1 =  -1 ~, \hspace{30pt}
	{\Lambda^1}_2 =  0 ~, \hspace{30pt}
	{\Lambda^2}_1 =  0 ~, \hspace{30pt}
	{\Lambda^2}_2 =  1 ~. 
\end{equation}
Applying equation \eqref{eq:lambda-a1-a2-swap} to the definition of an isotropic tensor \eqref{eq:t-isotropic-def} leads to
\begin{equation} \label{eq:t-a1-a2-swap}
	R^{1112} = 0 ~, \hspace{30pt}
	R^{1121} = 0 ~, \hspace{30pt}
	R^{1211} = 0 ~, \hspace{30pt}
	R^{2111} = 0 ~,
\end{equation}
reducing the remaining eight variables in $R^{\alpha \beta \gamma \mu}$ to four.
\textspace

The final transformation we consider is a rotation of the coordinate axes by $\pi/4$ radians about the direction of the normal vector $\bm{n}$.
In this case, the transformation tensor ${\Lambda^\nu}_\sigma$ is given by
\begin{equation} \label{eq:lambda-45-rotation}
		{\Lambda^1}_1 =  1/\sqrt{2} ~, \hspace{25pt}
		{\Lambda^1}_2 =  -1/\sqrt{2} ~, \hspace{25pt}
		{\Lambda^2}_1 =  1/\sqrt{2} ~, \hspace{25pt}
		{\Lambda^2}_2 =  1/\sqrt{2} ~. 
\end{equation}
Applying equation \eqref{eq:lambda-45-rotation} to equation \eqref{eq:t-isotropic-def} yields
\begin{equation} \label{eq:t-45-rotation}
	R^{1111} = R^{1122} + R^{1212} + R^{1221} ~,
\end{equation}
thereby reducing the four remaining degrees of freedom in $R^{\alpha \beta \gamma \mu}$ to three.
\textspace

Given the independent variables of $R^{\alpha \beta \gamma \mu}$, we now determine its functional form.
Due to the linear independence of $R^{1122}$, $R^{1212}$, and $R^{1221}$, we may specify each term arbitrarily.
Consider the case where $R^{1122} \ne 0$ and $R^{1212} = R^{1221} = 0$.
From equations \eqref{eq:t-45-rotation} and \eqref{eq:t-90-rotation}, $R^{1111} = R^{2222} = R^{1122} = R^{2211}$.
In Cartesian coordinates, the form of $R^{\alpha \beta \gamma \mu}$ would be expressed with Kronecker delta functions \cite{aris}.
In curvilinear coordinates, the contravariant form of $\delta^\alpha_\beta$ is $\delta^\alpha_\mu a^{\mu \beta} = a^{\alpha \beta}$ and so we may write $R^{\alpha \beta \gamma \mu} = \tfrac{1}{2}\lambda \, a^{\alpha \beta} a^{\gamma \mu}$, where $\lambda$ is a constant and the factor of $1/2$ is included for convenience.
The next case to consider is when $R^{1212} = R^{2121} = R^{1111} = R^{2222} \ne 0$ and $R^{1122} = R^{1221} = 0$.
In a similar fashion to the previous case, with curvilinear coordinates we find $R^{\alpha \beta \gamma \mu} = \zeta_1 \, a^{\alpha \gamma} a^{\beta \mu}$, where $\zeta_1$ is another constant.
Finally, if $R^{1221} = R^{2112} \ne 0$ and $R^{1212} = R^{1122} = 0$, we obtain $R^{\alpha \beta \gamma \mu} = \zeta_2 \, a^{\alpha \mu} a^{\beta \gamma}$, where $\zeta_2$ is yet another constant.
Due to the linear independence of $R^{1122}$, $R^{1212}$, and $R^{1221}$, the general form of $R^{\alpha \beta \gamma \mu}$ is given by the sum
\begin{equation} \label{eq:t-isotropic-general}
	R^{\alpha \beta \gamma \mu}
	= \zeta_1 \, a^{\alpha \gamma} a^{\beta \mu}
	+ \zeta_2 \, a^{\alpha \mu} a^{\beta \gamma}
	+ \dfrac{1}{2} \lambda \, a^{\alpha \beta} a^{\gamma \mu} ~.
\end{equation}
Equation \eqref{eq:t-isotropic-general} can be considered as providing the general form of a fourth-order isotropic tensor in curvilinear coordinates.
\textspace

Substituting the form of $R^{\alpha \beta \gamma \mu}$ in equation \eqref{eq:t-isotropic-general} into the expression for $\pi^{\alpha \beta}$ \eqref{eq:pi-phenom-no-cross}, we obtain
\begin{equation} \label{eq:pi-alpha-beta-reduce-step-1}
	\begin{split}
		\pi^{\alpha \beta}
		&= \Big(
			\zeta_1 \, a^{\alpha \gamma} a^{\beta \mu}
			+ \zeta_2 \, a^{\alpha \mu} a^{\beta \gamma}
			+ \tfrac{1}{2} \lambda \, a^{\alpha \beta} a^{\gamma \mu}
		\Big) \dot{a}_{\gamma \mu} \\[3pt]
		&= \zeta_1 \, a^{\alpha \gamma} a^{\beta \mu} \dot{a}_{\gamma \mu} 
		+ \zeta_2 \, a^{\alpha \mu} a^{\beta \gamma} \dot{a}_{\gamma \mu} 
		+ \tfrac{1}{2} \lambda \, a^{\alpha \beta} a^{\gamma \mu} \dot{a}_{\gamma \mu} \\[5pt]
		&= \zeta_1 \, a^{\alpha \gamma} a^{\beta \mu} \dot{a}_{\gamma \mu} 
		+ \zeta_2 \, a^{\alpha \gamma} a^{\beta \mu} \dot{a}_{\gamma \mu} 
		+ \tfrac{1}{2} \lambda \, a^{\alpha \beta} a^{\gamma \mu} \dot{a}_{\gamma \mu} \\[5pt]
		&= \big( \zeta_1 + \zeta_2 \big) a^{\alpha \gamma} a^{\beta \mu} \dot{a}_{\gamma \mu} 
		+ \tfrac{1}{2} \lambda a^{\alpha \beta} a^{\gamma \mu} \dot{a}_{\gamma \mu} ~,
	\end{split}
\end{equation}
where in the third equality we use the symmetry of $\dot{a}_{\gamma \mu}$.
Using equation \eqref{eq:jacobian-determinant-derivative} and defining $\zeta = \zeta_1 + \zeta_2$ reduces equation \eqref{eq:pi-alpha-beta-reduce-step-1} to
\begin{equation} \label{eq:pi-alpha-beta}
	\pi^{\alpha \beta}
	= \zeta \, a^{\alpha \gamma} a^{\beta \mu} \dot{a}_{\gamma \mu}
	+ \lambda \, a^{\alpha \beta} \big( v^\mu_{; \mu} - 2 v H \big) ~.
\end{equation}
\eqnspace

With the form of $\pi^{\alpha \beta}$ in equation \eqref{eq:pi-alpha-beta} and $M^{\alpha \beta}$ in equation \eqref{eq:thermo-m-alpha-beta}, the internal entropy production \eqref{eq:internal-entropy-complete} simplifies to
\begin{equation} \label{eq:internal-entropy-simplified-step-1}
	\begin{split}
		\rho \eta_{\mathrm{i}} 
		&= - \dfrac{\jq^{ \, \alpha} T_{, \alpha}}{T^2} 
		+ \dfrac{\pi^{\alpha \beta} \dot{a}_{\alpha \beta}}{2T} \\[5pt]
		&= - \dfrac{\jq^{ \, \alpha} T_{, \alpha}}{T^2} 
		+ \dfrac{1}{2T} \bigg\{
			\, \zeta \, a^{\alpha \gamma} a^{\beta \mu} \dot{a}_{\gamma \mu} \dot{a}_{\alpha \beta}
			+ \lambda \, a^{\alpha \beta} \big( v^\mu_{; \mu} - 2 v H \big) \dot{a}_{\alpha \beta} \,
		\bigg\}
		\ge 0 ~.
	\end{split}
\end{equation}
In the absence of temperature gradients, equation \eqref{eq:internal-entropy-simplified-step-1} simplifies to
\begin{equation} \label{eq:internal-entropy-simplified-step-2}
	\begin{split}
		\rho \eta_{\mathrm{i}} 
		= \dfrac{1}{2T} \bigg\{
			\, \zeta \, a^{\alpha \gamma} a^{\beta \mu} \dot{a}_{\gamma \mu} \dot{a}_{\alpha \beta}
			+ 2 \lambda \, \big( v^\mu_{; \mu} - 2 v H \big)^2 \,
		\bigg\}
		\ge 0 ~,
	\end{split}
\end{equation}
where use has been made of equation \eqref{eq:jacobian-determinant-derivative}.
For the inequality in equation \eqref{eq:internal-entropy-simplified-step-2} to hold, we require $\zeta \ge 0$ and $\lambda \ge 0$.
Physically, $\zeta$ describes the internal entropy production from velocity gradients and represents the in-plane shear viscosity, while $\lambda$ describes internal entropy production due to the fluid compressing or expanding and represents the in-plane bulk viscosity.
The final form of the total internal entropy production in the linear irreversible regime is given by
\begin{equation} \label{eq:internal-entropy-total-1c}
	\rho \eta_{\mathrm{i}}
	= \dfrac{\kappa \big(
		T^{, \alpha} \, T_{, \alpha}
	\big)} {T^2}
	+ \dfrac{1}{2T} \bigg\{
		\, \zeta \, a^{\alpha \gamma} a^{\beta \mu} \dot{a}_{\gamma \mu} \dot{a}_{\alpha \beta}
		+ 2 \lambda \, \big( v^\mu_{; \mu} - 2 v H \big)^2 \,
	\bigg\}
	\ge 0 ~,
\end{equation}
where $\kappa$, $\zeta$, and $\lambda$ are all non-negative.
\textspace

In this section, we determined how the stresses and moments of the membrane are related to the Helmholtz energy density, and the results can be summarized as
\begin{align}
	\sigma^{\alpha \beta}
	&= \rho \bigg(
		\dfrac{\partial \psi}{\partial a_{\alpha \beta}}
		+ \dfrac{\partial \psi}{\partial a_{\beta \alpha}}
	\bigg) + \pi^{\alpha \beta} ~,
	\label{eq:sigma-summary}
	\\[3pt]
	M^{\alpha \beta}
	&= \dfrac{\rho}{2} \bigg(
		\dfrac{\partial \psi}{\partial b_{\alpha \beta}}
		+ \dfrac{\partial \psi}{\partial b_{\beta \alpha}}
	\bigg) ~,
	\label{eq:m-summary}
	\\[5pt]
	N^{\alpha \beta} &= \sigma^{\alpha \beta} + b^\beta_\mu M^{\mu \alpha} ~, \\
	\shortintertext{and}
	S^\alpha &= - M^{\beta \alpha}_{; \beta} ~, \label{eq:s-summary}
\end{align}
with $\pi^{\alpha \beta}$ given by equation \eqref{eq:pi-alpha-beta}.
\textspace

At this stage, we derive the Gibbs equation for the single-component membrane system.
The Gibbs equation in general relates infinitesimal changes in thermodynamic state functions, and consequently will not account for any dissipation in the system.
It is therefore useful to define $(\sigma^{\alpha \beta})^{\text{el}}$ to be the reversible, elastic component of the in-plane stress $\sigma^{\alpha \beta}$ \eqref{eq:sigma-summary} given by 
\begin{equation} \label{eq:sigma-alpha-beta-elastic}
	(\sigma^{\alpha \beta})^{\text{el}}
	= \rho \bigg(
		\dfrac{\partial \psi}{\partial a_{\alpha \beta}}
		+ \dfrac{\partial \psi}{\partial a_{\beta \alpha}}
	\bigg) 
	~.
\end{equation}
To derive the Gibbs equation for our membrane system, we start with equation \eqref{eq:entropy-balance-reformulated} and substitute the material derivative of $\psi$ \eqref{eq:psi-dot}, the local equilibrium assumption \eqref{eq:local-equilibrium-entropy-sc}, the moment tensor $M^{\alpha \beta}$ \eqref{eq:m-summary}, and the elastic component of the in-plane traction $(\sigma^{\alpha \beta})^{\text{el}}$ \eqref{eq:sigma-alpha-beta-elastic} to obtain 
\begin{equation} \label{eq:gibbs-eqn-dt-sc}
	\rho \, T \dot{s}
	= \rho \, \dot{u}
	- \dfrac{1}{2} (\sigma^{\alpha \beta})^{\text{el}} \,\, \dot{a}_{\alpha \beta}
	- M^{\alpha \beta} \, \dot{b}_{\alpha \beta}
	~.
\end{equation}
Equation \eqref{eq:gibbs-eqn-dt-sc} relates the rates of change of thermodynamic state functions, and by multiplying both sides of the equation by $\mathrm{d}t$ we find
\begin{equation} \label{eq:gibbs-eqn-sc}
	\rho \, T \, \mathrm{d} s
	= \rho \, \mathrm{d} u
	- \dfrac{1}{2} (\sigma^{\alpha \beta})^{\text{el}} \,\, \mathrm{d} a_{\alpha \beta}
	- M^{\alpha \beta} \, \mathrm{d} b_{\alpha \beta}
	~.
\end{equation}
Equation \eqref{eq:gibbs-eqn-sc} is the Gibbs equation for a two-dimensional membrane surface with out-of-plane elastic bending and in-plane elastic compression and stretching.


\subsubsection{Helmholtz Free Energy---Change of Variables}

We have so far developed general equations of how the membrane stresses depend on a Helmholtz free energy density $\psi$, which in turn depends on the metric tensor $a_{\alpha \beta}$, the curvature tensor $b_{\alpha \beta}$, and the temperature $T$ \eqref{eq:psi-constitutive-elastic-compressible}.
The energy density $\psi$, in being an absolute scalar field, must be invariant to Galilean transformations.
For a fluid film, under Galilean invariance the Helmholtz free energy density may depend on $a_{\alpha \beta}$ and $b_{\alpha \beta}$ only through the density $\rho$, the mean curvature $H$, and the Gaussian curvature $K$, which are functions of the invariants of the metric and curvature tensors \cite{steigmann-fluid-film-arma-1999}.
This relationship is written as
\begin{equation} \label{eq:helmholtz-energy-change-variables}
	\psi (a_{\alpha \beta}, b_{\alpha \beta}, T)
	= \bar{\psi} (\rho, H, K, T) ~.
\end{equation}
Note equation \eqref{eq:helmholtz-energy-change-variables} can also be shown using material symmetry arguments as presented in \cite{jenkins-siam-1977}.
\textspace

When substituting $\bar{\psi}$ into the stress and moment relations \eqref{eq:sigma-summary}--\eqref{eq:s-summary}, we encounter terms like 
\begin{align} 
	\dfrac{\partial \bar{\psi}}{\partial a_{\alpha \beta}}
	&= \bar{\psi}_{, \rho} \dfrac{\partial \rho}{\partial a_{\alpha \beta}}
	+ \bar{\psi}_{, H} \dfrac{\partial H}{\partial a_{\alpha \beta}}
	+ \bar{\psi}_{, K} \dfrac{\partial K}{\partial a_{\alpha \beta}} \label{eq:psi-partial-a-expansion} \\
	\shortintertext{and}
	\dfrac{\partial \bar{\psi}}{\partial b_{\alpha \beta}}
	&= \bar{\psi}_{, \rho} \dfrac{\partial \rho}{\partial b_{\alpha \beta}}
	+ \bar{\psi}_{, H} \dfrac{\partial H}{\partial b_{\alpha \beta}}
	+ \bar{\psi}_{, K} \dfrac{\partial K}{\partial b_{\alpha \beta}} ~, \label{eq:psi-partial-b-expansion}
\end{align}
where subscripts including a comma denote a partial derivative---for example, $\bar{\psi}_{, \rho} = \partial \bar{\psi} / \partial \rho$.
The variations in $\rho$, $H$, and $K$ due to changes in $a_{\alpha \beta}$ and $b_{\alpha \beta}$ can be easily calculated, and are summarized in Table \ref{partial-derivatives}.
Substituting the partial derivatives given in Table \ref{partial-derivatives} into equations \eqref{eq:psi-partial-a-expansion}--\eqref{eq:psi-partial-b-expansion}, we find the stress and moment tensors \eqref{eq:sigma-summary}--\eqref{eq:s-summary} expressed in terms of $\bar{\psi} (\rho, H, K, T)$ to be given by
\begin{table}[!t]
	\centering
		\vspace{5pt}
		\setlength{\tabcolsep}{26pt}
		\renewcommand{\arraystretch}{2}
		\begin{tabular}{c|ccc}
		\hline
		\hline
			~ & $\rho$ & $H$ & $K$ \\
			\cline{1-4}
			$a_{\alpha \beta}$
			& $-\tfrac{1}{2} \rho a^{\alpha \beta}$
			& $-\tfrac{1}{2} b^{\alpha \beta}$
			& $-K a^{\alpha \beta}$
			\\
			$b_{\alpha \beta}$
			& 0
			& $\tfrac{1}{2} a^{\alpha \beta}$
			& $\bar{b}^{\alpha \beta}$
			\\[8pt]
		\hline
		\hline
		\end{tabular}
		\captionsetup{width=0.80\textwidth}
		\caption{The partial derivatives of the areal mass density $\rho$, mean curvature $H$, and Gaussian curvature $K$ with respect to the metric tensor $a_{\alpha \beta}$ and curvature tensor $b_{\alpha \beta}$.
		Each table entry is the partial derivative of the column header with respect to the row header.
		For example, the third column of the first row indicates $\partial K / \partial a_{\alpha \beta} = -K a^{\alpha \beta}$.
		}
		\label{partial-derivatives}
\end{table}
\begin{align}
	\sigma^{\alpha \beta}
	&= -\rho \big(
		\rho \, \bar{\psi}_{, \rho}
		+ 2H \bar{\psi}_{, H}
		+ 2K \bar{\psi}_{, K}
	\big) a^{\alpha \beta}
	+ \rho \, \bar{\psi}_{, H} \, \bar{b}^{\alpha \beta}
	+ \pi^{\alpha \beta} ~,
	\label{eq:sigma-alpha-beta-1c-general}
	\\[5pt]
	M^{\alpha \beta}
	&= \tfrac{1}{2} \rho \, \bar{\psi}_{, H} \, a^{\alpha \beta}
	+ \rho \, \bar{\psi}_{, K} \, \bar{b}^{\alpha \beta} ~,
	\label{eq:m-alpha-beta-1c-general}
	\\[5pt]
	N^{\alpha \beta}
	&= -\rho \big(
		\rho \bar{\psi}_{, \rho}
		+ H \bar{\psi}_{, H}
		+ K \bar{\psi}_{, K}
	\big) a^{\alpha \beta}
	+ \tfrac{1}{2} \, \rho \, \bar{\psi}_{, H} \, \bar{b}^{\alpha \beta}
	+ \pi^{\alpha \beta} ~,
	\label{eq:n-alpha-beta-1c-general}
	\\
	\shortintertext{and}
	S^{\alpha}
	&= - \tfrac{1}{2} (\rho \bar{\psi}_{, H})_{; \beta} \, a^{\alpha \beta}
	- ( \rho \bar{\psi}_{, K} )_{; \beta} \, \bar{b}^{\alpha \beta} ~.
	\label{eq:s-alpha-1c-general}
\end{align}
Equations \eqref{eq:sigma-alpha-beta-1c-general}--\eqref{eq:s-alpha-1c-general} are identical to those found in \cite{kranthi-bmm-2012}, obtained without using the formulation of an irreversible thermodynamic framework. 
When substituting equations \eqref{eq:sigma-alpha-beta-1c-general}--\eqref{eq:s-alpha-1c-general} in the equations of motion, it is useful to write the viscous stresses $\pi^{\alpha \beta}$ in a different form which contains the mean and Gaussian curvatures.
Substituting equation \eqref{eq:a-alpha-beta-dot} into equation \eqref{eq:pi-alpha-beta} and using the definition of the cofactor of curvature \eqref{eq:cofactor-curvature}, we obtain
\begin{equation} \label{eq:viscous-stress}
	\pi^{\alpha \beta}
	= 2 \zeta \, \Big(
		d^{\alpha \beta}
		- 2 v H a^{\alpha \beta} + v \, \bar{b}^{\alpha \beta}
	\Big)
	+ \lambda \, a^{\alpha \beta} \Big(
		v^\mu_{; \mu} - 2 v H
	\Big) ~,
\end{equation}
where $d^{\alpha \beta}$ is the symmetric part of the in-plane velocity gradients defined as 
\begin{equation} \label{eq:d-alpha-beta}
	d^{\alpha \beta} = \dfrac{1}{2} \big( v^{\alpha ; \beta} + v^{\beta ; \alpha} \big) ~.
\end{equation}


\subsubsection{Helfrich Energy Density}

For single-component lipid bilayers, the Helmholtz free energy density contains an energetic cost for bending and an energetic cost for areal compressions and dilations.
The energetic cost of bending, called the Helfrich energy \cite{helfrich-1973} and denoted $w_{\textrm{h}}$, is given by
\begin{equation} \label{eq:w-helfrich}
	w_{\textrm{h}}
	= \kb \big( H - C \big)^2
	+ \kg K ~.
\end{equation}
The constants $\kb$ and $\kg$ are the mean and Gaussian bending moduli, respectively.
In equation \eqref{eq:w-helfrich}, $C$ is the spontaneous curvature induced by proteins or lipids which the membrane would like to conform to.
The compression energy density $w_{\textrm{c}}$ equally penalizes areal compression and dilation, and is given by
\begin{equation} \label{eq:w-compress}
	w_{\textrm{c}}
	= \dfrac{1}{J} \kc \big( 1 - J \big)^2 ~,
\end{equation}
where $\kc$ is the compression modulus.
As $\kc$ tends to infinity, the membrane becomes incompressible.
The factor of $1/J$ in $w_{\textrm{c}}$ is necessary as areal compressions and dilations are calculated with respect to the reference patch $\scp_0$ at time $t_0$.
To see this explicitly, consider the total compression energy $W_{\textrm{c}}$ in the reference frame given by
\begin{equation} \label{eq:w-compress-convected-total}
	\begin{split}
		W_{\textrm{c}}
		= \int_\scp w_{\mathrm{c}} ~\mathrm{d}a 
		&= \int_{\scp_0} \!\! J \, w_{\mathrm{c}} ~\mathrm{d}A \\[5pt]
		&= \int_{\scp_0} \!\! \kc \big( 1 - J \big)^2 ~\mathrm{d}A ~.
	\end{split}
\end{equation}
The total Helmholtz free energy per unit area is given by
\begin{equation} \label{eq:helmholtz-total-1c}
	\rho \bar{\psi}
	= w_{\textrm{h}}
	+ w_{\textrm{c}} 
	+ \rho f(T)
	~,
\end{equation}
where $f(T)$ is a function of the temperature such that the entropy $s$ can be calculated using equation \eqref{eq:local-equilibrium-entropy-sc}.
\textspace

Given the total Helmholtz energy per unit area $\rho \bar{\psi}$ \eqref{eq:helmholtz-total-1c} and the forms of the Helfrich \eqref{eq:w-helfrich} and compression \eqref{eq:w-compress} energies, we calculate the partial derivatives 
\begin{align}
	\rho \bar{\psi}_{, H}
	&= 2 \kb \big(
		H - C
	\big) ~,
	\label{eq:bar-psi-h}
	\\[5pt]
	\rho \bar{\psi}_{, K} &= \kg ~,
	\label{eq:bar-psi-k}
	\\
	\shortintertext{and}
	\rho^2 \bar{\psi}_{, \rho}
	&= -\kb \big(
		H - C
	\big)^2
	- \kg K
	- 2 \kc \big(
		J - 1
	\big) ~.
	\label{eq:bar-psi-rho}
\end{align}
In obtaining equation \eqref{eq:bar-psi-rho} it is important to realize $J$ is a function of $\rho$, as shown in equation \eqref{eq:jacobian-rho}.
Substituting the partial derivates \eqref{eq:bar-psi-h}--\eqref{eq:bar-psi-rho} into the stress and moment tensors \eqref{eq:sigma-alpha-beta-1c-general} - \eqref{eq:s-alpha-1c-general} yields 
\begin{align}
	\begin{split}
		\sigma^{\alpha \beta}
		&= \kb \Big[
			\Big(
				-3 H^2
				+ 2 H C
				+ C^2
			\Big) a^{\alpha \beta}
			+ 2 \Big(
				H - C
			\Big) \bar{b}^{\alpha \beta}
		\Big] \\[2pt]
		&\hspace{70pt} - \kg K a^{\alpha \beta}
		+ 2 \kc \big( J - 1 \big) a^{\alpha \beta}
		+ \pi^{\alpha \beta}
		~,
		\label{eq:sigma-alpha-beta-1c-helfrich}
	\end{split}
	\\[9pt]
	M^{\alpha \beta}
	&= \kb \big(
		H - C
	\big) a^{\alpha \beta}
	+ \kg \, \bar{b}^{\alpha \beta} ~,
	\label{eq:m-alpha-beta-1c-helfrich}
	\\[9pt]
	\begin{split}
		N^{\alpha \beta}
		&= \kb \Big[
			\Big(
				-H^2
				+ C^2
			\Big) a^{\alpha \beta}
			+ \Big(
				H - C
			\Big) \bar{b}^{\alpha \beta}
		\Big]
		+ 2 \kc \big( J - 1 \big) a^{\alpha \beta}
		+ \pi^{\alpha \beta}
		~,
		\label{eq:n-alpha-beta-1c-helfrich}
	\end{split}
	\\
	\intertext{and}
	S^\alpha
	&= - \kb \big(
		H - C
	\big)^{, \alpha} ~,
	\label{eq:s-alpha-1c-helfrich}
\end{align}
with $\pi^{\alpha \beta}$ given by equation \eqref{eq:viscous-stress}.
Equations \eqref{eq:sigma-alpha-beta-1c-helfrich}--\eqref{eq:s-alpha-1c-helfrich} describe the stresses and moments of an elastic, compressible membrane with Helfrich energy density and viscous in-plane flow.
The protein- or lipid-induced spontaneous curvature $C$ appears in both the in-plane stresses and the out-of-plane shear.
Therefore, proteins or lipids with preferred spontaneous curvature are able to affect both in-plane flow and out-of-plane bending, an important observation noted in other theoretical and computational studies \cite{kranthi-bmm-2012, sauer-liquid-shell-corr-2016, kranthi-biophys-2014}.


\subsection{Equations of Motion}

In this section, we provide the equations of motion for a membrane with elastic out-of-plane bending and intra-membrane viscous flow.
The four unknowns in the membrane system are the areal mass density $\rho$ and the three components of the velocity $\bm{v}$.
The four equations to be solved for the four unknowns are the local mass balance \eqref{eq:local-mass-balance}, two in-plane momentum balances \eqref{eq:tangential-general-eoms-elastic-compressible}, and the shape equation \eqref{eq:normal-general-eom-elastic-compressible}.
We refer to the equations necessary to solve for all the unknowns as the governing equations.
Substituting the stress and moment tensors \eqref{eq:sigma-alpha-beta-1c-helfrich}--\eqref{eq:s-alpha-1c-helfrich} into equations \eqref{eq:tangential-general-eoms-elastic-compressible}--\eqref{eq:normal-general-eom-elastic-compressible}, we find the governing equations describing the motion of the membrane to be
\begin{gather}
	\rho \Big(
		v_{, t}
		+ v^\alpha w_\alpha
	\Big)
	= p + \pi^{\alpha \beta} b_{\alpha \beta}
	- 2 \kb \big( H-C \big) \big( H^2 + HC - K \big)
	- \kb \, \Delta \big( H - C \big)
	+ 4 H \kc (J - 1)
	~,
	\label{eq:normal-helfrich}
	\\[-3pt]
	\rho \Big(
		v^\alpha_{, t}
		- v \, w^\alpha
		+ v^\mu {w_\mu}^\alpha
	\Big) 
	= \rho b^\alpha
	+ \pi^{\mu \alpha}_{; \mu}
	- 2 \kb \big(
		H - C
	\big) C^{, \alpha}
	- 2 \kc \, J \, \dfrac{\rho^{, \alpha}}{\rho} 
	~,
	\label{eq:tangent-helfrich}
	\\
	\intertext{and}
	\rho_{, t} + \rho_{, \alpha} v^\alpha + \big( v^\alpha_{; \alpha} - 2 v H \big) \rho = 0 ~.
	\label{eq:mass-balance-expanded}
\end{gather}
The operator $\Delta$ is the surface Laplacian, and is defined by $\Delta ( \, \cdot \, ) = ( \, \cdot \, )_{; \beta \alpha} a^{\alpha \beta}$.
In deriving the shape equation \eqref{eq:normal-helfrich}, we have used the identity $\bar{b}^{\alpha \beta} b_{\alpha \beta} = 2K$.
We use the form of $\pi^{\alpha \beta}$ in equation \eqref{eq:viscous-stress} to calculate $\pi^{\mu \alpha}_{; \mu}$ and $\pi^{\alpha \beta} b_{\alpha \beta}$, which are found in equations \eqref{eq:tangent-helfrich} and \eqref{eq:normal-helfrich}, respectively, and are given by
\begin{align}
	&\pi^{\mu \alpha}_{; \mu}
	= 2 \zeta \Big(
		d^{\, \mu \alpha}_{; \mu}
		- v_{, \mu} b^{\mu \alpha}
		- 2 v H_{, \mu} a^{\mu \alpha}
	\Big)
	+ \lambda a^{\mu \alpha} \big(
		v^\beta_{; \beta} - 2 v H
	\big)_{; \mu} 
	\label{eq:viscosity-gradient} \\
	\intertext{and}
	&\pi^{\alpha \beta} b_{\alpha \beta}
	= 2 \zeta \Big(
		b^{\alpha \beta} d_{\alpha \beta}
		- v \big( 4 H^2 - 2 K \big)
	\Big)
	+ 2 \lambda H \big(
		v^\mu_{; \mu} - 2 v H
	\big) ~.
	\label{viscosity-contraction}
\end{align}
As expected, there are multiple modes of coupling among the equations of motion.
The spontaneous curvature and its gradients appear in both the in-plane momentum equations and the shape equation.
The viscosity provides a medium for coupling, as the viscous stress tensor $\pi^{\alpha \beta}$ appears in all three momentum equations (one out-of-plane \eqref{eq:normal-helfrich} and two in-plane \eqref{eq:tangent-helfrich}).
Finally, the presence of curvature and velocity components in the momentum equations couples them with the mass balance.


\subsection{Boundary and Initial Conditions} \label{sec:sc-boundary-conditions}

In this section, we provide suitable boundary conditions necessary for solving the governing equations \eqref{eq:normal-helfrich}--\eqref{eq:mass-balance-expanded}.
We begin by deriving the general form of the forces and moments along the membrane edge and then evaluate them for the Helmholtz energy density in equations \eqref{eq:helmholtz-total-1c}--\eqref{eq:w-compress}.
We conclude with a discussion of the possible boundary conditions for the membrane system.
A major portion of this discussion is developed in \cite{kranthi-bmm-2012, steigmann-mms-1998, steigmann-fluid-film-arma-1999}, but is provided here for completeness.
\textspace

The total force $\bm{f}$ at any arbitrary boundary on the membrane surface may be decomposed in the $\{\bm{\nu}, \bm{\tau}, \bm{n}\}$ basis as
\begin{equation} \label{eq:boundary-force-decomposition}
	\bm{f}
	= f_\nu \bm{\nu}
	+ f_\tau \bm{\tau}
	+ f_n \bm{n} ~,
\end{equation}
where $f_\nu$ and $f_\tau$ are the in-plane components of the force and $f_n$ is the out-of-plane shear force felt by the membrane in the normal direction.
The total force $\bm{f}$ at the membrane boundary is given by
\begin{equation} \label{eq:boundary-force}
	\bm{f}
	= \bm{T}^\alpha \nu_\alpha
	- \dfrac{\textrm{d}}{\textrm{d} \ell} \big( m_\nu \bm{n} \big) ~,
\end{equation}
where $\ell$ is the arc length parametrization of the boundary and $m_\nu$ is defined in equation \eqref{eq:m-nu-expanded} \cite{kranthi-bmm-2012, steigmann-mms-1998, steigmann-fluid-film-arma-1999, sauer-shell-2015}.
If the boundaries are piecewise continuous, and the discontinuities are indexed by $i$, the force $\bm{f}_i$ on the $i^{\text{th}}$ discontinuity is given by
\begin{equation} \label{eq:boundary-corner-force}
	\bm{f}_i = -[m_\nu]_i \, \bm{n} ~,
\end{equation}
where $[m_\nu]_i$ denotes the change in $m_\nu$ as the $i^{\text{th}}$ corner is traversed in the forward direction, as defined by the arc length parametrization $\ell$.
To decompose the force $\bm{f}$ in equation \eqref{eq:boundary-force} in the $\{ \bm{\nu}, \bm{\tau}, \bm{n} \}$ basis, we calculate $\mathrm{d} \bm{n} / \mathrm{d} \ell$.
By using the chain rule, the definition of $\bm{\tau}$ \eqref{eq:curve-unit-tangent}, the Weingarten equation \eqref{eq:weingarten}, and the decomposition of $\bm{a}_\alpha$ in equation \eqref{eq:a-alpha-decomposed-boundary}, we obtain
\begin{equation} \label{eq:d-n-d-l}
	\dfrac{\mathrm{d} \bm{n}}{\mathrm{d} \ell}
	= \bm{n}_{, \lambda} \dfrac{\mathrm{d} \theta^\lambda}{\mathrm{d} \ell}
	= \bm{n}_{, \lambda} \tau^\lambda
	= - b^\mu_\lambda \bm{a}_\mu \tau^\lambda
	= - b^{\mu \lambda} \tau_\lambda \tau_\mu \bm{\tau}
	- b^{\mu \lambda} \tau_\lambda \nu_\mu \bm{\nu} ~.
\end{equation}
With the result of equation \eqref{eq:d-n-d-l}, the decomposition of the stress vectors \eqref{eq:stress-vector-general}, and equation \eqref{eq:a-alpha-decomposed-boundary} once again, we find
\begin{align}
	f_\nu
	&= N^{\alpha \beta} \nu_\alpha \nu_\beta
	+ m_\nu \, b^{\mu \lambda} \, \tau_\lambda \, \nu_\mu ~,
	\label{eq:f-nu} \\[3pt]
	f_\tau
	&= N^{\alpha \beta} \nu_\alpha \tau_\beta
	+ m_\nu \, b^{\mu \lambda} \, \tau_\lambda \, \tau_\mu ~,
	\label{eq:f-tau} 
	\shortintertext{and}
	f_n
	&= S^\alpha \nu_\alpha - \dfrac{\mathrm{d} m_\nu}{\mathrm{d} \ell} ~.
	\label{eq:f-n}
\end{align}
\eqnspace

The moment $M$ which contributes to the elastic behavior of the membrane at the boundary by bending the unit normal in the direction of $\bm{\nu}$ is calculated as
\begin{equation} \label{eq:boundary-moment-b-c}
	M 
	= - m_\tau 
	= M^{\alpha \beta} \nu_\alpha \nu_\beta ~.
\end{equation}
The three components of the force provided in equations \eqref{eq:f-nu}--\eqref{eq:f-n} and the boundary moment $M$ \eqref{eq:boundary-moment-b-c} are the general forms of the forces and moments on the membrane boundary.
We now determine these quantities for more specific physical situations.
\textspace

First, consider a Helmholtz energy density of the form $\bar{\psi} (\rho, H, K, T)$ for which the stresses and moments are given by equations \eqref{eq:sigma-alpha-beta-1c-general}--\eqref{eq:s-alpha-1c-general}.
The moment $M$ is obtained by substituting equation \eqref{eq:m-alpha-beta-1c-general} into equation \eqref{eq:boundary-moment-b-c}, which yields
\begin{equation} \label{eq:moment-bc-general}
	M 
	= \dfrac{1}{2} \, \rho \, \bar{\psi}_{, H}
	+ \rho \, \bar{\psi}_{, K} \, \bar{b}^{\alpha \beta} \, \nu_\alpha \nu_\beta ~.
\end{equation}
We next define the normal curvatures in the $\bm{\nu}$ and $\bm{\tau}$ directions, $\kappa_\nu$ and $\kappa_\tau$, as $\kappa_\nu = b^{\alpha \beta} \nu_\alpha \nu_\beta$ and $\kappa_\tau = b^{\alpha \beta} \tau_\alpha \tau_\beta$, as well as the twist $\xi = b^{\alpha \beta} \nu_\alpha \tau_\beta$ \cite{kranthi-bmm-2012}.
Using these definitions along with the identities $H = (\kappa_\nu + \kappa_\tau)/2$ and $K = \kappa_\nu \kappa_\tau - \xi^2$, we substitute the constitutive forms of the stresses \eqref{eq:sigma-alpha-beta-1c-general}--\eqref{eq:s-alpha-1c-general} into equations \eqref{eq:f-nu}--\eqref{eq:f-n} to obtain
\begin{align}
	f_\nu
	&= -\rho^2 \, \bar{\psi}_{, \rho}
	- \rho \, \bar{\psi}_{, K} \, \kappa_\nu \kappa_\tau
	- \dfrac{1}{2} \rho \, \bar{\psi}_{, H} \, \kappa_\nu
	+ \pi^{\alpha \beta} \nu_\alpha \nu_\beta ~,
	\label{eq:f-nu-general} \\
	f_\tau
	&= -\dfrac{1}{2} \rho \, \bar{\psi}_{, H} \, \xi
	- \rho \, \bar{\psi}_{, K} \, \xi \, \kappa_\tau
	+ \pi^{\alpha \beta} \tau_\alpha \nu_\beta ~,
	\label{eq:f-tau-general} \\
	\shortintertext{and}
	f_n
	&= -\dfrac{1}{2} (\rho \, \bar{\psi}_{, H})_{, \nu}
	- (\rho \, \bar{\psi}_{, K})_{, \beta} \, \bar{b}^{\alpha \beta} \nu_\alpha
	+ \dfrac{\mathrm{d}}{\mathrm{d} \ell} (\rho \, \bar{\psi}_{, K} \, \xi) ~.
	\label{eq:f-n-general}
\end{align}
For the specific case of the Helmholtz energy density provided in equations \eqref{eq:helmholtz-total-1c}--\eqref{eq:w-compress}, we substitute the partial derivatives calculated in equations \eqref{eq:bar-psi-h}--\eqref{eq:bar-psi-rho} into equations \eqref{eq:moment-bc-general}--\eqref{eq:f-n-general} to find
\begin{align} 
	M &= \kb \big( H - C \big) 
	+ \kg \, \kappa_\tau
	~,
	\label{eq:moment-bc-helfrich} \\[4pt]
	f_\nu
	&= \kb \Big[
		(H - C)^2
		- (H - C) \kappa_\nu
	\Big]
	- \kg \, \xi^2
	+ 2 \kc (J - 1)
	+ \pi^{\alpha \beta} \nu_\alpha \nu_\beta ~,
	\label{eq:f-nu-helfrich} \\[5pt]
	f_\tau
	&= - \xi \Big[
		\kb (H - C)
		+ \kg \, \kappa_\tau
	\Big]
	+ \pi^{\alpha \beta} \tau_\alpha \nu_\beta ~,
	\label{eq:f-tau-helfrich} \\
	\shortintertext{and}
	f_n
	&= -\kb (H - C)_{, \nu}
	+ \kg \dfrac{\mathrm{d} \xi}{\mathrm{d} \ell} ~.
	\label{eq:f-n-helfrich}
\end{align}
\eqnspace

Now that the boundary force $\bm{f}$ and the boundary moment $M$ have been evaluated for our assumed form of the Helmholtz free energy density \eqref{eq:helmholtz-total-1c}, we consider the boundary conditions for the governing equations \eqref{eq:normal-helfrich}--\eqref{eq:mass-balance-expanded}.
As the membrane behaves as a fluid in-plane, we may specify either the tangential velocities $v^\alpha$ or the in-plane components of the force, $f_\nu$ \eqref{eq:f-nu-helfrich} and $f_\tau$ \eqref{eq:f-tau-helfrich}, at the patch boundary.
The shape equation, on the other hand, is an elastic bending equation, and therefore two boundary conditions need to be specified at every point along the boundary.
The simplest way to do so is to specify the membrane position and its gradient in the $\bm{\nu}$ direction, or to specify the moment $M$ and the shear force $f_n$ at the boundary.
\textspace

We have now provided the governing equations with possible boundary conditions for both the tangential and shape equations, and close the problem by providing possible initial conditions. 
To this end, we specify the initial position $\bm{x}$ and the initial velocity $\bm{v}$ everywhere on the membrane surface, and furthermore assume the density $\rho$ is initially constant, the Jacobian $J = 1$ everywhere, and the spontaneous curvature $C = 0$ for problems initially without proteins.
With these example initial conditions, and the suggested boundary conditions, the governing equations \eqref{eq:normal-helfrich}--\eqref{eq:mass-balance-expanded} are mathematically well-posed.


\subsection{Coupling to Bulk Fluid} \label{sec:sc-coupling}

We conclude our theoretical developments for single-component lipid membranes by discussing the coupling of the membrane equations of motion with the surrounding fluid. 
While the bulk fluid provides an additional dissipative mechanism via the bulk viscosity and dominates dissipation for long wavelength undulations \cite{zilman-safran-pre-2004,sapp-maibaum-pre-2016,brown-bpj-2003,cai-lubensky-prl-1994}, the bulk fluid can contribute negligibly in systems with membrane deformations on the order of five microns or less \cite{arroyo-pre-2009, rahimi-arroyo-pre-2012, seifert-epl-1993}.
Consequently, the surrounding fluid can sometimes be excluded when studying small length scale phenomena.
\textspace

In cases where bulk dissipation is non-negligible, the membrane and surrounding fluid should be modeled together and coupled through interface conditions.
The equations of motion for the surrounding bulk fluid are the Navier-Stokes equations.
We denote quantities in the bulk with a subscript `b' and label the fluid above and below the membrane with a superscript `$+$' and `$-$', respectively.
Thus, $\bm{v}_{\mathrm{b}}^+$ and $\bm{v}_{\mathrm{b}}^-$ are the bulk fluid velocities on either side of the membrane, and we make the no-slip assumption between the membrane and fluid such that
\begin{align} 
	\bm{v}_{\mathrm{b}}^+
	= \bm{v}
	\label{eq:bulk-coupling-plus}
	\\
	\shortintertext{and}
	\bm{v}_{\mathrm{b}}^-
	= \bm{v}
	\label{eq:bulk-coupling-minus}
\end{align}
at the membrane-fluid interface.
Additionally, the stresses in the bulk fluid, denoted by the tensors $\bm{\sigma}_{\mathrm{b}}^+$ and $\bm{\sigma}_{\mathrm{b}}^-$, enter the membrane equations of motion through the body force $\bm{b}$.
A force balance on the membrane yields
\begin{equation} \label{eq:bulk-body-force}
	\rho \bm{b}
	= \big(
		\bm{\sigma}_{\mathrm{b}}^+
		- \bm{\sigma}_{\mathrm{b}}^-
	\big) \, \bm{n}
	~,
\end{equation}
where $\bm{n}$ is the membrane normal.
By decomposing the body force in the $\{ \bm{a}_\alpha, \bm{n} \}$ basis as in equation \eqref{eq:body-force}, we find the pressure drop $p$ across the membrane to be given by
\begin{align}
	p
	&= \bm{n} \cdot \big(
		\bm{\sigma}_{\mathrm{b}}^+
		- \bm{\sigma}_{\mathrm{b}}^-
	\big) \, \bm{n}
	~,
	\label{eq:bulk-body-force-pressure}
	\\
	\intertext{and the in-plane body force components $b^\alpha$ to be given by}
	\rho b^\alpha
	&= \bm{a}^\alpha \cdot \big(
		\bm{\sigma}_{\mathrm{b}}^+
		- \bm{\sigma}_{\mathrm{b}}^-
	\big) \, \bm{n}
	~.
	\label{eq:bulk-body-force-in-plane}
\end{align}
The pressure drop and in-plane body forces in equations \eqref{eq:bulk-body-force-pressure}--\eqref{eq:bulk-body-force-in-plane} enter the equations of motion of the membrane \eqref{eq:normal-helfrich}--\eqref{eq:tangent-helfrich}, and the velocities in both the bulk fluid and membrane are solved in a self-consistent manner to satisfy equations \eqref{eq:bulk-coupling-plus}--\eqref{eq:bulk-body-force-in-plane}.
A computational implementation of the aforementioned conditions at a fluid-structure interface is provided in \cite{roger-cmame-2017} in the context of liquid menisci and elastic membranes.
For small deformations of the membrane, one can use the Oseen tensor to couple the bulk fluid and the membrane, as done in \cite{brown-bpj-2003,sapp-maibaum-pre-2016}.


\section{Intra-membrane Phase Transitions} \label{sec:intramembrane}

\begin{figure}[!b]
	\centering
		\includegraphics[width=0.80\textwidth]{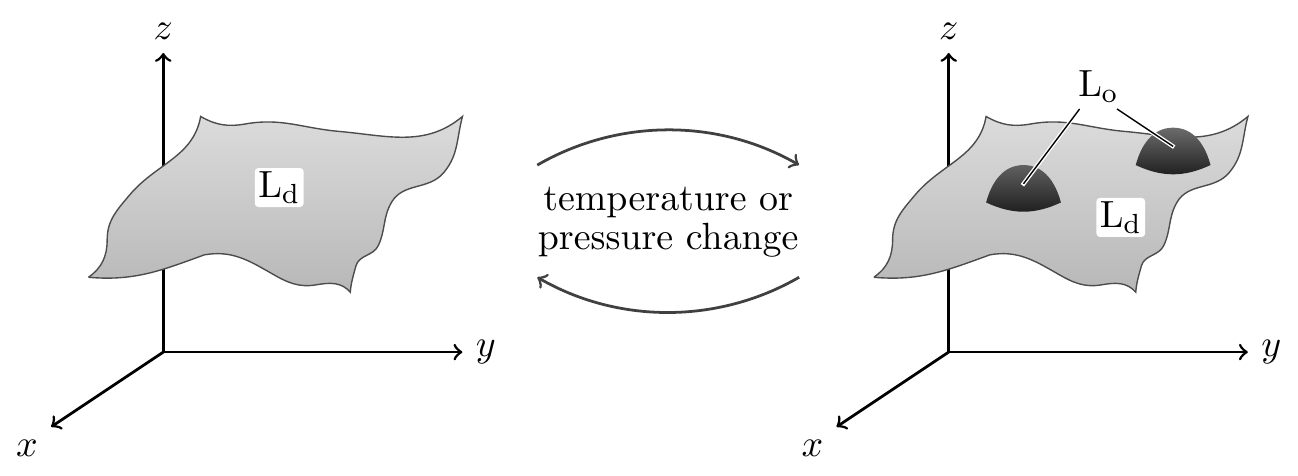}
	\captionsetup{width=0.80\textwidth}
	\caption{
		A schematic of a phase transition in lipid membranes.
		The membrane on the left is initially in the liquid-disordered state (L$_\text{d}$, depicted in light gray).
		Under a suitable change of external conditions, such as a temperature or pressure change, the membrane undergoes a phase transition and contains both liquid-ordered domains (L$_\text{o}$, shown in dark gray) and liquid-disordered domains.
		Such phase transitions are reversible, and the membrane can be returned to its initial disordered state.
	}
	\label{fig:diffusion}
\end{figure}

Biological membranes may consist of hundreds of different protein and lipid constituents and exhibit complex behavior in which species diffuse to form heterogeneous domains, which then take part in important biological phenomena \cite{carlson-mahadevan-plos-comp-bio-2015, qi-pnas-2001}.
In order to better understand such processes, many experiments have been carried out on artificially created giant unilamellar vesicles (GUVs), the composition of which may be precisely controlled.
Of particular interest is the phase transition in which a membrane initially in a liquid-disordered (\ld) phase develops, under a suitable change of external conditions, phase coexistence between \ld\ phases and liquid-ordered (\lo) phases.
The nature of these phases is governed by their concentrations---the \ld\ phase generally consists of a low melting temperature phospholipid such as Dioleoylphosphatidylcholine (DOPC) while the \lo\ phase generally primarily consists of a high melting temperature phospholipid such as Dipalmitoylphosphatidylcholine (DPPC) as well as cholesterol.
A line tension exists at the \lo--\ld\ phase boundary, and imaging experiments have shown one of the two phases may bulge out to reduce this line tension \cite{veatch-keller-bpj-2003, baumgart-hess-nature-2003}, thus indicating a coupling between membrane bending, diffusion, and flow (Figure \ref{fig:diffusion}).
\textspace

In this section, we extend the irreversible thermodynamic framework of the single-component model to have multiple phospholipid constituents and determine the new forms of the balances of mass, linear momentum, angular momentum, energy, and entropy.
We then propose a general constitutive form of the Helmholtz free energy density and determine the constitutive relations for stresses and moments as well as phenomenological relations for the diffusive fluxes.
We conclude this general treatment by determining restrictions on the Helmholtz free energy due to invariance postulates. 
Finally, we study a two-component membrane model in order to better understand the coupling between bending, diffusion, and flow in membranes which exhibit \lo--\ld\ phase coexistence, as described above.
For these cases, we provide the equations of motion and suitable boundary conditions.
While we focus on multi-component phospholipid membranes, all the analysis in this section can be applied to the diffusion and segregation of transmembrane proteins as well.


\subsection{Kinematics} \label{sec:mc-kinematics}

The membrane is modeled as having $N$ types of phospholipid constituents, which are able to diffuse in the plane of the membrane.
We index the phospholipid constituents by $k$, where $k \in \{1, 2, \dots, N \}$.
In treating multi-component systems, we follow the procedure outlined in \cite{degroot-mazur}.
\textspace

The mass density of species $k$ is denoted by $\rho_k (\theta^\alpha, t)$, and the total mass density $\rho$ is given by
\begin{align}
	\rho(\theta^\alpha, t) = \sum_{k = 1}^N \rho_k(\theta^\alpha, t) ~.
	\label{eq:total-density}
\end{align}
The barycentric velocity $\bm{v}$, also called the center-of-mass velocity, is defined by the relation
\begin{equation} \label{eq:barycentric-velocity}
	\rho \bm{v} := \sum_{k = 1}^N \rho_k \bm{v}_k ~,
\end{equation}
where $\bm{v}_k = v_k^{\, \alpha} \bm{a}_\alpha + v_k \bm{n}$ is velocity of species $k$.
While different species may have different in-plane velocity components $v^{\, \alpha}_k$, the normal velocity components $v_k$ must be the same for all species.
\textspace

As we track an infinitesimal membrane patch over time, our reference frame moves with the barycentric velocity $\bm{v}$.
There exists an in-plane diffusive flux of species $k$ across the patch boundary due to the difference in the species and barycentric velocities.
The diffusive flux of species $k$, $\bm{j}_k$, is given by 
\begin{equation} \label{eq:diffusive-species-flux}
	\bm{j}_k = \rho_k \big(
		\bm{v}_k - \bm{v}
	\big) ~.
\end{equation}
This flux lies in the plane of the membrane and can be written in component form as 
\begin{equation} \label{eq:diffusive-species-flux-component}
	\begin{split}
		\bm{j}_k 
		&= j_k^{\, \alpha} \bm{a}_\alpha \\
		&= \rho_k \big(
			v_k^{\, \alpha} - v^\alpha
		\big) \bm{a}_\alpha
		~.
	\end{split}
\end{equation}
Summing all the diffusive fluxes in equation \eqref{eq:diffusive-species-flux}, we obtain
\begin{equation} \label{eq:sum-diff-flux}
	\sum_{k = 1}^N \bm{j}_k = \bm{0} ~,
\end{equation}
using equation \eqref{eq:barycentric-velocity}.
We define the mass fraction $c_k$ of each species $k$ as 
\begin{equation} \label{eq:concentration-definition}
	c_k = \dfrac{\rho_k}{\rho} ~,
\end{equation}
such that
\begin{equation} \label{eq:sum-concentration}
	\sum_{k = 1}^N c_k = 1 ~.
\end{equation}
As only $N - 1$ of the mass fractions are linearly independent, we can completely define the composition of the membrane by specifying the mass density $\rho$ and mass fraction $c_k$ for $k = 2, 3, 4, \dots, N$.


\subsection{Balance Laws}

We follow the same general procedure as in the single-component case to determine local forms of the mass, linear momentum, and angular momentum balances.
We employ mixture theory, as presented in de Groot \& Mazur \cite{degroot-mazur}, to develop a wholistic model of membrane behavior without modeling the individual forces between different phospholipid species. 
We assume no chemical reactions and defer their study to the last part of this work (Section \ref{sec:binding}).


\subsubsection{Mass Balance}

The mass of a single species $k$ in a membrane patch $\scp$ may only change due to a diffusive mass flux at the boundary.
In this case, the global form of the conservation of mass of species $k$ is given by
\begin{equation} \label{eq:global-mass-balance-k}
	\dfrac{\mathrm{d}}{\mathrm{d}t} \bigg(
		\int_\scp \rho_k ~\mathrm{d}a
	\bigg)
	= - \int_{\partial \scp} \!\! \bm{j}_k \cdot \bm{\nu} ~\mathrm{d}s ~.
\end{equation}
Applying the Reynolds transport theorem \eqref{eq:rtt} and the surface divergence theorem \eqref{eq:divergence-theorem} to equation \eqref{eq:global-mass-balance-k}, we obtain
\begin{equation} \label{eq:global-mass-balance-k-simplified}
	\int_\scp \Big(
		\dot{\rho}_k + \big(
			v^\alpha_{; \alpha} - 2 v H
		\big) \rho_k
	\Big) ~\mathrm{d}a
	= - \int_\scp j_{k \, ; \alpha}^{\, \alpha} ~\mathrm{d}a ~.
\end{equation}
In equation \eqref{eq:global-mass-balance-k-simplified} and from now on, $v^\alpha$ and $v$ refer to the tangential and normal components of the barycentric velocity $\bm{v}$, respectively, as decomposed in equation \eqref{eq:velocity-euler}.
Since the membrane patch $\scp$ in equation \eqref{eq:global-mass-balance-k-simplified} is arbitrary, we obtain the local form of the conservation of mass of species $k$ as
\begin{align} 
	\dot{\rho}_k
	+ \big( v^\alpha_{; \alpha} - 2 v H \big) \rho_k
	= - j^{\, \alpha}_{k \, ; \alpha} ~.
	\label{eq:local-mass-balance-k}
\end{align}

Summing equation \eqref{eq:local-mass-balance-k} over all $k$ and realizing the right hand side is zero due to equation \eqref{eq:sum-diff-flux}, we recover the local form of the total mass balance to be given by
\begin{equation} \label{eq:local-mass-balance-mc} 
	\dot{\rho}
	+ \big( v^\alpha_{; \alpha} - 2 v H \big) \rho
	= 0 ~.
\end{equation}
Using the local forms of the species mass balance \eqref{eq:local-mass-balance-k} and total mass balance \eqref{eq:local-mass-balance-mc} as well as the definition of $c_k$ \eqref{eq:concentration-definition}, we obtain the balance of the mass fraction of species $k$ as
\begin{align}
	\rho \dot{c}_k
	= - j_{k \, ; \alpha}^{\, \alpha} ~.
	\label{eq:concentration-dot}
\end{align}


\subsubsection{Linear Momentum Balance} \label{sec:linear-momentum-mc}

For a membrane with multiple components, each species $k$ is acted on by a body force $\bm{b}_k (\theta^\alpha, t)$.
The entire membrane patch is subjected to a traction $\bm{T} (\bmxb, t; \bm{\nu})$ at the boundary, and the total linear momentum of the membrane patch is the sum of the linear momentum of each species as shown in equation \eqref{eq:barycentric-velocity}.
In this case, the global form of the linear momentum balance is given by
\begin{equation} \label{eq:linear-momentum-balance-mc}
	\dfrac{\mathrm{d}}{\mathrm{d}t} \bigg[
		\int_\scp \Big(
			\sum_{k = 1}^N 
			\rho_k \bm{v}_k
		\Big) ~\mathrm{d}a 
	\bigg] 
	= \int_\scp \Big(
		\sum_{k = 1}^N \rho_k \bm{b}_k
	\Big) ~\mathrm{d}a
	+ \int_{\partial \scp} \!\! \bm{T} ~\mathrm{d}s
	~.
\end{equation}
Defining the mass-weighted body force $\bm{b}$ as
\begin{equation} \label{eq:body-force-barycentric}
	\rho \bm{b} = \sum_{k = 1}^N \rho_k \bm{b}_k ~,
\end{equation}
and using the barycentric velocity $\bm{v}$ in equation \eqref{eq:barycentric-velocity}, the global linear momentum balance \eqref{eq:linear-momentum-balance-mc} can be reduced to
\begin{equation} \label{eq:global-linear-momentum-balance-step-2-mc}
	\dfrac{\mathrm{d}}{\mathrm{d}t} \bigg(
		\int_\scp \rho \bm{v} ~\mathrm{d}a
	\bigg) 
	= \int_\scp \rho \bm{b} ~\mathrm{d}a
	+ \int_{\bscp} \!\! \bm{T} ~\mathrm{d}s
	~.
\end{equation}
Equation \eqref{eq:global-linear-momentum-balance-step-2-mc} has the same form as the single-component global linear momentum balance \eqref{eq:euler}, where now the body force and velocity are mass-averaged quantities.
The boundary traction $\bm{T}$ may once again be decomposed into a linear combination of stress vectors as in equation \eqref{eq:traction-decomposition}, and by using the same set of procedures as in the single-component section we find the local form of the linear momentum balance is given by
\begin{equation} \label{eq:linear-momentum-balance-local-mc}
	\rho \dot{\bm{v}} 
	= \rho \bm{b} + \bm{T}^\alpha_{; \alpha} 
	~.
\end{equation}
As equation \eqref{eq:linear-momentum-balance-local-mc} is identical to the single-component result \eqref{eq:local-momentum-balance}, equations \eqref{eq:stress-vector-general}--\eqref{eq:cauchy-stress-tensor} continue to be valid descriptions of the membrane stresses.
\textspace

We have so far seen the local mass balance and local linear momentum balance are identical for single-component and multi-component membranes.
The director traction $\bm{M}$ may also be decomposed in the same manner as in the single-component case \eqref{eq:distributed-moment-decomposition}, and thus the results of the single-component angular momentum balance---namely the symmetry of $\sigma^{\alpha \beta}$ \eqref{eq:angular-momentum-symmetric} and the form of $S^{\alpha}$ \eqref{eq:angular-momentum-s}---are appropriate for multi-component membranes as well.
Furthermore, the mechanical power balance is unchanged from the single-component case \eqref{eq:global-power-balance-solution} since it only depends on the linear momentum balance.


\subsection{Thermodynamics}

In this section, we follow the same procedure as in the single-component section to develop the local form of the first law of thermodynamics.
The local form of the entropy balance and the second law of thermodynamics are unchanged.
We end with the expression for the internal entropy production.


\subsubsection{First Law---Energy Balance} \label{sec:first-law-mc}

The global form of the first law of thermodynamics for the multi-component membrane system is given by
\begin{equation} \label{eq:global-energy-balance-mc}
	\dfrac{\mathrm{d}}{\mathrm{d}t} \bigg( \int_\scp \rho e ~\mathrm{d}a \bigg)
	= \int_\scp \rho r ~\mathrm{d}a 
	- \int_{\partial \scp} \!\! \bmjq \cdot \bm{\nu} ~\mathrm{d}s
	+ \int_\scp \Big( \sum_{k = 1}^N \rho_k \bm{b}_k \cdot \bm{v}_k \Big) ~\mathrm{d}a
	+ \int_{\partial \scp} \!\! \Big(
		\bm{v} \cdot \bm{T} 
		+ \bm{M} \cdot \dot{\bm{n}}
	\Big) ~\mathrm{d}s ~.
\end{equation}
Equation \eqref{eq:global-energy-balance-mc} is identical to the one-component result \eqref{eq:global-energy-balance}, except for the force term which is now a sum over the individual species. 
The difference in body force contribution between the global form of the first law of thermodynamics \eqref{eq:global-energy-balance-mc} and the mechanical power balance \eqref{eq:global-power-balance-solution} is 
\begin{equation} \label{eq:body-force-diff-mc}
	\begin{split}
		\int_\scp \Big(
			\sum_{k = 1}^N \rho_k \bm{b}_k \cdot \bm{v}_k
		\Big) ~\mathrm{d}a
		- \int_\scp \big(
			\rho \bm{b} \cdot \bm{v}
		\big) ~\mathrm{d}a
		&= \int_\scp \Big(
			\sum_{k = 1}^N \rho_k \bm{b}_k \cdot \bm{v}_k
			- \sum_{k = 1}^N \rho_k \bm{b}_k \cdot \bm{v}
		\Big) ~\mathrm{d}a \\[4pt]
		&= \int_\scp \Big(
			\sum_{k = 1}^N \rho_k \bm{b}_k \cdot \big( \bm{v}_k - \bm{v} \big)
		\Big) ~\mathrm{d}a \\[4pt]
		&= \int_\scp \Big(
			\sum_{k = 1}^N \bm{b}_k \cdot \bm{j}_k
		\Big) ~\mathrm{d}a ~,
	\end{split}
\end{equation}
where we have used the definition of the mass averaged body force $\bm{b}$ \eqref{eq:body-force-barycentric} and the definition of the in-plane diffusive flux $\bm{j}_k$ \eqref{eq:diffusive-species-flux}.
With equation \eqref{eq:body-force-diff-mc}, the mechanical power balance \eqref{eq:global-power-balance-solution}, and assuming the decomposition of total energy $e$ into internal energy $u$ and kinetic energy given in equation \eqref{eq:energy-def}, the global form of the first law of thermodynamics \eqref{eq:global-energy-balance-mc} simplifies to
\begin{equation} \label{eq:global-energy-balance-step-2-mc}
	\begin{split}
		\int_\scp \rho \dot{u}  ~\mathrm{d}a
		= \int_\scp \bigg(
			\rho r
			- J_{\mathrm{q} \, ; \alpha}^{\, \alpha} 
			+ \sum_{k = 1}^N (b_k)_\alpha \, j_k^{\, \alpha}
			+ \dfrac{1}{2} \sigma^{\alpha \beta} \dot{a}_{\alpha \beta} 
			+ M^{\alpha \beta} \dot{b}_{\alpha \beta} 
		\bigg) ~\mathrm{d}a ~.
	\end{split}	
\end{equation}
In writing the covariant component of the in-plane body force acting on species $k$, $(b_k)_\alpha$, in equation \eqref{eq:global-energy-balance-step-2-mc}, we have explicitly included parenthesis to avoid confusion with the curvature tensor.
Comparing equation \eqref{eq:global-energy-balance-step-2-mc} with the single-component result \eqref{eq:global-energy-balance-step-2} shows the only difference is the $\bm{b}_k \cdot \bm{j}_k$ term calculated in equation \eqref{eq:body-force-diff-mc}.
Since the membrane patch $\scp$ is arbitrary, the local form of the first law of thermodynamics is given by
\begin{equation} \label{eq:first-law-thermo-ip}
	\rho \dot{u}
	= \rho r
	- J_{\mathrm{q} \, ; \alpha}^{\, \alpha} 
	+ \sum_{k = 1}^N (b_k)_\alpha \, j_k^{\, \alpha}
	+ \dfrac{1}{2} \sigma^{\alpha \beta} \dot{a}_{\alpha \beta} 
	+ M^{\alpha \beta} \dot{b}_{\alpha \beta}  ~.
\end{equation}
Comparing equation \eqref{eq:first-law-thermo-ip} to equation \eqref{eq:local-energy-balance}, one additional term has appeared---namely the sum of $(b_k)_\alpha \, j_k^{\, \alpha}$.


\subsubsection{Entropy Balance \& Second Law}

We define the in-plane entropy flux $\bmjs$, the external entropy supply $\rho \eta_{\mathrm{e}}$, and the internal entropy production $\rho \eta_{\mathrm{i}}$ as in the single-component model.
The global entropy balance is unchanged from the single-component case \eqref{eq:global-entropy-balance}, and consequently the local form of the entropy balance is given by equation \eqref{eq:local-entropy-balance}.
The second law of thermodynamics is unchanged as well, and is given by equation~\eqref{eq:second-law-thermo-local}.
\\[-7pt]

At this stage, we invoke the Helmholtz free energy density $\psi$ for the multi-component membrane system, which is of the form given in equation \eqref{eq:helmholtz}.
Following an identical procedure to the single-component case described in Section \ref{sec:sc-choice-thermo-potential}, we find the local form of the entropy balance can be expressed as
\begin{equation} \label{eq:entropy-energy-balance-mc}
	\begin{split}
		\rho \dot{s}
		&= - J_{\mathrm{s} \, ; \alpha}^{\, \alpha}
		+ \rho \eta_{\mathrm{e}}
		+ \rho \eta_{\mathrm{i}} \\[4pt]
		&= \dfrac{1}{T} \bigg(
			\rho r
			- J_{\mathrm{q} \, ; \alpha}^{\, \alpha}
			+ \sum_{k = 1}^N (b_k)_\alpha \, j_k^{\, \alpha}
			+ \dfrac{1}{2} \sigma^{\alpha \beta} \dot{a}_{\alpha \beta}
			+ M^{\alpha \beta} \dot{b}_{\alpha \beta}
			- \rho \dot{T} s
			- \rho \dot{\psi}
		\bigg) ~.
		\vspace{4pt}
	\end{split}
\end{equation}


\subsection{Constitutive Relations} \label{sec:is-ce}

In this section, we extend the framework developed in the single-component section.
When we apply linear irreversible thermodynamics to relate thermodynamic forces to their corresponding fluxes, we find diffusive species fluxes are driven by gradients in chemical potential and non-uniform body forces across different species.
We learn that the form of the stresses and moments are identical to the single-component case.


\subsubsection{General Thermodynamic Variables} 

With $N$ types of phospholipids composing the lipid membrane, there are an additional $N - 1$ degrees of freedom relative to the single-component case, which are the $N - 1$ mass fractions $\{ c_k \}_{k = 2, \dots , N} := \{ c_2, c_3, c_4, \dots, c_N \}$.
For this case, the Helmholtz free energy density per unit mass for the membrane system is formally written as
\begin{equation} \label{eq:psi-constitutive-ip}
	\psi = \psi(a_{\alpha \beta}, b_{\alpha_\beta}, T, \{c_k\}_{k = 2, \dots , N}) ~.
\end{equation}
We define the chemical potential of species $k$, $\mu_k$, as
\begin{equation} \label{eq:chemical-potential-def}
	\mu_k := \dfrac{\partial \psi}{\partial c_k} ~, \hspace{15pt} k \in \{2, 3, 4, \dots, N\} ~,
\end{equation}
where the partial derivative holds $a_{\alpha \beta}$, $b_{\alpha \beta}$, and $c_{j \ne k}$ constant, with $j$ in $\{2, 3, 4, \dots, N\}$.
The chemical potential $\mu_1$ is not defined in equation \eqref{eq:chemical-potential-def} as $c_1$ is not an independent variable due to the relation \eqref{eq:sum-concentration}.
For the membrane system, the chemical potential describes the change in Helmholtz free energy when species $k$ is exchanged with species one, such that all other mass fractions remain constant.
\textspace

Taking the material derivative of equation \eqref{eq:psi-constitutive-ip} and multiplying by $\rho$, we obtain
\begin{equation} \label{eq:rho-dot-psi-ip}
	\begin{split}
		\rho \dot{\psi}
		&= \dfrac{\rho}{2} \Big(
			\dfrac{\partial \psi}{\partial a_{\alpha \beta}}
			+ \dfrac{\partial \psi}{\partial a_{\beta \alpha}}
		\Big) \dot{a}_{\alpha \beta}
		+ \dfrac{\rho}{2} \Big(
			\dfrac{\partial \psi}{\partial b_{\alpha \beta}}
			+ \dfrac{\partial \psi}{\partial b_{\beta \alpha}}
		\Big) \dot{b}_{\alpha \beta}
		+ \rho \dot{T} \, \dfrac{\partial \psi}{\partial T} 
		+ \sum_{k = 2}^N \rho \dot{c}_k \, \dfrac{\partial \psi}{\partial c_k} \\
		&= \dfrac{\rho}{2} \Big(
			\dfrac{\partial \psi}{\partial a_{\alpha \beta}}
			+ \dfrac{\partial \psi}{\partial a_{\beta \alpha}}
		\Big) \dot{a}_{\alpha \beta}
		+ \dfrac{\rho}{2} \Big(
			\dfrac{\partial \psi}{\partial b_{\alpha \beta}}
			+ \dfrac{\partial \psi}{\partial b_{\beta \alpha}}
		\Big) \dot{b}_{\alpha \beta}
		- \rho s \dot{T} 
		- \sum_{k = 2}^N \mu_k \, j_{k \, ; \alpha}^{\, \alpha} ~,
	\end{split}
\end{equation}
where in obtaining the second equality we have substituted equations \eqref{eq:concentration-dot} and \eqref{eq:chemical-potential-def} and used the local equilibrium assumption given by 
\begin{equation} \label{eq:local-equilibrium-entropy-mc}
	s
	= -\Big(
		\dfrac{\partial \psi}{\partial T}
	\Big)_{a_{\alpha \beta}, \, b_{\alpha \beta}, \, \{c_k\}_{k = 2, \dots , N}}
	~.
\end{equation}
We substitute equation \eqref{eq:rho-dot-psi-ip} into the entropy balance \eqref{eq:entropy-energy-balance-mc} to find 
\begin{equation} \label{eq:second-law-thermo-local-expanded-mc}
	\begin{split}
		\rho \dot{s}
		&= - J_{\mathrm{s} \, ; \alpha}^{\, \alpha} 
		+ \rho \eta_{\mathrm{e}}
		+ \rho \eta_{\mathrm{i}} \\[3pt]
		&= \dfrac{1}{T} \bigg\{
			\rho r
			- J_{\mathrm{q} \, ; \alpha}^{\, \alpha}
			+ \sum_{k = 1}^N (b_k)_\alpha \, j_k^{\, \alpha}
			+ \sum_{k = 2}^N \mu_k \, j_{k \, ; \alpha}^{\, \alpha}
		\\[3pt]
		& \hspace{40pt} + 
			\dfrac{1}{2} \bigg[
				\sigma^{\alpha \beta}
				- \rho \Big(
					\dfrac{\partial \psi}{\partial a_{\alpha \beta}}
					+ \dfrac{\partial \psi}{\partial a_{\beta \alpha}}
				\Big)
			\bigg] \dot{a}_{\alpha \beta}
			+ \bigg[
				M^{\alpha \beta}
				- \dfrac{\rho}{2} \Big(
					\dfrac{\partial \psi}{\partial b_{\alpha \beta}}
					+ \dfrac{\partial \psi}{\partial b_{\beta \alpha}}
				\Big)
			\bigg] \dot{b}_{\alpha \beta} 
		\bigg\} ~.
	\end{split}
\end{equation}
We rewrite the gradient terms in equation \eqref{eq:second-law-thermo-local-expanded-mc} to obtain
\begin{equation} \label{eq:second-law-thermo-local-step-2-mc}
	\begin{split}
		\rho \dot{s}
		&= - J_{\mathrm{s} \, ; \alpha}^{\, \alpha} 
		+ \rho \eta_{\mathrm{e}}
		+ \rho \eta_{\mathrm{i}} \\[3pt]
		&= - \Big(
			\dfrac{\jq^{\, \alpha}
			- \sum_{k = 2}^N \mu_k \, j_k^{\, \alpha}}{T}
		\Big)_{; \alpha}
	 	+ \dfrac{\rho r}{T}
		- \dfrac{\jq^{ \, \alpha} \, T_{, \alpha}}{T^2}
		+ \sum_{k = 1}^N \dfrac{(b_k)_\alpha \, j_k^{\, \alpha}}{T} 
		- \sum_{k = 2}^N \Big(
			\dfrac{\mu_k}{T}
		\Big)_{\!\! , \alpha} j_k^{\, \alpha}
		\\[3pt]
		&\hspace{20pt} + \dfrac{1}{T} \bigg\{
			\dfrac{1}{2} \bigg[
				\sigma^{\alpha \beta}
				- \rho \Big(
					\dfrac{\partial \psi}{\partial a_{\alpha \beta}}
					+ \dfrac{\partial \psi}{\partial a_{\beta \alpha}}
				\Big)
			\bigg] \dot{a}_{\alpha \beta}
			+ \bigg[
				M^{\alpha \beta}
				- \dfrac{\rho}{2} \Big(
					\dfrac{\partial \psi}{\partial b_{\alpha \beta}}
					+ \dfrac{\partial \psi}{\partial b_{\beta \alpha}}
				\Big)
			\bigg] \dot{b}_{\alpha \beta} 
		\bigg\} ~.
	\end{split}
\end{equation}
From equation \eqref{eq:second-law-thermo-local-step-2-mc}, we see only the first term contributes to the in-plane entropy flux $J_{\mathrm{s}}^{\, \alpha}$, given by
\begin{equation} \label{eq:entropy-flux-ip}
	J_{\mathrm{s}}^{\, \alpha}
	= \dfrac{1}{T} \bigg(
		\jq^{\, \alpha}
		- \sum_{k = 2}^N \mu_k \, j_k^{\, \alpha}
	\bigg) ~.
\end{equation}
Again, the heat source per unit mass is the only term which contributes to the external entropy supply $\rho \eta_{\mathrm{e}}$, given by
\begin{align}
	\rho \eta_{\mathrm{e}}
	= \dfrac{\rho r}{T} ~.
	\label{eq:external-entropy-ip}
\end{align}
The terms on the right hand side of equation \eqref{eq:second-law-thermo-local-step-2-mc} which contribute to neither the entropy flux nor the external entropy contribute to the rate of internal entropy production per unit area $\rho \eta_{\mathrm{i}}$, yielding
\begin{equation} \label{eq:internal-entropy-ip}
	\begin{split}
		\rho \eta_{\mathrm{i}}
		&= - \dfrac{\jq^{ \, \alpha} \, T_{, \alpha}}{T^2}
		+ \sum_{k = 1}^N \dfrac{(b_k)_\alpha \, j_k^{\, \alpha}}{T} 
		- \sum_{k = 2}^N \Big(
			\dfrac{\mu_k}{T}
		\Big)_{\!\! , \alpha} j_k^{\, \alpha} \\[3pt]
		&\hspace{10pt} + \dfrac{1}{T} \bigg\{
			\dfrac{1}{2} \bigg[
				\sigma^{\alpha \beta}
				- \rho \Big(
					\dfrac{\partial \psi}{\partial a_{\alpha \beta}}
					+ \dfrac{\partial \psi}{\partial a_{\beta \alpha}}
				\Big)
			\bigg] \dot{a}_{\alpha \beta}
			+ \bigg[
				M^{\alpha \beta}
				- \dfrac{\rho}{2} \Big(
					\dfrac{\partial \psi}{\partial b_{\alpha \beta}}
					+ \dfrac{\partial \psi}{\partial b_{\beta \alpha}}
				\Big)
			\bigg] \dot{b}_{\alpha \beta} 
		\bigg\} \ge 0 ~.
	\end{split}
\end{equation}
Using equation \eqref{eq:sum-diff-flux}, we simplify the second term on the right hand side of equation \eqref{eq:internal-entropy-ip} as
\begin{equation} \label{eq:body-diff-flux-manipulation}
	\begin{split}
		\sum_{k = 1}^N \dfrac{(b_k)_\alpha \, j_k^{\, \alpha}}{T} 
		&=  \dfrac{(b_1)_\alpha \, j_1^{\, \alpha}}{T} 
		+ \sum_{k = 2}^N \dfrac{(b_k)_\alpha \, j_k^{\, \alpha}}{T} \\
		&= \sum_{k = 2}^N \bigg(
			\dfrac{(b_k)_\alpha - (b_1)_\alpha}{T}
		\bigg) j_k^{\, \alpha} ~.
	\end{split}
\end{equation}
Substituting equation \eqref{eq:body-diff-flux-manipulation} into the rate of internal entropy production per unit area \eqref{eq:internal-entropy-ip}, we obtain
\begin{equation} \label{eq:internal-entropy-ip-step-2}
	\begin{split}
		\rho \eta_{\mathrm{i}}
		&= - \dfrac{\jq^{ \, \alpha} \, T_{, \alpha}}{T^2}
		+ \sum_{k = 2}^N \bigg(
			\dfrac{(b_k)_\alpha - (b_1)_\alpha}{T}
			- \Big(
				\dfrac{\mu_k}{T}
			\Big)_{\!\! , \alpha}
		\bigg) j_k^{\, \alpha} \\[3pt]
		&\hspace{15pt} + \dfrac{1}{T} \bigg\{
			\dfrac{1}{2} \bigg[
				\sigma^{\alpha \beta}
				- \rho \Big(
					\dfrac{\partial \psi}{\partial a_{\alpha \beta}}
					+ \dfrac{\partial \psi}{\partial a_{\beta \alpha}}
				\Big)
			\bigg] \dot{a}_{\alpha \beta}
			+ \bigg[
				M^{\alpha \beta}
				- \dfrac{\rho}{2} \Big(
					\dfrac{\partial \psi}{\partial b_{\alpha \beta}}
					+ \dfrac{\partial \psi}{\partial b_{\beta \alpha}}
				\Big)
			\bigg] \dot{b}_{\alpha \beta} 
		\bigg\} \ge 0 ~.
	\end{split}
\end{equation}

Each term on the right hand side of equation \eqref{eq:internal-entropy-ip-step-2} is the product of a thermodynamic force and a corresponding flux, as in equation \eqref{eq:thermo-force-rate}.
In the linear irreversible regime, one can now assume linear phenomenological relations between the thermodynamic forces and fluxes as in equation \eqref{eq:phenomenological}.
We invoke the Curie principle \cite{curie, degroot-mazur} and assume the fluxes $\jq^{\, \alpha}$ and $j_k^{\, \alpha}$ are independent from the tensorial thermodynamic forces $\dot{a}_{\alpha \beta}$ and $\dot{b}_{\alpha \beta}$, and the tensorial fluxes are independent from the vectorial thermodynamic forces.
Therefore, in the linear irreversible regime the most general relations between the vectorial fluxes and forces are given by 
\begin{align}
	\jq^{\, \alpha}
	&= - \kappa^{\alpha \beta} \, T_{, \beta}
	+ \sum_{k = 2}^N F_k^{\, \alpha \beta} \bigg(
		\dfrac{(b_k)_\beta - (b_1)_\beta}{T}
		- \Big(
			\dfrac{\mu_k}{T}
		\Big)_{\!\! , \beta}
	\bigg) \\
	\shortintertext{and}
	j_k^{\, \alpha}
	&= - G^{\alpha \beta} T_{, \beta}
	+ \sum_{\ell = 2}^N D_\ell^{\, \alpha \beta} \bigg(
		\dfrac{(b_\ell)_\beta - (b_1)_\beta}{T}
		- \Big(
			\dfrac{\mu_\ell}{T}
		\Big)_{\!\! , \beta}
	\bigg) ~,
\end{align}
where the phenomenological coefficients $F_k^{\, \alpha \beta}$ and $G^{\alpha \beta}$ describe interference between the heat flow and in-plane species diffusion.
For the lipid bilayers under consideration, we assume these to be zero.
In the case where there is no cross-coupling between the heat and diffusive fluxes, the in-plane fluxes simplify to
\begin{gather}
	\jq^{\, \alpha}
	= - \kappa^{\alpha \beta} \, T_{, \beta} \\
	\shortintertext{and}
	j_k^{\, \alpha}
	= \sum_{\ell = 2}^N D_\ell^{\, \alpha \beta} \bigg(
		\dfrac{(b_\ell)_\beta - (b_1)_\beta}{T}
		- \Big(
			\dfrac{\mu_\ell}{T}
		\Big)_{\!\! , \beta}
	\bigg) ~.
\end{gather}
We assume, as before, the thermal conductivity tensor $\kappa^{\alpha \beta}$ is isotropic and obtain
\begin{equation} \label{eq:heat-flux-mc}
	\jq^{\, \alpha}
	= - \kappa \, T^{, \alpha} ~.
\end{equation}
As noted in the single-component case, we assume lipid bilayers do not sustain in-plane temperature gradients and equation \eqref{eq:heat-flux-mc} is not of relevance in our modeling of irreversible processes.
Assuming further the diffusion tensors $D_\ell^{\, \alpha \beta}$ only affect $j_k^{\, \alpha}$ when $k = \ell$ and $D_k^{\, \alpha \beta}$ is isotropic, we obtain $D_\ell^{\, \alpha \beta} = D_k \, a^{\alpha \beta}$ when $k = \ell$ and $D_\ell^{\, \alpha \beta} = 0$ otherwise.
With these assumptions, the linear phenomenological relation for the in-plane species flux $\bm{j}_k$ simplifies to 
\begin{equation} \label{eq:species-flux-ip}
	j_k^{\, \alpha}
	= D_k \bigg(
		\dfrac{(b_k)^\alpha - (b_1)^\alpha}{T}
		- \Big(
			\dfrac{\mu_k}{T}
		\Big)^{, \alpha}
	\bigg) ~.
\end{equation}
Equation \eqref{eq:species-flux-ip} indicates gradients in chemical potentials and differences in the in-plane forces $\bm{b}_k$ between individual species drive the diffusive fluxes.
\textspace

Finally, the tensorial terms in the rate of internal entropy production $\rho \eta_{\mathrm{i}}$ are identical to those of the single-component case \eqref{eq:internal-entropy-complete}.
Therefore, following similar arguments to the single-component case, we find the form of the stresses and moments in equations \eqref{eq:sigma-summary}--\eqref{eq:s-summary} remain unchanged for multi-component systems.
\textspace

For lipid bilayers, it can again be argued that Galilean invariance \cite{steigmann-fluid-film-arma-1999} or material symmetry arguments \cite{jenkins-siam-1977} lead to a change in the fundamental variables of the Helmholtz free energy density, as given by
\begin{equation} \label{eq:psi-bar-intramembrane}
	\psi (a_{\alpha \beta}, b_{\alpha \beta}, T, \{c_k\}_{k = 2, \dots, N})
	= \bar{\psi} (\rho, H, K, T, \{c_k\}_{k = 2, \dots, N}) ~.
\end{equation}
Given equation \eqref{eq:psi-bar-intramembrane}, we note there is again no change to the form of the stresses and moments given by equations \eqref{eq:sigma-alpha-beta-1c-general}--\eqref{eq:viscous-stress}, in terms of their depedendence on $\rho$, $H$, and $K$.
\textspace

We have now determined the governing equations for a general multi-component membrane patch, for which the unknowns are the density $\rho$, the three components of the velocity $\bm{v}$, and the $N - 1$ mass fractions $c_k$ for $k \in \{2, 3, \dots, N \}$.
The corresponding equations are given by the total mass balance \eqref{eq:local-mass-balance-mc}, the three linear momentum balances \eqref{eq:tangential-general-eoms-elastic-compressible}--\eqref{eq:normal-general-eom-elastic-compressible} in which we know the equations for the stresses, and the $N - 1$ linearly independent mass fraction balances for $c_k$ \eqref{eq:concentration-dot}, where $k \in \{ 2, 3, \dots, N \}$ and $j_k^{\, \alpha}$ is given by equation \eqref{eq:species-flux-ip}.
\textspace

As the form of the tractions $\bm{T}$ and director tractions $\bm{M}$ on the patch boundary are unchanged, the boundary conditions outlined for the single-component system in Section \ref{sec:sc-boundary-conditions} are again appropriate for the mechanical equations of motion of multi-component systems.
As the total mass balance \eqref{eq:local-mass-balance-mc} and mass fraction balances \eqref{eq:concentration-dot} contain only first order derivatives in time, we need to provide their initial values to solve for the dynamics of the membrane patch.
The required boundary conditions on $c_k$ will depend on the form of the Helmholtz free energy, as will be seen in the following example.


\subsection{Application to L\textsubscript{o}--L\textsubscript{d} Phase Transitions}

Now that we have found the governing equations for a general multi-component membrane, we return to the specific case of L\textsubscript{o}--L\textsubscript{d} phase coexistence observed in GUVs \cite{veatch-keller-bpj-2003, baumgart-hess-nature-2003}.
As mentioned earlier, a membrane initially in an L\textsubscript{d} phase can form interfaces between \lo\ and \ld\ regions when, for example, the temperature is quenched (Figure \ref{fig:diffusion}).
Experimentally, such a phase transition has been observed in three-component systems \cite{veatch-keller-bpj-2003, veatch-keller-prl-2002, veatch-keller-bba-2005}.
\textspace

We consider a scenario in which the three components of a membrane patch are DOPC, DPPC, and cholesterol as in \cite{veatch-keller-bpj-2003}.
The \ld\ phase consists of DOPC, which is a low melting temperature lipid, while the \lo\ phase consists of DPPC, which is a high melting temperature lipid, as well as cholesterol.
Because the mass fraction of DOPC is an order parameter in that it can be used to specify which phase a region is in, we define the dimensionless order parameter $\phi$ as
\begin{equation} \label{eq:phi-two-component-is}
	\phi = c_{\raisebox{-1pt}{\text{\tiny{DOPC}}}} ~. 
\end{equation}
In doing so, we choose to model our three-component system with only two components, in which one component is DOPC while the other component is the lumped DPPC and cholesterol.

%
%

\subsubsection{Helmholtz Free Energy}

With two membrane components, the Helmholtz free energy density contains four terms.
The first of these terms captures the energetic cost of bending, through a modified form of the single-component Helfrich energy density $w_{\text{h}}$ \eqref{eq:w-helfrich} in which the bending rigidity $\kb$ and spontaneous curvature $C$, if it exists, depend on the local mass fraction $\phi$.
Such modifications reflect experimental observations that the bending modulus of \lo\ and \ld\ phases differ \cite{baumgart-hess-nature-2003, duwe-physica-a-1990}, and also that local species concentrations may affect the spontaneous curvature of the membrane \cite{bacia-nature-2011}.
The Helfrich energy density for our two-component membrane system is 
\begin{equation} \label{eq:w-helfrich-mc}
	w_{\text{h}}
	= \kb(\phi) \, \big[
		H - C(\phi)
	\big]^2
	+ \kg K ~.
\end{equation}
\eqnspace

The Helmholtz free energy density also contains a modified form of the single-component cost of areal dilation and compression, $w_{\text{c}}$ \eqref{eq:w-compress}, in which the compression modulus $\kc$ is again modified to depend on the local mass fraction $\phi$.
Different phases may be relatively more or less compressible, and the multi-component compression energy $w_{\text{c}}$ is written as
\begin{equation} \label{eq:w-compress-mc}
	w_{\text{c}}
	= \dfrac{\kc (\phi)}{J} \big(
		1 - J
	\big)^2 ~.
\end{equation}

Next, for our Helmholtz free energy to give rise to phase separation, we include a double-well potential $w_{\text{dw}}$ by considering a mean-field model of molecular interactions \cite{cahn-hilliard-jcp-1958, hill, thanos}, given by 
\begin{equation} \label{eq:w-dw}
	w_{\textrm{dw}}
	= \dfrac{k_\phi}{J} \Big(
		\chi \phi \big( 1 - \phi \big)
		+ k_{\textrm{B}} \, T \, \Big[
			\phi \ln \phi
			+ (1 - \phi) \ln (1 - \phi)
		\Big]
	\Big) ~.
\end{equation}
The double-well potential contains a competition between the energetic term $\chi \phi (1 - \phi)$ and the entropic term $k_{\textrm{B}} \, T \, [ \phi \ln \phi + (1 - \phi) \ln (1 - \phi) ]$, where $k_{\textrm{B}}$ is Boltzmann's constant and $\chi$, which has units of energy, describes the mean-field interaction energies of the two components.
In equation \eqref{eq:w-dw}, we assume the temperature is low enough such that the energetic term dominates and the potential has two minima---for our model, this assumption is valid for $\chi > 2 \, k_{\textrm{B}} \, T$.
The parameter $k_\phi$ tunes the relative energetic penalty of concentrations which are not energetically favorable.
The factor of $1/J$ is required for $w_{\text{dw}}$ to have no density dependence when convected to the reference patch $\scp_0$, in the same way as the compression energy $w_{\text{c}}$.
\textspace

The final term in our two-component Helmholtz free energy accounts for the line tension at the interface between two phases.
We consider a simplified view in which $\phi = 1$ in the \ld\,phase and $\phi = 0$ in the \lo\,phase, such that the gradient of $\phi$ is a delta function which is nonzero at every point along the phase boundary.
This simple example motivates the gradient term in the Helmholtz free energy, $w_{\text{g}}$, written as 
\begin{equation} \label{eq:w-g}
	w_{\mathrm{g}}
	= \dfrac{\gamma}{2} \big(
		\phi_{, \mu} \, \phi^{, \mu}
	\big) ~.
\end{equation}
In equation \eqref{eq:w-g}, we have chosen the simplest form of $w_{\text{g}}$ which penalizes phase boundaries and is invariant to coordinate transformations.
The constant $\gamma$ describes the energetic penalty for phase boundaries, and equation \eqref{eq:w-g} is the curvilinear form of the energetic contribution in standard Cartesian coordinates \cite{cahn-hilliard-jcp-1958}.
\textspace

Combining all of the energy contributions, the total Helmholtz free energy density $\psi$ for our two-component membrane system is 
\begin{equation} \label{eq:total-helmholtz-2c-is}
	\begin{split}
		\rho \psi
		&= w_{\textrm{h}}
		+ w_{\textrm{c}}
		+ w_{\textrm{g}}
		+ w_{\textrm{dw}} \\[3pt]
		&= \kb(\phi) \big[ H - C(\phi) \big]^2
		+ \kg K
		+ \dfrac{\kc (\phi)}{J} \big( 1 - J \big)^2 \\[3pt]
		&\hspace{20pt} + \dfrac{\gamma}{2} \phi_{, \mu} \phi^{, \mu}
		+ \dfrac{k_\phi}{J} \Big(
			\chi \phi (1 - \phi)
			+ k_{\mathrm{B}} \, T \, \big[
				\phi \ln \phi
				+ (1 - \phi) \ln (1 - \phi)
			\big]
		\Big) ~.
	\end{split}
\end{equation}
To determine the stresses, we note that $w_{\text{g}} = \tfrac{\gamma}{2} \phi_{, \mu} \phi^{, \mu} = \tfrac{\gamma}{2} \phi_{, \mu} \phi_{, \lambda} a^{\lambda \mu}$ and thus $\psi$ may not be completely expressed as a function of only $\rho$, $H$, $K$, $T$, and $\phi$---consequently, in equation \eqref{eq:total-helmholtz-2c-is} we have written $\psi$ instead of $\bar{\psi}$. 
We therefore use the equations for the stresses in terms of the more general variables $a_{\alpha \beta}$, $b_{\alpha \beta}$, $T$, and $\phi$ \eqref{eq:sigma-summary}--\eqref{eq:s-summary}.
Using the partial derivatives found in Table \ref{partial-derivatives}, we find the membrane stresses and couple-stresses are given by
\begin{align}
	\begin{split}
		\sigma^{\alpha \beta}
		&= \kb(\phi) \Big(
			\Big[
				-3 H^2
				+ 2 H C(\phi)
				+ \big( C(\phi) \big)^2
			\Big] a^{\alpha \beta}
			+ 2 \Big[
				H - C(\phi)
			\Big] \bar{b}^{\alpha \beta}
		\Big) \\[2pt]
		&\hspace{10pt} - \kg K a^{\alpha \beta}
		+ 2 \kc(\phi) \Big( J - 1 \Big) a^{\alpha \beta}
		+ \gamma \Big(
			\dfrac{1}{2} \phi_{, \mu} \phi^{, \mu} \, a^{\alpha \beta}
			- \phi^{, \alpha} \phi^{, \beta}
		\Big)
		+ \pi^{\alpha \beta}
		~,
		\label{eq:sigma-alpha-beta-im-helfrich}
	\end{split}
	\\[9pt]
	\begin{split}
		M^{\alpha \beta}
		&= \kb (\phi) \Big[
			H - C (\phi)
		\Big] a^{\alpha \beta}
		+ \kg \bar{b}^{\alpha \beta} ~,
		\label{eq:m-alpha-beta-im-helfrich}
	\end{split}
	\\[12pt]
	\begin{split}
		N^{\alpha \beta}
		&= \kb(\phi) \Big(
			\Big[
				- H^2
			+ \big( C(\phi) \big)^2
			\Big] a^{\alpha \beta}
			+ \Big[
				H - C(\phi)
			\Big] \bar{b}^{\alpha \beta}
		\Big) \\[2pt]
		&\hspace{10pt} + 2 \kc (\phi) \Big( J - 1 \Big) a^{\alpha \beta}
		+ \gamma \Big(
			\dfrac{1}{2} \phi_{, \mu} \phi^{, \mu} \, a^{\alpha \beta}
			- \phi^{, \alpha} \phi^{, \beta}
		\Big)
		+ \pi^{\alpha \beta}
		~,
		\label{eq:n-alpha-beta-im-helfrich}
	\end{split}
	\\
	\intertext{and}
	\begin{split}
		S^\alpha
		&= - \Big( \kb (\phi) \Big[ H - C (\phi) \Big] \Big)^{\! , \alpha} ~.
		\label{eq:s-alpha-im-helfrich}
	\end{split}
\end{align}
In equations \eqref{eq:sigma-alpha-beta-im-helfrich}--\eqref{eq:s-alpha-im-helfrich}, we can determine which components of the total free energy $\rho \psi$ \eqref{eq:total-helmholtz-2c-is} contribute to the in-plane and out-of-plane stresses through the coefficients $\kb (\phi)$, $\kg$, $\kc (\phi)$, $\gamma$, and $k_\phi$.
For example, the couple-stresses $M^{\alpha \beta}$ \eqref{eq:m-alpha-beta-im-helfrich} and shear stress $S^\alpha$ \eqref{eq:s-alpha-im-helfrich} are only affected by the Helfrich bending energy $w_{\text{h}}$ \eqref{eq:w-helfrich-mc}, while the in-plane stresses $N^{\alpha \beta}$ \eqref{eq:n-alpha-beta-im-helfrich} contain contributions from the bending energy $w_{\text{h}}$, the compression energy $w_{\text{c}}$ \eqref{eq:w-compress-mc}, and the gradient energy $w_{\text{g}}$ \eqref{eq:w-g}.
The gradient energy contributions are Korteweg-like, and describe in-plane momentum transfer due to line tensions and concentration gradients in the system \cite{korteweg-1901, joseph-eur-j-mech-b-1990, kostin-esaim-2003, truzzolillo-soft-matter-2017}.
\textspace

The double-well potential $w_{\text{dw}}$ \eqref{eq:w-dw} does not enter the stresses, but affects the equations of motion through the chemical potential $\mu_\phi = \partial \psi / \partial \phi$, calculated as 
\begin{equation} \label{eq:mu-phi}
	\begin{split}
		\mu_\phi
		= \dfrac{1}{\rho} \bigg\{
			\kb' (\phi) \Big[ H - C(\phi) \Big]^2
			&- 2 \kb (\phi) C'(\phi) \Big[ H - C(\phi) \Big] \\[3pt]
			+ \dfrac{\kc' (\phi)}{J} \big( 1 - J \big)^2 
			&- \gamma \, \Delta \phi
			+ \dfrac{k_\phi}{J} \Big[
				\chi \big( 1 - 2 \phi \big)
				+ k_{\textrm{B}} \, T \, \ln \Big( \dfrac{\phi}{1 - \phi} \Big)
			\Big]
		\bigg\} ~.
	\end{split}
\end{equation}
The chemical potential $\mu_\phi$ thus contains contributions from the bending energy, compression energy, and gradient energy.


\subsubsection{Equations of Motion}

For our two-component model, there are five unknowns: the total mass density $\rho$, the mass fraction $\phi$, and the three components of the velocity $\bm{v}$.
The five corresponding governing equations are the total mass balance, the mass fraction balance of $\phi$, and the three components of the linear momentum balance.
The total mass balance \eqref{eq:local-mass-balance-mc} is given by
\begin{equation} \label{eq:continuity-two-component-dwp-general-c-ip}
	\rho_{, t}
	+ \rho_{, \alpha} v^\alpha
	+ (v^\alpha_{; \alpha} - 2 v H) \rho
	= 0 ~,
\end{equation}
where we have expanded the material derivative to show how the in-plane velocities enter the mass balance.
The mass fraction balance of $\phi$ is given by equation \eqref{eq:concentration-dot}, where we substitute the form of the diffusive flux found in equation \eqref{eq:species-flux-ip} to obtain
\begin{equation} \label{eq:species-flux-two-component-dwp-general-c-ip}
	\rho \big(
		\phi_{, t} + \phi_{, \alpha} v^{\alpha}
	\big)
	+ \bigg(
		D \bigg[
			\dfrac{(b_2)^\alpha - (b_1)^\alpha}{T}
			- \Big(
				\dfrac{\mu_\phi}{T}
			\Big)^{\!, \alpha}
		\bigg]
	\bigg)_{; \alpha} 
	= 0 ~,
\end{equation}
where the chemical potential $\mu_\phi$ is given by equation \eqref{eq:mu-phi}. 
The chemical potential contains information from all four components of the total Helmholtz free energy density, and is coupled to the concentration and curvatures.
Substituting the stresses and moments \eqref{eq:sigma-alpha-beta-im-helfrich}--\eqref{eq:s-alpha-im-helfrich} into the tangential equations \eqref{eq:tangential-general-eoms-elastic-compressible} and the shape equation \eqref{eq:normal-general-eom-elastic-compressible}, we obtain 
\begin{gather}
	\begin{split} \label{eq:normal-eom-two-component-dwp-general-c-ip}
		\rho \Big(
			v_{, t} + v^\alpha w_\alpha
		\Big)
		= p
		&+ \pi^{\alpha \beta} b_{\alpha \beta}
		- 2 \kb(\phi) \big[
			H - C(\phi)
		\big] \Big(
			H^2 + H C(\phi) - K
		\Big) \\[4pt]
		&+ 4 H \kc(\phi) \big[ J - 1 \big] 
		+ \gamma \, \phi_{, \alpha} \, \phi_{, \beta} \Big(
			H a^{\alpha \beta}
			- b^{\alpha \beta}
		\Big)
		- \Delta \Big(
			\kb(\phi) \big[ H - C \big]
		\Big)
	\end{split} 
\end{gather}
and
\begin{gather}
	\begin{split} \label{eq:tangential-eom-two-component-dwp-general-c-ip}
		\rho \Big(
			v^\alpha_{, t} &- w^\alpha v + v^\lambda {w_\lambda}^\alpha
		\Big) 
		= \rho b^\alpha
		+ \pi^{ \mu \alpha}_{; \mu}
		- 2 \kc(\phi) \dfrac{J}{\rho} \rho^{, \alpha} \\[4pt]
		&+ \phi^{, \alpha} \bigg(
			\kb'(\phi) \big[
				H - C(\phi)
			\big]^2
			- 2 \, \kb(\phi) \, C'(\phi) \big[
				H - C(\phi) 
			\big]
			+ 2 \, \kc'(\phi) \big[ J - 1 \big] 
			- \gamma \Delta \phi
		\bigg) 
		.
	\end{split}
\end{gather}
From the governing equations \eqref{eq:continuity-two-component-dwp-general-c-ip}--\eqref{eq:tangential-eom-two-component-dwp-general-c-ip}, we can clearly see the highly nonlinear coupling between species diffusion, bending, and intra-membrane viscous flow.
We note a similar model was proposed by Agrawal and Steigmann \cite{steigmann-agrawal-zamp-2011}, in which viscous stresses were neglected.
As the presence of multiple components introduces diffusive fluxes, evident from equation \eqref{eq:species-flux-two-component-dwp-general-c-ip}, and therefore velocity gradients, our model indicates viscous stresses cannot be omitted in describing multi-component membranes.


\subsubsection{Boundary and Initial Conditions} \label{sec:bc-mc}

In this section, we specify boundary conditions on the governing equations \eqref{eq:continuity-two-component-dwp-general-c-ip}--\eqref{eq:tangential-eom-two-component-dwp-general-c-ip}, as appropriate, for them to be mathematically well-posed.
\textspace

The mass fraction balance for $\phi$ \eqref{eq:species-flux-two-component-dwp-general-c-ip} has a fourth-order spatial derivative, as it contains the surface Laplacian of the chemical potential, which itself contains the surface Laplacian of the mass fraction $\phi$ \eqref{eq:mu-phi}.
This is similar in structure to the shape equation, and for a well-posed problem we must specify two boundary conditions in $\phi$ at every point on the boundary.
For example, we may specify $\phi$ and its gradient in the $\bm{\nu}$ direction, or the chemical potential $\mu_\phi$ and the diffusive flux in the $\bm{\nu}$ direction, $j_\phi^{\, \alpha} \, \nu_\alpha$, where the diffusive flux $j_\phi^{\, \alpha}$ is given by equation \eqref{eq:species-flux-ip}.
Such boundary conditions are studied for systems described by Cartesian coordinates in \cite{wells-kuhl-jcp-2006}.
\textspace

The momentum equations are of the same general form as in the single-component case.
For the in-plane momentum equations \eqref{eq:tangential-eom-two-component-dwp-general-c-ip}, we may specify either the velocities $v^\alpha$ or the in-plane force components $f_\nu$ and $f_\tau$ of the total force $\bm{f}$ at the boundary, given in equations \eqref{eq:boundary-force-decomposition}--\eqref{eq:boundary-force}.
For the shape equation \eqref{eq:normal-eom-two-component-dwp-general-c-ip}, we may specify the position and its gradient or the shear force $f_n$ and the moment $M$ \eqref{eq:boundary-moment-b-c} at the boundary.
We find the functional form of the forces and moments by substituting the stresses \eqref{eq:sigma-alpha-beta-im-helfrich}--\eqref{eq:s-alpha-im-helfrich} into the general relations \eqref{eq:f-nu}--\eqref{eq:boundary-moment-b-c} and obtain
\begin{align}
	\begin{split}
		f_\nu
		&= \kb(\phi) \Big[
			H - C(\phi)
		\Big] \Big(
			H - C(\phi) - \kappa_\nu
		\Big)
		+ 2 \kc (\phi) \big[
			J - 1
		\big] \\[3pt]
		&\hspace{63pt} + \gamma \Big(
			\dfrac{1}{2} \phi_{, \mu} \phi^{, \mu}
			- \phi^{, \alpha} \phi^{, \beta} \nu_\alpha \nu_\beta
		\Big)
		- \kg \, \xi^2
		+ \pi^{\alpha \beta} \nu_\alpha \nu_\beta ~,
	\end{split}
	\label{eq:f-nu-helfrich-double-well} \\[8pt]
	\begin{split}
		f_\tau
		&= - \gamma \, \phi^{, \alpha} \phi^{, \beta} \nu_\alpha \tau_\beta
		- \kb (\phi) \, \xi \, \big[
			H - C(\phi)
		\big]
		- \kg \, \xi \, \kappa_\tau
		+ \pi^{\alpha \beta} \nu_\alpha \tau_\beta ~,
	\end{split}
	\label{eq:f-tau-helfrich-double-well} \\[5pt]
	\begin{split}
		f_n
		&= - \Big(
			\kb (\phi) \big[
				H - C(\phi)
			\big]
		\Big)_{\! , \nu}
		+ \kg \dfrac{\mathrm{d} \xi}{\mathrm{d} \ell} ~,
	\end{split}
	\label{eq:f-n-helfrich-double-well} \\
	\shortintertext{and}
	\begin{split}
		M
		&= \kb (\phi) \big[
			H - C(\phi)
		\big]
		+ \kg \, \kappa_\tau
		~.
	\end{split}
	\label{eq:boundary-moment-b-c-helfrich-double-well}
\end{align}
\eqnspace

We close the problem by providing example initial conditions.
As in the single-component case, we specify the initial membrane position $\bm{x}$, velocity $\bm{v}$, and assume initially  the density $\rho$ is constant and the Jacobian $J = 1$ everywhere.
One may also specify an initial distribution of the mass fraction $\phi$ everywhere, which determines the spontaneous curvature $C(\phi)$.
With these initial conditions and the aforementioned boundary conditions, our problem is well-posed.


\section{Peripheral Proteins---Chemical Reactions} \label{sec:binding}

The binding and unbinding of peripheral proteins plays a crucial role in many cellular processes involving lipid membranes.
In vitro experiments have shown dramatic shape changes can occur as a result of protein binding and unbinding events involving the membrane and the surrounding fluid \cite{harrison-nsmb-2008, yang-tamm-nat-comm-2016, barlowe-cell-1994, bacia-nature-2011, hurley-nrmcb-2016, roux-cell-2015}.
Once proteins bind to the membrane, they are able to diffuse in-plane as well.
The complex interplay between protein binding, in-plane diffusion, in-plane lipid flow, and membrane bending is currently not well-understood.
In this section, we extend the multi-component model to include binding and unbinding chemical reactions of proteins onto membranes.
We determine the new form of the balance laws and equations of motion, and learn the thermodynamic driving force governing the binding and unbinding reactions.
We motivate our theoretical development with the following example.


\subsection{Chemistry}

\begin{figure}[!b]
	\centering
	\includegraphics[width=0.65\textwidth]{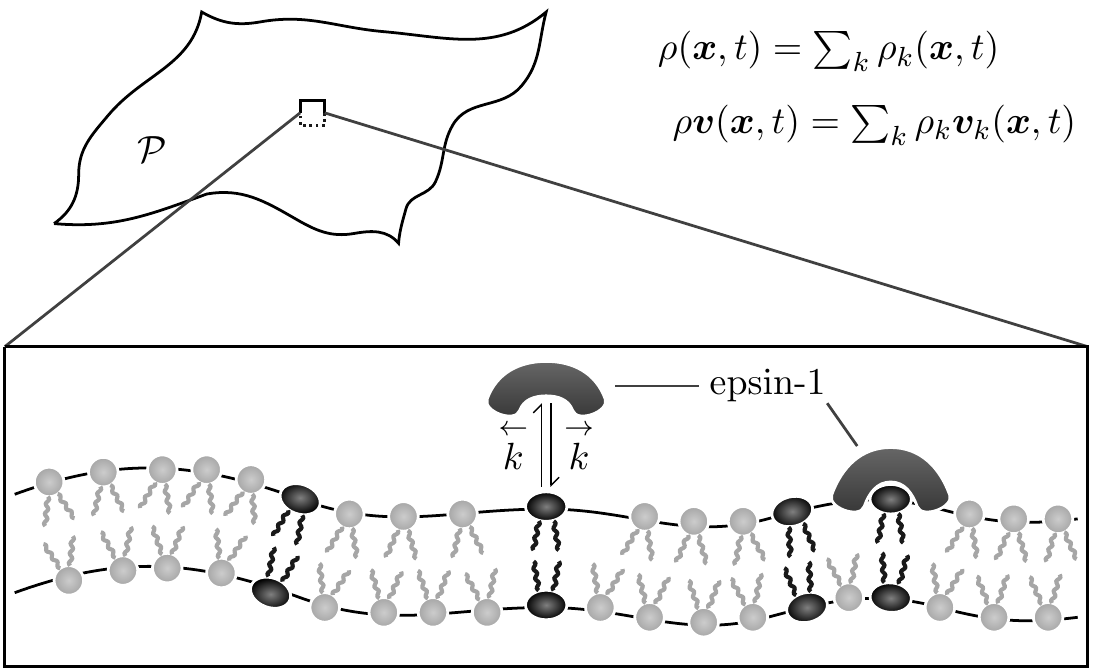}
	\vspace{10pt}
	\captionsetup{width=0.75\textwidth}
	\caption{
		A schematic depicting the binding and unbinding of epsin-1 proteins to and from PI(4,5)P$_2$ lipids (black) in the membrane patch $\scp$.
		The rate constants for binding and unbinding are denoted by $\fvect{k}$ and $\bvect{k}$, respectively.
		Epsin-1 proteins cannot bind to the DOPC lipids (light gray).
		All interactions are modeled as part of the continuum of the membrane, where changes in concentration are accounted for in terms of changes to the local species density $\rho_k (\bm{x}, t)$.
	}
	\label{fig:binding}
\end{figure}
Consider the membrane to consist of two types of phospholipids: DOPC and Phosphatidylinositol-4,5-biphosphate (PI(4,5)P\textsubscript{2}).
In the fluid surrounding the membrane, there is a reservoir of epsin-1 proteins which bind only to PI(4,5)P\textsubscript{2} lipids \cite{stachowiak-natcellbio-2012, schmid-nature-2007}.
The concentration of PI(4,5)P\textsubscript{2} is usually around 2\% by mass under physiological conditions.
Because we model the membrane as a continuum, at every point in space there may be multiple binding and unbinding events (Figure \ref{fig:binding}).
In this entire theoretical development, electrostatic effects are neglected.
While we study this scenario in detail, another motivating example is the binding of BAR proteins onto membranes, which has been extensively studied in \cite{baumgart-bpj-2015}.
\textspace

\begin{wraptable}{r}{0.44\textwidth}
	\small
	\centering
	\vspace{-03pt}
	\setlength{\tabcolsep}{18pt}
	\renewcommand{\arraystretch}{1.4}
	\begin{tabular}{l l}
	\hline
	\hline
	k & Species \\
	\hline
	1 & DOPC \\
	2 & PI(4,5)P\textsubscript{2} \\
	3 & PI(4,5)P\textsubscript{2}--epsin-1 \\
	\hline
	\hline
	\end{tabular}
	\vspace{-03pt}
	\captionsetup{width=0.39\textwidth}
	\caption{
		The chemical species in our example membrane system undergoing protein binding and unbinding reactions, indexed by $k$.
	}
	\label{binding-species}
	\vspace{-05pt}
\end{wraptable}
Our system includes the membrane patch $\scp$ and any epsin-1 proteins bound to it.
In the foregoing example, there exist three chemical species in our system.
They are labeled by the index $k$, and are provided in Table~\ref{binding-species}.
The index $k = 3$ refers to an epsin-1 protein bound to a PI(4,5)P\textsubscript{2} lipid.
The binding and unbinding reaction between epsin-1 and PI(4,5)P\textsubscript{2} lipids can be written as 
\begin{equation} \label{eq:binding}
	\textrm{PI(4,5)P\textsubscript{2} + epsin-1}
	\rightleftharpoons
	\textrm{ PI(4,5)P\textsubscript{2}--epsin-1} ~.
\end{equation}
The overall reaction rate $\mathcal{R}$ for equation \eqref{eq:binding} has units of a molar flux and is given by
\begin{equation} \label{eq:reaction-rate-forward-backward-difference}
	\mathcal{R} 
	= \fvect{\mathcal{R}} 
	- \bvect{\mathcal{R}} 
	~,
\end{equation}
where $\fvect{\mathcal{R}}$ and $\bvect{\mathcal{R}}$ are the forward and reverse reaction rates, respectively.
Defining $\scmp$ as the molar mass of epsin-1 proteins, $\scmp \mathcal{R}$ is the mass flux from the reservoir of proteins either above or below the membrane, herein called the bulk phase, to the membrane surface.
\textspace

The concentrations $n_k$ on the membrane patch, with units of moles per area, are defined as
\begin{equation} \label{eq:c-k-binding}
	n_k
	:= \dfrac{\rho_k}{\mathscr{M}_k}
	~,
\end{equation}
where $\rho_k$ is the mass density of species $k$ and $\mathscr{M}_k$ is the molar mass of species $k$.
The concentration of espin-1 proteins in the bulk phase is denoted $\npb$ and has units of moles per volume.
The forward and reverse chemical reactions in equation \eqref{eq:binding} are assumed to be elementary steps, for which the forward and reverse reaction rates may be written as
\begin{align}
	\fvect{\mathcal{R}} 
	&= \fvect{k} \, n_2 \, \npb 
	\label{eq:forward-mass-action} 
	\\
	\shortintertext{and}
	\bvect{\mathcal{R}} 
	&= \bvect{k} \, n_3 
	~, 
	\label{eq:backward-mass-action}
\end{align}
where $\fvect{k}$ and $\bvect{k}$ are the forward and reverse rate constants, respectively.
At this stage, we introduce the stoichiometric coefficients $\alpha_k$, which are the coefficients of the $k^{\mathrm{th}}$ species in the chemical reaction \eqref{eq:binding} and from inspection are given by $\alpha_1 = 0$, $\alpha_2 = -1$, and $\alpha_3 = 1$.
It is notationally convenient to define the stoichiometric coefficient of the epsin-1 protein in the chemical reaction \eqref{eq:binding} as $\alpha_{\mathrm{p}} = -1$.
A summary of the new variables introduced in this section is provided in Table~\ref{surface-protein-definitions}.
\begin{table}[!t]
	\small
	\centering
		\vspace{5pt}
		\setlength{\tabcolsep}{16pt}
		\renewcommand{\arraystretch}{2}
		\begin{tabular}{ccl}
		\hline
		\hline
			Symbol & Units & Description \\
			\hline
			$n_k$
			& mol / m$^2$
			& concentration of component $k$.
			\\
			$\npb$
			& mol / m$^3$
			& concentration of proteins in the surrounding fluid.
			\\
			$\mathscr{M}_k$
			& g / mol
			& Molar mass of component $k$.
			\\
			$\scmp$
			& g / mol
			& Molar mass of epsin-1 protein.
			\\
			$\mathcal{R}$
			& mol / m$^{2}$ s 
			& Rate of reaction per unit area.
			\\
			$\fvect{\mathcal{R}}$, $\bvect{\mathcal{R}}$
			& mol / m$^{2}$ s 
			& Forward and reverse reaction rates, respectively.
			\\
			$\fvect{k}$
			& m$^{3}$ / mol s 
			& Forward reaction rate constant. 
			\\
			$\bvect{k}$
			& 1 / s 
			& Reverse reaction rate constant. 
			\\
			$\alpha_k$
			& --
			& The coefficient of the $k^{\mathrm{th}}$ species in the chemical reaction. \\
			$\alpha_{\mathrm{p}}$
			& --
			& The coefficient of the epsin-1 protein in the chemical reaction.
			\\[6pt]
		\hline
		\hline
		\end{tabular}
		\caption{New quantities relevant for reactions within the membrane system.}
		\vspace{00pt}
		\label{surface-protein-definitions}
\end{table}


\subsection{Balance Laws}

We continue to use our continuum mechanical framework, including mixture theory, to describe the membrane patch.
All of the kinematic results derived in Section \ref{sec:mc-kinematics} continue to hold.
The binding and unbinding of proteins, however, affects the overall mass balance of the membrane patch and consequently the global forms of the linear and angular momentum balances as well.


\subsubsection{Mass Balance}

For a membrane patch $\scp$, the mass of species $k$ can change due to a diffusive mass flux $\bm{j}_k$ at the patch boundary or the binding and unbinding of proteins along the membrane surface, which is a mass flux from the bulk phase to the membrane surface.
In this case, the global form of the conservation of mass of species $k$ is given by
\begin{equation} \label{eq:mass-conservation-surface-species-global}
	\dfrac{\textrm{d}}{\textrm{d}t} \bigg(
		\int_\scp \rho_k ~\textrm{d}a
	\bigg)
	= - \int_{\partial \scp} \bm{j}_k \cdot \bm{\nu} ~\textrm{d}s
	+ \int_\scp \alpha_k \mathscr{M}_k \mathcal{R} ~\textrm{d}a ~.
\end{equation}
The form of the diffusive flux of species $k$ is given in equation \eqref{eq:diffusive-species-flux-component} such that the sum of diffusive fluxes satisfies equation \eqref{eq:sum-diff-flux}.
Comparing equation \eqref{eq:mass-conservation-surface-species-global} to its multi-component analog \eqref{eq:global-mass-balance-k}, we see the binding and unbinding of proteins affects the total mass of each species through the reaction rate $\mathcal{R}$.
Again, we use the Reynolds transport theorem \eqref{eq:rtt} and the surface divergence theorem \eqref{eq:divergence-theorem} to obtain
\begin{equation} \label{eq:mass-conservation-surface-species-global-step-2}
	\int_\scp \Big(
		\dot{\rho}_k + \big( v^\alpha_{; \alpha} - 2 v H \big) \rho_k
	\Big) ~\mathrm{d}a
	= \int_\scp \Big(
		- j_{k \, ; \alpha}^{\, \alpha}
		+ \alpha_k \mathscr{M}_k \mathcal{R}
	\Big) ~\mathrm{d}a ~.
\end{equation}
As the membrane patch $\scp$ is arbitrary, the local form of the mass balance of species $k$ can be obtained as 
\begin{equation} \label{eq:mass-conservation-surface-species-local}
	\dot{\rho}_k
	+ (v^\alpha_{; \alpha} - 2 v H) \rho_k
	= - j_{k \, ; \alpha}^{\, \alpha}
	+ \alpha_k \mathscr{M}_k \mathcal{R} ~.
\end{equation}
While we have accounted for bound proteins through the index $k = 3$, which refers to epsin-1 proteins bound to PI(4,5)P\textsubscript{2} lipids, it is possible to determine the continuity equation for proteins alone by multiplying equation \eqref{eq:mass-conservation-surface-species-local}, for $k = 3$, by $\scmp / \mathscr{M}_3$.
\textspace

Summing equation \eqref{eq:mass-conservation-surface-species-local} over all $k$, we find the local form of the total mass balance is given by
\begin{equation} \label{eq:mass-conservation-surface-species-total}
	\dot{\rho}
	+ (v^\alpha_{; \alpha} - 2 v H) \rho
	= \scmp \mathcal{R} ~,
\end{equation}
indicating binding and unbinding reactions change the mass of the membrane patch only due to the addition or removal of proteins.
When proteins are binding, $\mathcal{R} > 0$ and the mass of the membrane patch increases, while when proteins are unbinding $\mathcal{R} < 0$ and the mass of the patch decreases.
Because the mass of an infinitesimal patch changes over time due to protein binding, the relation $J = \rho_0 / \rho$ from equation \eqref{eq:jacobian-rho} is no longer true. 
Consequently, $a_{\alpha \beta}$ determines the Jacobian determinant $J$, but does not determine the areal mass density  $\rho$. 
\textspace

To calculate the appropriate form of the modified Reynolds transport theorem for the case of binding and unbinding proteins, we apply the Reynolds transport theorem \eqref{eq:rtt} where $f = \rho u$ and substitute the local mass balance \eqref{eq:mass-conservation-surface-species-total} to obtain
\begin{equation} \label{eq:rtt-density-ss}
	\dfrac{\textrm{d}}{\textrm{d}t} \bigg(
		\int_{\scp} \rho u ~\textrm{d}a
	\bigg)
	= \int_{\scp} \Big(
		\rho \dot{u}
		+ u \scmp \mathcal{R}
	\Big)
	~\textrm{d}a ~.
\end{equation}
As before, in equation \eqref{eq:rtt-density-ss} $u$ can be a scalar-, vector-, or tensor-valued function.


\subsubsection{Linear Momentum Balance}

As our system now consists of the membrane patch and any proteins bound to it, the binding and unbinding of proteins will affect the total linear momentum of the membrane patch.
The mass flux of proteins from the bulk to the membrane, $\scmp \mathcal{R}$, carries momentum to and from the membrane surface.
Assuming a no-slip condition between the membrane surface and the adjacent fluid, the proteins travel at an average velocity $\bm{v}$ just before binding and just after unbinding.
Correspondingly, the momentum flux from the bulk to the membrane is $\bm{v} \scmp \mathcal{R}$, and the global form of the linear momentum balance is given by
\begin{equation} \label{eq:momentum-balance-global-surface-species}
	\dfrac{\textrm{d}}{\textrm{d}t} \bigg(
		\int_\scp \rho \bm{v} ~\textrm{d}a
	\bigg) 
	= \int_\scp \bigg( \sum_{k = 1}^3 \rho_k \bm{b}_k \bigg) ~\textrm{d}a
	+ \int_{\partial \scp} \!\! \bm{T} ~\textrm{d}s
	+ \int_\scp \bm{v} \scmp \mathcal{R} ~\textrm{d}a
	~.
\end{equation}
Using the Reynolds transport theorem \eqref{eq:rtt-density-ss} and the definition of the mass-weighted body force provided in equation \eqref{eq:body-force-barycentric} with $N = 3$, equation \eqref{eq:momentum-balance-global-surface-species} can be reduced to
\begin{equation} \label{eq:momentum-balance-global-ss-step-2}
	\int_\scp \rho \dot{\bm{v}} ~\mathrm{d}a 
	= \int_\scp \rho \bm{b} ~\mathrm{d}a
	+ \int_{\bscp} \!\! \bm{T} ~\mathrm{d}s
	~.
\end{equation}
Equation \eqref{eq:momentum-balance-global-ss-step-2} is identical to the global form of the linear momentum balance of a multi-component system, found in Section \ref{sec:linear-momentum-mc} to be given by equation \eqref{eq:euler}.
Since the membrane patch $\scp$ is arbitrary, the local form of the linear momentum balance may be written as 
\begin{equation} \label{eq:momentum-balance-local-surface-species}
	\rho \dot{\bm{v}} 
	= \rho \bm{b}
	+ \bm{T}^\alpha_{; \alpha}
	~.
\end{equation}


\subsubsection{Angular Momentum Balance}

Similar to the linear momentum analysis, the additional change in the angular momentum due to the chemical reactions is $\bm{x} \times \bm{v} \scmp \mathcal{R}$.
Therefore, the global form of the angular momentum balance is given by
\begin{equation} \label{eq:angular-momentum-balance-global-surface-species}
	\dfrac{\mathrm{d}}{\mathrm{d}t} \bigg(
		\int_\scp \rho \bm{x} \times \bm{v} ~\mathrm{d}a
	\bigg) 
	= \int_\scp \rho \bm{x} \times \bm{b}~\mathrm{d}a
	+ \int_{\partial \scp} \Big(
		\bm{x} \times \bm{T} 
		+ \bm{m}
	\Big) ~\mathrm{d}s
	+ \int_\scp \bm{x} \times \bm{v} \, \scmp \, \mathcal{R} ~\mathrm{d}a
	~.
\end{equation}
Applying the Reynolds transport theorem \eqref{eq:rtt-density-ss} to the left hand side, we obtain
\begin{equation} \label{eq:angular-momentum-balance-global-ss-step-2}
	\int_\scp \rho \bm{x} \times \dot{\bm{v}} ~\mathrm{d}a 
	= \int_\scp \rho \bm{x} \times \bm{b}~\mathrm{d}a
	+ \int_{\partial \scp} \Big(
		\bm{x} \times \bm{T} 
		+ \bm{m} 
	\Big) ~\mathrm{d}s
	~,
\end{equation}
which is identical to the global form of the multi-component angular momentum balance found in Section \ref{sec:linear-momentum-mc} to be given by equation \eqref{eq:angular-momentum-balance}.
The local form of the angular momentum balance for our membrane system is therefore identical in form to the single-component case \eqref{eq:angular-momentum-balance-local}.
As the local linear momentum balance \eqref{eq:momentum-balance-local-surface-species} is also unchanged, the symmetry of $\sigma^{\alpha \beta}$ \eqref{eq:angular-momentum-symmetric}, form of $S^\alpha$ \eqref{eq:angular-momentum-s}, and mechanical power balance \eqref{eq:global-power-balance-solution} are valid in the presence of chemical reactions as well.


\subsection{Thermodynamics}

In this section, we develop local forms of the first and second laws of thermodynamics as well as the local entropy balance.
Thermodynamic balance laws must now take into account energy and entropy changes due to chemical reactions.


\subsubsection{First Law---Energy Balance}

In addition to the mechanisms described in the multi-component case, the total energy of the membrane patch can change due to the flux of proteins from the bulk phase to the membrane surface.
Assuming the changes in energy due to the binding and unbinding of proteins can be decomposed into kinetic and internal energy components and there is no slip between the membrane surface and the surrounding fluid, these changes are given by $(\tfrac{1}{2} \bm{v} \cdot \bm{v} + u) \scmp \mathcal{R}$.
Correspondingly, the global form of the energy balance can be written as
\begin{align}
	\begin{split} \label{eq:first-law-thermo-global-surface-species}
		\dfrac{\mathrm{d}}{\mathrm{d}t} \bigg(
			\int_\scp \rho e ~\mathrm{d}a
		\bigg)
		&= \int_\scp \sum_{k = 1}^3 \Big(
			\rho_k \bm{b}_k \cdot \bm{v}_k
		\Big) ~\mathrm{d}a
		+ \int_\scp \rho r ~\mathrm{d}a 
		+ \int_{\partial \scp} \!\! \bm{v} \cdot \bm{T} ~\mathrm{d}s \\[3pt]
		&\hspace{30pt} - \int_{\partial \scp} \!\! \bmjq \cdot \bm{\nu} ~\mathrm{d}s
		+ \int_{\partial \scp} \!\! \bm{M} \cdot \dot{\bm{n}} ~\mathrm{d}s
		+ \int_\scp \Big(
			\dfrac{1}{2} \bm{v} \cdot \bm{v} + u
		\Big)
		\scmp \mathcal{R} ~\mathrm{d}a ~.
	\end{split}
\end{align}
In equation \eqref{eq:first-law-thermo-global-surface-species}, only the last integral term is new compared to the multi-component form \eqref{eq:global-energy-balance-mc}.
Applying the Reynolds transport theorem \eqref{eq:rtt-density-ss} and substituting the definition of the total energy \eqref{eq:energy-def}, we obtain
\begin{align}
	\begin{split} \label{eq:first-law-thermo-global-ss-step-2}
		\int_\scp \rho \dot{u} ~\mathrm{d}a 
		&= \int_\scp \sum_{k = 1}^3 \Big(
			\rho_k \bm{b}_k \cdot \bm{v}_k
		\Big) ~\mathrm{d}a
		+ \int_\scp \rho r ~\mathrm{d}a 
		+ \int_{\partial \scp} \!\! \Big(
			\bm{v} \cdot \bm{T} 
			- \bmjq \cdot \bm{\nu} 
			+ \bm{M} \cdot \dot{\bm{n}} 
		\Big) ~\mathrm{d}s 
		~.
	\end{split}
\end{align}
The right hand side of equation \eqref{eq:first-law-thermo-global-ss-step-2} is equal to the right hand side of the multi-component global energy balance \eqref{eq:global-energy-balance-mc}. 
Following similar arguments to those of the multi-component case described in Section \ref{sec:first-law-mc}, the local form of the first law of thermodynamics is found to be 
\begin{equation} \label{eq:first-law-thermo-local-surface-species}
	\rho \dot{u}
	= \sum_{k = 1}^3 \, j_k^\alpha \, (b_k)_\alpha
	+ \rho r
	- J_{\mathrm{q} \, ; \alpha}^{\, \alpha}
	+ \dfrac{1}{2} \sigma^{\alpha \beta} \dot{a}_{\alpha \beta}
	+ M^{\alpha \beta} \dot{b}_{\alpha \beta}
	~.
\end{equation}
Equation \eqref{eq:first-law-thermo-local-surface-species} is identical to the local energy balance of the multi-component membrane \eqref{eq:first-law-thermo-ip}, for $N = 3$.


\subsubsection{Entropy Balance \& Second Law}

As in the momentum balances and the energy balance, in the global entropy balance we introduce a new term to account for the entropy changes due to the chemical reaction.
To this end, the entropy is modified by $s \scmp \mathcal{R}$, and accordingly the global form of the entropy balance is given by
\begin{equation} \label{eq:second-law-thermo-global-surface-species}
	\dfrac{\mathrm{d}}{\mathrm{d}t} \bigg(
		\int_\scp \rho s ~\mathrm{d}a
	\bigg)
	= - \int_{\partial \scp} \bmjs \cdot \bm{\nu} ~\mathrm{d}s
	+ \int_\scp \Big(
		\rho \eta_{\mathrm{e}}
		+ \rho \eta_{\mathrm{i}}
		+ s \scmp \mathcal{R}
	\Big) ~\mathrm{d}a
	~.
\end{equation}
Applying the Reynolds transport theorem \eqref{eq:rtt-density-ss} and the surface divergence theorem \eqref{eq:divergence-theorem} to equation \eqref{eq:second-law-thermo-global-surface-species}, we obtain
\begin{equation} \label{eq:second-law-thermo-global-ss-step-2}
	\int_\scp 
		\rho \dot{s} 
	~\mathrm{d}a
	= \int_\scp \Big(
		-J_{\mathrm{s} \, ; \alpha}^{\, \alpha}
		+ \rho \eta_{\mathrm{e}}
		+ \rho \eta_{\mathrm{i}}
	\Big) ~\mathrm{d}a ~.
\end{equation}
As the membrane patch $\scp$ is arbitrary, the local form of the entropy balance is 
\begin{equation} \label{eq:local-entropy-balance-surface-species}
	\rho \dot{s}
	= - J_{\mathrm{s} \, ; \alpha}^{\, \alpha}
	+ \rho \eta_{\mathrm{e}}
	+ \rho \eta_{\mathrm{i}} ~.
\end{equation}
The second law of thermodynamics is still given by equation \eqref{eq:second-law-thermo-local}.


\subsubsection{Choice of Thermodynamic Potential}

As in the previously considered single- and multi-component membranes, we express the entropy balance in terms of the Helmholtz free energy density $\psi$. 
Using equations \eqref{eq:helmholtz} and \eqref{eq:local-entropy-balance-surface-species} we obtain the total rate of change of entropy to be 
\begin{equation} \label{eq:entropy-energy-ss-step-1}
	\rho \dot{s}
	= -J^{\, \alpha}_{\mathrm{s} \, ; \alpha}
	+ \rho \eta_{\mathrm{e}}
	+ \rho \eta_{\mathrm{i}}
	= \dfrac{1}{T} \Big(
		\rho \dot{u}
		- \rho \dot{T} s
		- \rho \dot{\psi}
	\Big) 
	~.
\end{equation}
Substituting the local form of the first law of thermodynamics \eqref{eq:first-law-thermo-local-surface-species} into equation \eqref{eq:entropy-energy-ss-step-1}, we obtain
\begin{equation} \label{eq:entropy-energy-ss}
	\begin{split}
		\rho \dot{s}
		&= -J^{\, \alpha}_{\mathrm{s} \, ; \alpha}
		+ \rho \eta_{\mathrm{e}}
		+ \rho \eta_{\mathrm{i}} \\[3pt]
		&= \dfrac{1}{T} \bigg\{
			\rho r
			- J_{\mathrm{q} \, ; \alpha}^{\, \alpha}
			+ \sum_{k = 1}^3 j_k^\alpha \, (b_k)_\alpha
			+ \dfrac{1}{2} \sigma^{\alpha \beta} \dot{a}_{\alpha \beta}
			+ M^{\alpha \beta} \dot{b}_{\alpha \beta} 
			- \rho \dot{T} s
			- \rho \dot{\psi}
		\bigg\} ~.
	\end{split}
\end{equation}
Equation \eqref{eq:entropy-energy-ss} is identical to the multi-component entropy production, given by equation \eqref{eq:entropy-energy-balance-mc}, for $N = 3$.


\subsection{Constitutive Relations}

In this section, we extend the framework developed in the multi-component analysis to determine the internal entropy production equation.
We again apply linear irreversible thermodynamics to relate thermodynamic forces and fluxes, and determine the stresses, moments, and diffusive species fluxes. 
We also determine the thermodynamic driving force governing the binding and unbinding of proteins in the linear irreversible regime, and then extend our analysis to apply to systems arbitrarily far from chemical equilibrium.


\subsubsection{General Thermodynamic Variables}

As mentioned earlier, due to the binding and unbinding of proteins, the metric tensor $a_{\alpha \beta}$ determines the Jacobian determinant $J$ but not the areal mass density $\rho$.
Consequently, the Helmholtz free energy per unit mass $\psi$ may depend on all the species densities $\rho_1$, $\rho_2$, and $\rho_3$ in addition to $a_{\alpha \beta}$, and may be written as
\begin{equation} \label{eq:free-energy-form-surface-species} 
	\psi = \psi(a_{\alpha \beta}, \, b_{\alpha \beta}, \, T, \, \{ \rho_k \}_{k \, = \, 1, \, 2, \, 3}) ~.
\end{equation}
\eqnspace

Taking the material derivative of $\psi$ in equation \eqref{eq:free-energy-form-surface-species} and multiplying by $\rho$, we obtain
\begin{equation} \label{eq:psi-material-derivative-surface-species}
	\begin{split}
		\rho \dot{\psi}
		&= \dfrac{\rho}{2} \Big(
			\dfrac{\partial \psi}{\partial a_{\alpha \beta}}
			+ \dfrac{\partial \psi}{\partial a_{\beta \alpha}}
		\Big) \dot{a}_{\alpha \beta}
		+ \dfrac{\rho}{2} \Big(
			\dfrac{\partial \psi}{\partial b_{\alpha \beta}}
			+ \dfrac{\partial \psi}{\partial b_{\beta \alpha}}
		\Big) \dot{b}_{\alpha \beta}
		+ \rho \dfrac{\partial \psi}{\partial T} \dot{T} 
		+ \sum_{k = 1}^3 \rho \dfrac{\partial \psi}{\partial \rho_k} \dot{\rho}_k \\[4pt]
		&= \dfrac{\rho}{2} \Big(
			\dfrac{\partial \psi}{\partial a_{\alpha \beta}}
			+ \dfrac{\partial \psi}{\partial a_{\beta \alpha}}
		\Big) \dot{a}_{\alpha \beta}
		+ \dfrac{\rho}{2} \Big(
			\dfrac{\partial \psi}{\partial b_{\alpha \beta}}
			+ \dfrac{\partial \psi}{\partial b_{\beta \alpha}}
		\Big) \dot{b}_{\alpha \beta}
		- \rho s \dot{T} \\
		& \hspace{10pt} + \sum_{k = 1}^3 \mu_k \Big(
			- j_{k \, ; \alpha}^{\, \alpha}
			- \big(
				v^\alpha_{; \alpha} - 2 v H
			\big) \rho_k
			+ \alpha_k \mathscr{M}_k \mathcal{R} 
		\Big) ~,
	\end{split}
\end{equation}
where in obtaining the second equality, we used the local equilibrium assumption 
\begin{equation} \label{eq:local-equilibrium-entropy-binding}
	s
	= -\Big(
		\dfrac{\partial \psi}{\partial T}
	\Big)_{a_{\alpha \beta}, \, b_{\alpha \beta}, \, \rho_1, \, \rho_2, \, \rho_3}
	~,
\end{equation}
defined the chemical potential as
\begin{equation} \label{eq:mu-binding}
	\mu_k
	= \rho \Big(
		\dfrac{\partial \psi}{\partial \rho_k}
	\Big)_{a_{\alpha \beta}, \, b_{\alpha \beta}, \, \rho_{j \ne k}}
	~,
\end{equation}
and substituted the expression for $\dot{\rho}_k$ given by equation \eqref{eq:mass-conservation-surface-species-local}.
The chemical potential $\mu_k$ \eqref{eq:mu-binding} has units of energy per unit mass, and is now defined for all three species, as $\rho_1$, $\rho_2$, and $\rho_3$ are linearly independent. 
\textspace

Substituting equation \eqref{eq:psi-material-derivative-surface-species} into the entropy balance \eqref{eq:entropy-energy-ss} and rearranging terms yields
\begin{equation} \label{eq:second-law-thermo-local-step-2-ss}
	\begin{split}
		\rho \dot{s}
		&= - J_{\mathrm{s} \, ; \alpha}^{\, \alpha} 
		+ \rho \eta_{\mathrm{e}}
		+ \rho \eta_{\mathrm{i}} \\[5pt]
		&= - \Big(
			\dfrac{\jq^{\, \alpha}
			- \sum_{k = 1}^3 \mu_k \, j_k^{\, \alpha}}{T}
		\Big)_{; \alpha}
		+ \dfrac{\rho r}{T}
		- \dfrac{\jq^{ \, \alpha} \, T_{, \alpha}}{T^2}
		+ \sum_{k = 1}^3 \dfrac{(b_k)_\alpha \, j_k^{\, \alpha}}{T} 
		- \sum_{k = 1}^3 \Big(
			\dfrac{\mu_k}{T}
		\Big)_{\!\! , \alpha} j_k^{\, \alpha}
		\\[3pt]
		&\hspace{20pt} + \dfrac{1}{T} \bigg\{
			\dfrac{1}{2} \bigg[
				\sigma^{\alpha \beta}
				- \rho \Big(
					\dfrac{\partial \psi}{\partial a_{\alpha \beta}}
					+ \dfrac{\partial \psi}{\partial a_{\beta \alpha}}
				\Big)
			\bigg] \dot{a}_{\alpha \beta}
			+ \bigg[
				M^{\alpha \beta}
				- \dfrac{\rho}{2} \Big(
					\dfrac{\partial \psi}{\partial b_{\alpha \beta}}
					+ \dfrac{\partial \psi}{\partial b_{\beta \alpha}}
				\Big)
			\bigg] \dot{b}_{\alpha \beta} 
		\bigg\} \\[4pt]
		&\hspace{40pt} 
		- \dfrac{1}{T} \sum_{k = 1}^3 \mu_k \Big(
			\alpha_k \mathscr{M}_k \mathcal{R}
			- \big( v^\alpha_{; \alpha} - 2 v H \big) \rho_k 
		\Big) ~.
	\end{split}
\end{equation}
As outlined by Prigogine \cite{prigogine}, we introduce the chemical affinity $\mathscr{A}$ for the chemical reaction in equation \eqref{eq:binding}, which was first introduced by de Donder \cite{de-donder} as  
\begin{equation} \label{eq:affinity}
	\mathscr{A} 
	= -\alpha_{\mathrm{p}} \mupb \scmp 
	- \sum_{k = 1}^3 \alpha_k \mu_k \mathscr{M}_k ~,
\end{equation}
where $\mupb$ is the chemical potential of proteins in the bulk, again defined to have units of energy per unit mass.
As described by de Groot \& Mazur \cite{degroot-mazur}, because the diffusive fluxes $\bm{j}_k$ sum to zero \eqref{eq:sum-diff-flux}, for any scalar quantity $f_k$ we can write
\begin{equation} \label{eq:diffusive-flux-rearrangement}
	\sum_{k = 1}^3 f_k \, j_k^{\, \alpha}
	= \sum_{k = 2}^3 \big( f_k - f_1 \big) \, j_k^{\, \alpha}
	~.
\end{equation}
Equation \eqref{eq:diffusive-flux-rearrangement} also holds for vector or tensor quantities.
Substituting equation \eqref{eq:affinity} into equation \eqref{eq:second-law-thermo-local-step-2-ss}, reorganizing terms using equation \eqref{eq:diffusive-flux-rearrangement}, recognizing $(v^\alpha_{; \alpha} - 2 v H)$ contains $\dot{a}_{\alpha \beta}$ due to equation \eqref{eq:jacobian-determinant-derivative}, and rearranging terms gives 
\begin{equation} \label{eq:second-law-thermo-local-step-3-ss}
	\begin{split}
		\rho \dot{s}
		&= - J_{\mathrm{s} \, ; \alpha}^{\, \alpha} 
		+ \rho \eta_{\mathrm{e}}
		+ \rho \eta_{\mathrm{i}} \\[3pt]
		&= - \Big(
			\dfrac{\jq^{\, \alpha}
			- \sum_{k = 2}^3 \, ( \mu_k - \mu_1) \, j_k^{\, \alpha}}{T}
		\Big)_{; \alpha}
		+ \dfrac{\rho r}{T}
		+ \dfrac{\alpha_{\mathrm{p}} \, \mupb \, \scmp \mathcal{R}}{T} \\[3pt]
		&\hspace{20pt} - \dfrac{\jq^{ \, \alpha} \, T_{, \alpha}}{T^2}
		+ \sum_{k = 2}^3 \bigg(
			\dfrac{(b_k)_\alpha - (b_1)_\alpha}{T}
			- \Big(
				\dfrac{\mu_k - \mu_1}{T}
			\Big)_{\!\! , \alpha}
		\bigg) \, j_k^{\, \alpha}
		+ \dfrac{\mathscr{A} \mathcal{R}}{T}
		\\[4pt]
		&\hspace{40pt} + \dfrac{1}{T} \bigg\{
			\dfrac{1}{2} \bigg[
				\sigma^{\alpha \beta}
				- \rho \Big(
					\dfrac{\partial \psi}{\partial a_{\alpha \beta}}
					+ \dfrac{\partial \psi}{\partial a_{\beta \alpha}}
				\Big)
				+ a^{\alpha \beta} \sum_{k = 1}^3 \mu_k \, \rho_k
			\bigg] \dot{a}_{\alpha \beta}
			\\[3pt]
			&\hspace{188pt}+ \bigg[
				M^{\alpha \beta}
				- \dfrac{\rho}{2} \Big(
					\dfrac{\partial \psi}{\partial b_{\alpha \beta}}
					+ \dfrac{\partial \psi}{\partial b_{\beta \alpha}}
				\Big)
			\bigg] \dot{b}_{\alpha \beta} 
		\bigg\} 
		~.
	\end{split}
\end{equation}
\eqnspace

By inspection of equation \eqref{eq:second-law-thermo-local-step-3-ss}, the in-plane entropy flux $\bmjs$ is 
\begin{equation} \label{eq:entropy-flux-surface-species}
	J_{\mathrm{s}}^{\, \alpha}
	= \dfrac{1}{T} \bigg(
		\jq^{\, \alpha}
		- \sum_{k = 2}^3 \big( \mu_k - \mu_1 \big) \, j_k^{\, \alpha}
	\bigg) ~.
\end{equation}
\eqnspace

The external entropy supply now contains a new contribution from the bulk phase, namely $\alpha_{\mathrm{p}} \mupb \, \scmp \mathcal{R} / T$, in addition to $\rho r / T$.
The rate of change of mass of the membrane is given by $\scmp \mathcal{R}$ and is equivalent to the diffusive flux of proteins from the bulk in the direction normal to the membrane.
Therefore, $\alpha_{\mathrm{p}} \mupb \, \scmp \mathcal{R} / T$ is the entropy flux at the boundary of the bulk phase \cite{degroot-mazur}.
Accordingly, the external entropy supply $\rho \eta_{\mathrm{e}}$ is given by 
\begin{equation} \label{eq:external-entropy-surface-species}
	\rho \eta_{\mathrm{e}} 
	= \dfrac{\rho r}{T} 
	- \dfrac{\mupb \, \scmp \mathcal{R}}{T}
	~,
\end{equation}
where we have substituted $\alpha_{\mathrm{p}} = -1$.
\textspace

Next, the terms on the right hand side of equation \eqref{eq:second-law-thermo-local-step-3-ss} which have not contributed to the entropy flux or the external entropy contribute to the internal entropy production.
To this end, the rate of internal entropy production per unit area $\rho \eta_{\mathrm{i}}$ is given by
\begin{equation} \label{eq:internal-entropy-surface-species}
	\begin{split}
		\hspace{-7pt} \rho \eta_{\mathrm{i}}
		&= - \dfrac{\jq^{ \, \alpha} \, T_{, \alpha}}{T^2}
		+ \sum_{k = 2}^3 \bigg(
			\dfrac{(b_k)_\alpha - (b_1)_\alpha}{T}
			- \Big(
				\dfrac{\mu_k - \mu_1}{T}
			\Big)_{\!\! , \alpha}
		\bigg) j_k^{\, \alpha} 
		\\[3pt]
		&\hspace{18pt} + \dfrac{\mathscr{A} \mathcal{R}}{T}
		+ \dfrac{1}{T} \bigg\{
			\dfrac{1}{2} \bigg[
				\sigma^{\alpha \beta}
				- \rho \Big(
					\dfrac{\partial \psi}{\partial a_{\alpha \beta}}
					+ \dfrac{\partial \psi}{\partial a_{\beta \alpha}}
				\Big)
				+ a^{\alpha \beta} \sum_{k = 1}^3 \mu_k \, \rho_k
			\bigg] \dot{a}_{\alpha \beta}
			\\[3pt]
			&\hspace{160pt} + \bigg[
				M^{\alpha \beta}
				- \dfrac{\rho}{2} \Big(
					\dfrac{\partial \psi}{\partial b_{\alpha \beta}}
					+ \dfrac{\partial \psi}{\partial b_{\beta \alpha}}
				\Big)
			\bigg] \dot{b}_{\alpha \beta} 
		\bigg\}
		\, \ge \, 0 ~.
	\end{split}
\end{equation}
Equation \eqref{eq:internal-entropy-surface-species} is the sum of terms which are a product of a thermodynamic force and a corresponding flux, where the forces and fluxes are either scalars, vectors, or tensors.
We invoke the Curie principle \cite{curie, degroot-mazur} and assume the phenomenological coefficients connecting terms of different tensorial order are zero.
As the vectorial and tensorial terms in equation \eqref{eq:internal-entropy-surface-species} are similar to the multi-component internal entropy production \eqref{eq:internal-entropy-ip}, we follow the procedure described in Section \ref{sec:general-thermo-vars} to find the relations between the vectorial and tensorial thermodynamic forces and fluxes as 
\begin{align}
	\jq^{\, \alpha}
	&= - \kappa \, T^{, \alpha} 
	~,
	\label{eq:heat-flux-binding}
	\\[5pt]
	j_k^{\, \alpha}
	&= D_k \bigg(
		\dfrac{(b_k)^\alpha - (b_1)^\alpha}{T}
		- \Big(
			\dfrac{\mu_k - \mu_1}{T}
		\Big)^{, \alpha}
	\bigg) 
	~,
	\label{eq:species-flux-binding}
	\\[3pt]
	\sigma^{\alpha \beta}
	&= \rho \bigg(
		\dfrac{\partial \psi}{\partial a_{\alpha \beta}}
		+ \dfrac{\partial \psi}{\partial a_{\beta \alpha}}
	\bigg)
	- a^{\alpha \beta} \sum_{k = 1}^3 \mu_k \, \rho_k
	+ \pi^{\alpha \beta} ~,
	\label{eq:sigma-alpha-beta-binding}
	\\[5pt]
	M^{\alpha \beta}
	&= \dfrac{\rho}{2} \bigg(
		\dfrac{\partial \psi}{\partial b_{\alpha \beta}}
		+ \dfrac{\partial \psi}{\partial b_{\beta \alpha}}
	\bigg) ~,
	\label{eq:m-alpha-beta-binding}
	\\[8pt]
	N^{\alpha \beta} &= \sigma^{\alpha \beta} + b^\beta_\mu M^{\mu \alpha} ~, 
	\label{eq:n-alpha-beta-binding}
	\\
	\shortintertext{and}
	S^\alpha &= - M^{\beta \alpha}_{; \beta} ~, \label{eq:s-alpha-binding}
\end{align}
with $\pi^{\alpha \beta}$ given by equation \eqref{eq:viscous-stress}.
Equation \eqref{eq:species-flux-binding} holds for $k \in \{ 2, 3 \}$, and $j_1^{\, \alpha}$ is calculated as $j_1^{\, \alpha} = - j_2^{\, \alpha} - j_3^{\, \alpha}$.
It can be seen from equation \eqref{eq:sigma-alpha-beta-binding} that the in-plane stresses now contain additional contributions from the adsorption of proteins onto the membrane. 
\textspace

There is only a single scalar thermodynamic term in equation \eqref{eq:internal-entropy-surface-species}, namely that involving the chemical affinity $\mathscr{A}$, and the phenomenological relationship between the chemical affinity and the reaction rate is obtained as
\begin{gather}
	\mathcal{R} = \dfrac{\mathscr{A} L}{T} 
	~,
	\label{eq:j-linear-irreversible-surface-species}
\end{gather}
where $L$ is a positive constant.
Equation \eqref{eq:j-linear-irreversible-surface-species} is analagous to the result for chemical reactions in three-dimensional Cartesian systems, as determined by Prigogine \cite{prigogine}.
While linear irreversible thermodynamics may provide us with phenomenological relations with a wide range of validity, for chemical reactions nonlinear effects become significant soon after departure from equilibrium.
Established theory allows us to operate outside the linear irreversible regime \cite{prigogine, thanos}.
To this end, the chemical potential of an individual species $k$ is defined as
\begin{equation} \label{eq:chemical-potential-activity-expansion}
	\mathscr{M}_k \mu_k
	= \mathscr{M}_k \mu_k^\circ (T)
	+ RT \ln{a_k} ~.
\end{equation}
Here $\mu_k^\circ (T)$ is a chemical potential under a set of standard conditions, which is a function of only the temperature, and $R$ is the ideal gas constant.
The activity of component $k$, denoted as $a_k$, is simply the concentration $n_k$ for ideal systems. 
The equilibrium constant $K_{\mathrm{eq}} (T)$ for the reaction is given by
\begin{equation} \label{eq:equilibrium-constant}
	RT \ln{K_{\mathrm{eq}}(T)}
	= - \alpha_{\mathrm{p}} \scmp \mu_{\mathrm{p}}^\circ(T)
	- \sum_{k = 2}^3 \alpha_k \mathscr{M}_k \mu_k^\circ(T) ~,
\end{equation}
and is related to the forward and backward reaction rate constants, $\fvect{k}$ and $\bvect{k}$, through the equation 
\begin{equation} \label{eq:forward-backward-rates-equilibrium-constant}
	K_{\mathrm{eq}}(T) 
	= \dfrac{\fvect{k}}{\bvect{k}}
	~.
\end{equation}
We now have several relations which relate the reaction rates to rate constants \eqref{eq:reaction-rate-forward-backward-difference}--\eqref{eq:backward-mass-action}, chemical affinity \eqref{eq:affinity}, and thermodynamic quantities between the membrane and bulk \eqref{eq:chemical-potential-activity-expansion}--\eqref{eq:forward-backward-rates-equilibrium-constant}.
Combining these equations, we obtain
\begin{equation} \label{eq:reaction-rate-general}
	\mathcal{R}
	= \fvect{\mathcal{R}} \bigg(
		1 - \exp \bigg\{
			- \dfrac{\mathscr{A}}{RT}
		\bigg\}
	\bigg) ~.
\end{equation}
\eqnspace

In developing equation \eqref{eq:reaction-rate-general}, none of the arguments required the system be close to equilibrium.
The result is therefore generally valid and describes how the reaction rate is related to a thermodynamic force.
In the limit
\begin{equation} \label{eq:affinity-limit-linear}
	\left| \dfrac{\mathscr{A}}{RT} \right| \ll 1 ~,
\end{equation}
we Taylor expand the reaction rate \eqref{eq:reaction-rate-general} about equilibrium to find 
\begin{equation} \label{eq:reaction-rate-linear-irreversible-regime}
	\mathcal{R} = \fvect{\mathcal{R}}_{\mathrm{e}} \,\, \bigg(
		\dfrac{\mathscr{A}}{RT}
	\bigg) ~,
\end{equation}
where $\fvect{\mathcal{R}}_{\mathrm{e}}$ is the forward reaction rate at equilibrium ($\fvect{\mathcal{R}}_{\mathrm{e}} = \bvect{\mathcal{R}}_{\mathrm{e}}$).
Comparing equation \eqref{eq:reaction-rate-linear-irreversible-regime} with equation \eqref{eq:j-linear-irreversible-surface-species}, we find the phenomenological constant $L$ is given by $L = \fvect{\mathcal{R}}_{\mathrm{e}} / R$ in cases where equation \eqref{eq:affinity-limit-linear} is valid.


\subsubsection{Helmholtz Free Energy---Change of Variables}

As in the single- and multi-component cases, the requirement for the Helmholtz free energy density $\psi$ to be an absolute scalar field places restrictions on its functional form.
Under Galilean invariance the Helmholtz free energy density may depend on $a_{\alpha \beta}$ and $b_{\alpha \beta}$ only through $J$, $H$, and $K$, where $\partial J / \partial a_{\alpha \beta} = \tfrac{1}{2} J a^{\alpha \beta}$ and $\partial J / \partial b_{\alpha \beta} = 0$.
In this case, the Helmholtz free energy density may be written as 
\begin{equation} \label{eq:psi-bar-binding}
	\psi(a_{\alpha \beta}, \, b_{\alpha \beta}, \, T, \, \{ \rho_k \}_{k \, = \, 1, \, 2, \, 3})
	= \bar{\psi} (J, \, H, \, K, \, T, \, \{ \rho_k \}_{k \, = \, 1, \, 2, \, 3})
	~.
\end{equation}
\eqnspace

Substituting equation \eqref{eq:psi-bar-binding} into the stresses found in equations \eqref{eq:sigma-alpha-beta-binding}--\eqref{eq:s-alpha-binding} and using an analogous procedure to the derivation of equations \eqref{eq:sigma-alpha-beta-1c-general}--\eqref{eq:s-alpha-1c-general}, we obtain
\begin{align}
	\sigma^{\alpha \beta}
	&= \rho \big(
		J \, \bar{\psi}_{, J}
		- 2H \bar{\psi}_{, H}
		- 2K \bar{\psi}_{, K}
	\big) a^{\alpha \beta}
	- a^{\alpha \beta} \sum_{k = 1}^3 \mu_k \, \rho_k
	+ \rho \, \bar{\psi}_{, H} \, \bar{b}^{\alpha \beta}
	+ \pi^{\alpha \beta} ~,
	\label{eq:sigma-alpha-beta-binding-general}
	\\[5pt]
	M^{\alpha \beta}
	&= \tfrac{1}{2} \rho \, \bar{\psi}_{, H} \, a^{\alpha \beta}
	+ \rho \, \bar{\psi}_{, K} \, \bar{b}^{\alpha \beta} ~,
	\label{eq:m-alpha-beta-binding-general}
	\\[5pt]
	N^{\alpha \beta}
	&= \rho \big(
		J \bar{\psi}_{, J}
		- H \bar{\psi}_{, H}
		- K \bar{\psi}_{, K}
	\big) a^{\alpha \beta}
	- a^{\alpha \beta} \sum_{k = 1}^3 \mu_k \, \rho_k
	+ \tfrac{1}{2} \, \rho \, \bar{\psi}_{, H} \, \bar{b}^{\alpha \beta}
	+ \pi^{\alpha \beta} ~,
	\label{eq:n-alpha-beta-binding-general}
	\\
	\shortintertext{and}
	S^{\alpha}
	&= - \tfrac{1}{2} (\rho \bar{\psi}_{, H})_{; \beta} \, a^{\alpha \beta}
	- ( \rho \bar{\psi}_{, K} )_{; \beta} \, \bar{b}^{\alpha \beta} 
	~.
	\label{eq:s-alpha-binding-general}
\end{align}


\subsubsection{Helfrich Energy Density}

For the membrane system under consideration, we assume the bound proteins do not exhibit a phase transition.
Consequently, the Helmholtz free energy contains neither the gradient contribution \eqref{eq:w-g} nor the double-well contribution \eqref{eq:w-dw} present in the multi-component model.
We further assume the spontaneous curvature in the membrane, $C$, is only due to the presence of bound proteins, and may be written as $C = C(\rho_3)$.
Finally, we assume there is an energetic penalty for protein-bound and unbound PI(4,5)P\textsubscript{2} on the membrane patch. 
\\[-9pt]

Given the above assumptions about membrane energetics in the case of protein binding reactions, the total Helmholtz free energy consists of three terms.
The first is the now familiar Helfrich bending energy $w_{\text{h}}$ given by
\begin{equation} \label{eq:w-h-ss}
	w_{\text{h}}
	= \kb \big[
		H - C(\rho_3)
	\big]^2
	+ \kg K ~,
\end{equation}
where we assume the bending modulus $\kb$ is independent of concentration.
The cost of areal dilation and compression, $w_{\text{c}}$, is again written as
\begin{equation} \label{eq:w-c-ss}
	w_{\text{c}}
	= \dfrac{\kc}{J} \big( 1 - J \big)^2 ~,
\end{equation}
where we also assume the compression modulus $\kc$ is independent of concentration.
Finally, we model the energetic penalty of the minority species PI(4,5)P\textsubscript{2} and PI(4,5)P\textsubscript{2}--epsin-1 as
\begin{equation} \label{eq:w-sw}
	w_{\text{s}}
	= k_2 (\rho_2)^2
	+ k_3 (\rho_3)^2
	~,
\end{equation}
where $k_2$ and $k_3$ are constants.
Equation \eqref{eq:w-sw} may be considered as a simple model for understanding the energetic penalty of minority species, while a general derivation from microscopic considerations is left to future work.
The total energy density is given by
\begin{equation} \label{eq:helmholtz-surface-single-well}
	\begin{split}
		\rho \bar{\psi}
		&= w_{\text{h}}
		+ w_{\text{c}}
		+ w_{\text{s}} 
		+ \rho f(T)
		\\[3pt]
		&= \kb \big[
			H - C(\rho_3)
		\big]^2
		+ \kg K
		+ \dfrac{\kc}{J} \big( 1 - J \big)^2
		+ k_2 (\rho_2)^2
		+ k_3 (\rho_3)^2
		+ \rho f(T)
		~,
	\end{split}
\end{equation}
where as in the single-component case \eqref{eq:helmholtz-total-1c} $f(T)$ is a function of the temperature such that the local equilibrium assumption \eqref{eq:local-equilibrium-entropy-binding} holds.
\eqnspace

Given the Helmholtz free energy density in equation \eqref{eq:helmholtz-surface-single-well}, the chemical potentials $\mu_k$ defined in equation \eqref{eq:mu-binding} can be obtained as 
\begin{align}
	\begin{split}
		\mu_1
		&= - \dfrac{1}{\rho} \Big(
			\kb \big[
				H - C(\rho_3)
			\big]^2
			+ \kg K
			+ \dfrac{\kc}{J} \big( 1 - J \big)^2
			+ k_2 (\rho_2)^2 
			+ k_3 (\rho_3)^2 
		\Big)
		~,
		\label{eq:mu-1-ss}
	\end{split}
	\\[3pt]
	\begin{split}
		\mu_2
		&= - \dfrac{1}{\rho} \Big(
			\kb \big[
				H - C(\rho_3)
			\big]^2
			+ \kg K
			+ \dfrac{\kc}{J} \big( 1 - J \big)^2
			+ k_2 (\rho_2)^2 
			+ k_3 (\rho_3)^2 
		\Big)
		+ 2 \, k_2 \, \rho_2
		~,
		\label{eq:mu-2-ss}
	\end{split}
	\\
	\shortintertext{and}
	\begin{split}
		\hspace{-9.5pt}\mu_3
		&= - \dfrac{1}{\rho} \Big(
			\kb \big[
				H - C(\rho_3)
			\big]^2
			+ \kg K
			+ \dfrac{\kc}{J} \big( 1 - J \big)^2
			+ k_2 (\rho_2)^2 
			+ k_3 (\rho_3)^2 
		\Big)
		\\[2pt]
		&\hspace{106pt} - 2 \kb \big[ H - C(\rho_3) \big] C'(\rho_3)
		+ 2 \, k_3 \, \rho_3
		~.
		\label{eq:mu-3-ss}
		\\[3pt]
	\end{split}
\end{align}
Substituting the Helmholtz free energy density \eqref{eq:helmholtz-surface-single-well} and chemical potentials \eqref{eq:mu-1-ss}--\eqref{eq:mu-3-ss} into equations \eqref{eq:sigma-alpha-beta-binding-general}--\eqref{eq:s-alpha-binding-general}, we find the stresses and moments of the membrane to be 
\begin{align}
	\begin{split}
		\sigma^{\alpha \beta}
		&= \kb \bigg(
			\Big[
				-3 H^2
				+ 2 H C (\rho_3)
				+ \big( C (\rho_3) \big)^2
				+ 2 \, \rho_3 \, \big[ H - C(\rho_3) \big] C'(\rho_3)
			\Big] a^{\alpha \beta}
			+ 2 \Big[
				H - C(\rho_3)
			\Big] \bar{b}^{\alpha \beta}
		\bigg) 
		\\[3pt]
		&\hspace{20pt}- \kg K a^{\alpha \beta}
		+ 2 \kc \big( J - 1 \big) a^{\alpha \beta}
		- \Big[
			k_2 (\rho_2)^2
			+ k_3 (\rho_3)^2
		\Big] a^{\alpha \beta}
		+ \pi^{\alpha \beta}
		~,
		\label{eq:sigma-alpha-beta-ss-helfrich}
	\end{split}
	\\[0pt]
	M^{\alpha \beta}
	&= \kb \Big[
		H - C(\rho_3)
	\Big] a^{\alpha \beta}
	+ \kg \, \bar{b}^{\alpha \beta}
	~,
	\label{eq:m-alpha-beta-ss-helfrich}
	\\[11pt]
	\begin{split}
		N^{\alpha \beta}
		&= \kb \Big(
			\Big[
				- H^2
				+ \big( C (\rho_3) \big)^2
				+ 2 \, \rho_3 \, \big[ H - C(\rho_3) \big] C'(\rho_3)
			\Big] a^{\alpha \beta}
			+ \Big[
				H - C(\rho_3)
			\Big] \bar{b}^{\alpha \beta}
		\Big)
		\\[3pt]
		&\hspace{20pt}+ 2 \kc \Big( J - 1 \Big) a^{\alpha \beta}
		- \Big[
			k_2 (\rho_2)^2
			+ k_3 (\rho_3)^2
		\Big] a^{\alpha \beta}
		+ \pi^{\alpha \beta}
		~,
		\label{eq:n-alpha-beta-ss-helfrich}
	\end{split}
	\\
	\intertext{and}
	S^\alpha
	&= - \kb \Big[
		H
		- C(\rho_3) 
	\Big]^{, \alpha}
	~.
	\label{eq:s-alpha-ss-helfrich}
\end{align}


\subsection{Equations of Motion}

In this section, we provide the equations of motion for a membrane undergoing protein binding and unbinding reactions.
For the specific membrane patch under consideration, there are seven unknowns: the total mass density $\rho$, the species densities $\rho_2$ and $\rho_3$, the reaction rate $\mathcal{R}$, and the three components of the velocity $\bm{v}$.
We choose the total mass density $\rho$, rather than the DOPC mass density $\rho_1$, as a fundamental variable because we are interested in the overall membrane behavior rather than that of the chemically unreactive DOPC.
The reaction rate $\mathcal{R}$ may be expressed as
\begin{equation} \label{eq:reaction-rate-general-eom}
	\mathcal{R}
	= \fvect{k} \, n_2 \, \npb \bigg(
		1 - \exp \bigg\{
			- \dfrac{\mathscr{A}}{RT}
		\bigg\}
	\bigg) ~,
\end{equation}
where we have substituted equation \eqref{eq:forward-mass-action} for $\fvect{\mathcal{R}}$ into equation \eqref{eq:reaction-rate-general}.
In equation \eqref{eq:reaction-rate-general-eom}, we assume the concentration of bulk proteins $\npb$ is known, and $n_2$ is given by equation \eqref{eq:c-k-binding} for $k = 2$.
\textspace

The conservation of total mass is given by equation \eqref{eq:mass-conservation-surface-species-total}, where we expand the material derivative to obtain 
\begin{equation} \label{eq:total-mass-balance-eom-ss}
	\rho_{, t}
	+ \rho_{, \alpha} v^\alpha
	+ \big( v^\alpha_{; \alpha} - 2 v H \big) \rho
	= \scmp \mathcal{R} ~.
\end{equation}
The mass balance of species two and three are given by equation \eqref{eq:mass-conservation-surface-species-local}, where we can substitute the diffusive flux obtained in equation \eqref{eq:species-flux-binding}.
Since $\mu_2 - \mu_1 = 2 \, k_2 \, \rho_2$ from equations \eqref{eq:mu-1-ss}--\eqref{eq:mu-2-ss}, the mass balance for species two is given by
\begin{equation} \label{eq:species-2-mass-balance-eom-ss}
	\rho_{2, t}
	+ \rho_{2, \alpha} v^\alpha
	+ \big( v^\alpha_{; \alpha} - 2 v H \big) \rho_2
	+ \bigg(
		D_2 \bigg[
			\dfrac{(b_2)^\alpha - (b_1)^\alpha}{T}
			- 2 \, k_2 \Big(
				\dfrac{\rho_2}{T}
			\Big)^{\! , \alpha}
		\bigg]
	\bigg)_{; \alpha}
	= - \mathscr{M}_2 \mathcal{R} ~.
\end{equation}
Calculating the chemical potential difference $\mu_3 - \mu_1$ from equations \eqref{eq:mu-1-ss} and \eqref{eq:mu-3-ss}, we obtain the mass balance for species 3 as
\begin{equation} \label{eq:species-3-mass-balance-eom-ss}
	\begin{split}
		\rho_{3, t}
		&+ \rho_{3, \alpha} v^\alpha
		+ \big( v^\alpha_{; \alpha} - 2 v H \big) \rho_3
		\\[3pt]
		&+ \bigg(
			D_3 \bigg[
				\dfrac{(b_3)^\alpha - (b_1)^\alpha}{T}
				+ \Big(
					\dfrac{2 \, \kb}{T} \Big[ H - C(\rho_3) \Big] C'(\rho_3)
					- \dfrac{2 \, k_3 \, \rho_3}{T}
				\Big)^{\! , \alpha}
			\bigg]
		\bigg)_{; \alpha}
		= \mathscr{M}_3 \mathcal{R} ~.
	\end{split}
\end{equation}
\eqnspace

Substituting the stresses and moments \eqref{eq:sigma-alpha-beta-ss-helfrich}--\eqref{eq:s-alpha-ss-helfrich} into the tangential equations \eqref{eq:tangential-general-eoms-elastic-compressible} and the shape equation \eqref{eq:normal-general-eom-elastic-compressible}, we obtain
\begin{gather}
	\begin{split} \label{eq:tangential-eom-ss}
		\rho \Big(
			v^\alpha_{, t} - w^\alpha v + v^\mu {w_\mu}^\alpha
		\Big)
		= \rho b^\alpha
		&+ \pi^{ \mu \alpha}_{; \mu}
		+ 2 \, \kb \, \rho_3 \Big(
			C'(\rho_3) \Big[ H - C(\rho_3) \Big] 
		\Big)^{, \alpha}
		\\[2pt]
		&+ 2 \, \kc \, J^{, \alpha} 
		- 2 \Big(
			k_2 \, \rho_2 \, \rho_2^{, \alpha}
			+ k_3 \, \rho_3 \, \rho_3^{, \alpha}
		\Big)
	\end{split}
	\intertext{and}
	\begin{split} \label{eq:normal-eom-ss}
		\rho \Big(
			v_{, t} + v^\alpha w_\alpha
		\Big) 
		= p
		&+ \pi^{\alpha \beta} b_{\alpha \beta}
		- \kb \, \Delta \Big[
			H - C(\rho_3)
		\Big] 
		- 2 H \Big[
			k_2 (\rho_2)^2
			+ k_3 (\rho_3)^2
		\Big]
		\\[2pt]
		&+ 4 \kc H \big( J - 1 \big)
		- 2 \kb \Big[H - C(\rho_3)\Big] \Big(
			H^2 
			+ HC(\rho_3) 
			- K 
			-2 \rho_3 C'(\rho_3)
		\Big)
		~.
	\end{split} 
\end{gather}
\eqnspace

Equations \eqref{eq:reaction-rate-general-eom}--\eqref{eq:normal-eom-ss} solve for the seven unknowns of the membrane patch.
We note that in developing these equations, we have assumed a phase transition does not exist---and we enforced this assumption by our choice of $\psi$ in equation \eqref{eq:helmholtz-surface-single-well}.
If we were to consider a different system in which proteins exhibit phase separation, we would use the Helmholtz energy density provided in the multi-component case \eqref{eq:total-helmholtz-2c-is} with $\phi$ given by $\phi = \rho_3 / \rho$.
As we have already calculated the stresses and moments for such an energy density, we could easily determine the equations of motion and the consequences of a system with such a phase separation.

%
%

\subsection{Boundary and Initial Conditions}

We conclude the problem of protein binding and unbinding reactions by providing possible boundary conditions.
As the momentum equations \eqref{eq:tangential-eom-ss}--\eqref{eq:normal-eom-ss} are of the same structure as in the multi-component example \eqref{eq:normal-eom-two-component-dwp-general-c-ip}--\eqref{eq:tangential-eom-two-component-dwp-general-c-ip}, the momentum balance boundary conditions from the multi-component analysis of Section \ref{sec:bc-mc} are appropriate for a membrane patch with chemical reactions.
The tangential equations require either the in-plane velocities $v^\alpha$ or the in-plane components of the force, $f_\nu$ and $f_\tau$, to be specified at the boundary.
The shape equation becomes mathematically well-posed if we specify the position and its gradient in the $\bm{\nu}$ direction, or the moment $M$ and the shear force $f_n$, at the boundary.
We determine the force and moment at the boundary by substituting the stresses \eqref{eq:sigma-alpha-beta-ss-helfrich}--\eqref{eq:s-alpha-ss-helfrich} into equations \eqref{eq:f-nu}--\eqref{eq:boundary-moment-b-c} to obtain
\begin{align}
	\begin{split}
		f_\nu
		&= \kb \Big[
			H - C(\rho_3)
		\Big] \Big(
			H - C(\rho_3)
			- \kappa_\nu
			+ 2 \, \rho_3 \, C'(\rho_3)
		\Big)
		\\[2pt]
		&\hspace{40pt} + 2 \kc \big(
			J - 1
		\big)
		- \Big[
			k_2 (\rho_2)^2
			+ k_3 (\rho_3)^2
		\Big] 
		- \kg \, \xi^2
		+ \pi^{\alpha \beta} \nu_\alpha \nu_\beta ~,
	\end{split}
	\label{eq:f-nu-helfrich-single-well} \\[8pt]
	\begin{split}
		f_\tau
		&= - \kb \, \xi \, \Big[
			H - C(\rho_3)
		\Big]
		- \kg \, \xi \, \kappa_\tau
		+ \pi^{\alpha \beta} \nu_\alpha \tau_\beta ~,
	\end{split}
	\label{eq:f-tau-helfrich-single-well} \\[5pt]
	\begin{split}
		f_n
		&= - \kb \Big[
			H - C(\rho_3)
		\Big]_{\! , \nu}
		+ \kg \dfrac{\mathrm{d} \xi}{\mathrm{d} \ell} ~,
	\end{split}
	\label{eq:f-n-helfrich-single-well} \\
	\shortintertext{and}
	\begin{split}
		M
		&= \kb \Big[
			H - C(\rho_3)
		\Big]
		+ \kg \, \kappa_\tau
		~.
	\end{split}
	\label{eq:boundary-moment-b-c-helfrich-single-well}
\end{align}
\eqnspace

The species balances for PI(4,5)P\textsubscript{2} and PI(4,5)P\textsubscript{2}--epsin-1 are given by equations \eqref{eq:species-2-mass-balance-eom-ss} and \eqref{eq:species-3-mass-balance-eom-ss}, respectively. 
Both species balances contain the Laplacian of the corresponding mass density.
We therefore specify either $\rho_k$ or the species flux at the boundary in the $\bm{\nu}$ direction, $j_k^{\alpha} \, \nu_\alpha$, where $k \in \{ 2, 3 \}$.
\textspace

Finally, we note the equation for the reaction rate $\mathcal{R}$ \eqref{eq:reaction-rate-general-eom} does not contain any spatial derivatives, and we only need to know the initial concentrations to determine the value of $\mathcal{R}$ at any later time.
Given these initial concentrations, the initial membrane configuration and velocity, and boundary conditions, our problem is mathematically well-posed.


\section{Conclusions} \label{sec:conclusion}

In this paper, we developed an irreversible thermodynamic framework for arbitrarily curved lipid membranes.
We began by modeling a compressible single-component lipid membrane, including out-of-plane elastic bending and in-plane viscous fluid flow.
Using the balances of mass, linear and angular momentum, energy, and entropy, we determined the entropy production for single-component membranes in terms of the viscous stresses and in-plane velocity gradients through the framework of linear irreversible thermodynamics.
This framework provided a natural way to develop constitutive laws including the viscous stresses, and the resulting equations of motion are identical to the results of earlier studies which proposed constitutive forms of the viscous stresses \cite{arroyo-pre-2009, kranthi-bmm-2012}.
We then extended the model to include multiple components that could diffuse in the plane of the membrane.
We modeled phase transitions between \lo\ and \ld\ domains, and learned how phase transitions are coupled to fluid flow, diffusion, and bending.
Finally, we extended our multi-component model further to include the binding and unbinding of peripheral proteins in a biologically relevant example.
We found how bending, flow, diffusion, and binding are coupled, and determined the thermodynamic driving force governing protein binding.
\textspace

The current theory could be expanded in several ways to develop a more complete description of lipid membranes.
First, we could model lipid membranes as two monolayer leaflets rather than a single sheet as in \cite{rahimi-soft-matter-2013, seifert-epl-1993, rahimi-arroyo-pre-2012}.
By accounting for individual monolayer leaflet behavior, we would understand how inter-monolayer friction and asymmetries between the monolayers affect the equations of motion.
Furthermore, we could model the bulk viscosity of the fluid surrounding the membrane and understand how an additional dissipative mode affects membrane dynamics as in \cite{seifert-long, kraus-prl-1996}.
Finally, we could model phase transitions involving proteins in which proteins, once bound, can separate into high-density and low-density phases.
While this is a simple theoretical extension from our current \lo--\ld\ phase transition model, phase transitions due to proteins are important in many biological phenomena.
Computational microscopic studies have recently demonstrated the importance of protein-lipid and protein-protein interactions in understanding membrane bending \cite{brannigan-bpj-2007, west-bpj-2009, blood-voth-pnas-2006, atzberger-jcp-2013}, and have also shown a force of assembly between proteins which favor the disordered phase, but which have been placed in the ordered phase \cite{katira-elife-2016}.
To this end, it may be of interest to understand how protein interactions in the presence of \lo\ and \ld\ phases can affect bending, flow, and the collective reorganization of the membrane.
\textspace

In addition to further theoretical advances, the theoretical framework presented in this paper can be used to develop numerical methods such as finite element methods to simulate membrane behavior in different biological processes such as endocytosis, intra-cellular trafficking, and cell-cell signaling.
An initial effort along these lines is presented in \cite{sauer-liquid-shell-corr-2016}.
\textspace

Finally, we note the theoretical framework presented here may be generally applied to systems beyond lipid membranes.
For example, we could apply this procedure to understand the dynamics of the cell wall, which plays a central role in providing structure as well as interacting with the cell membrane.
As the cell wall has elastic in-plane behavior in addition to elastic out-of-plane bending, the constitutive assumptions will differ from those of lipid membranes but can be easily included into the framework developed here.
\vspace{10pt}

%
%

\section*{Acknowledgements}

We thank Eva Schmid, David Limmer, and Clay Radke for useful discussions and Joel Tchoufag for carefully reading the manuscript.
K.K.M. acknowledges the support of the University of California, Berkeley and National Institutes of Health Grant No. R01-GM110066.
R.A.S. acknowledges the support of the German Research Foundation (DFG) through Grant No. GSC 111.
A.S. is supported by the Computational Science Graduate Fellowship from the U.S. Department of Energy.\\
\vspace{02pt}

%
%

\begin{appendices}

	\renewcommand{\theequation}{\thesection.\arabic{equation}}
	\setcounter{equation}{0}

	%
	%

	\section{Surface parametrizations} \label{sec:appendix-surface-param}

	%
	%

	\subsection{Convected Coordinates} \label{sec:appendix-convected-coordinates}

	To better understand the connection between kinematic quantities on the current and reference membrane configurations, as done in \cite{kranthi-bmm-2012, powers-rmp-2010}, we begin by introducing the convected coordinates $\xi^\alpha$ which parameterize the reference configuration---defined to be the membrane patch at some fixed time $t_0$.
	The convected coordinates $\xi^\alpha$ are defined as
	\begin{equation}
		\xi^\alpha
		= \theta^\alpha \big\rvert_{t = t_0}
		~.
	\end{equation}
	At later times the membrane patch will in general occupy a different configuration, yet any such configuration can be mapped back to the reference patch.
	It is therefore appropriate to talk about how a point with constant $\xi^\alpha$ moves in time, as these coordinates are convected along with material points---hence our choice of the name ``convected coordinates.''
	Since a point with constant convected coordinates $\xi^\alpha$ always refers to the same material point, $\xi^\alpha$ can also be called a Lagrangian or referential parametrization.
	\textspace

	To understand how a point with constant fixed surface coordinates $\theta^\alpha$ moves in time, we consider the relationship between the two parametrizations $\theta^\alpha$ and $\xi^\alpha$.
	The fixed surface coordinates $\theta^\alpha$ change over time as the membrane deforms, and can be formally written as
	\begin{equation}
		\theta^\alpha
		= \theta^\alpha (\xi^\beta, t)
		~.
	\end{equation}
	The above equation allows us to relate velocities in terms of the two parametrizations.
	To begin, the membrane position $\bm{x}$ is expressed in terms of the surface coordinates as $\bm{x} (\theta^\alpha, t)$ and in terms of the convected coordinates as $\hat{\bm{x}} (\xi^\alpha, t)$, with
	\begin{equation}
		\bm{x} (\theta^\alpha, t)
		= \bm{x} (\theta^\alpha(\xi^\beta, t), t)
		= \hat{\bm{x}} (\xi^\beta, t)
		~,
	\end{equation}
	where the `hat' denotes a quantity expressed in terms of the convected coordinates $\xi^\alpha$.
	The velocity $\bm{v}$ is formally defined as the rate of change of position of a material point, written as
	\begin{equation} \label{eq:appendix-v-def}
		\bm{v}
		= \dfrac{\partial \hat{\bm{x}}}{\partial t} \Big\rvert_{\xi^\alpha}
		~.
	\end{equation}
	By expressing $\hat{\bm{x}}(\xi^\beta, t)$ in terms of the fixed surface coordinates as $\bm{x} (\theta^\alpha, t)$ and applying the chain rule, we obtain
	\begin{equation}
		\bm{v} \label{eq:appendix-v-decomposed}
		= \dfrac{\partial \bm{x}}{\partial t} \Big\rvert_{\theta^\alpha}
		+ \dfrac{\partial \theta^\alpha}{\partial t} \Big\rvert_{\xi^\beta} \, \bm{a}_\alpha
		~.
	\end{equation}
	We define the in-plane contravariant velocity components $v^\alpha$ as the rate of change of the fixed surface coordinates for a given $\xi^\beta$, given by
	\begin{equation} \label{eq:appendix-v-alpha}
		v^\alpha
		:= \dfrac{\partial \theta^\alpha}{\partial t} \Big\rvert_{\xi^\beta} 
		~,
	\end{equation}
	and moreover require the fixed surface coordinates $\theta^\alpha$ to be chosen such that
	\begin{equation} \label{eq:appendix-v-n}
		\dfrac{\partial \bm{x}}{\partial t} \Big\rvert_{\theta^\alpha} 
		= v \bm{n}
		~,
	\end{equation}
	so points of constant $\theta^\alpha$ move only in the direction normal to the membrane.
	The velocity can accordingly be decomposed as
	\begin{equation}
		\bm{v} = v \bm{n} + v^\alpha \bm{a}_\alpha
		~,
	\end{equation}
	as done in equation~\eqref{eq:velocity-euler}, on substitution of equations \eqref{eq:appendix-v-alpha}--\eqref{eq:appendix-v-n} into equation \eqref{eq:appendix-v-decomposed}.
	\textspace

	To further explain the relationship between the fixed surface and convected parametrizations, we consider a scalar function $f$ which can be written as $f(\theta^\alpha, t)$ or $\hat{f}(\xi^\beta, t)$.
	The material derivative of $f$ is formally defined as
	\begin{equation}
		\dfrac{\mathrm{d} f}{\mathrm{d}t}
		= \dfrac{\partial \hat{f}}{\partial t}\Big\rvert_{\xi^\beta}
		~,
	\end{equation}
	which on substitution of $f(\theta^\alpha, t)$ and application of the chain rule leads to
	\begin{equation}
		\begin{split}
			\dfrac{\mathrm{d} f}{\mathrm{d}t}
			&= \dfrac{\partial f}{\partial t}\Big\rvert_{\theta^\alpha}
			+ \dfrac{\partial f}{\partial \theta^\alpha}
			\dfrac{\partial \theta^\alpha}{\partial t}\Big\rvert_{\xi^\beta}
			\\[4pt]
			&= f_{, t} + f_{, \alpha} v^\alpha
			~,
		\end{split}
	\end{equation}
	in agreement with equation~\eqref{eq:material-derivative}.

	%
	%

	\subsection{Material Time Derivative of In-Plane Basis Vectors} \label{sec:appendix-material-derivative}

	In what follows, we derive explicitly the expression for the material time derivative of $\bm{a}_\alpha$ using the convected coordinates $\xi^\alpha$.
	We first express $\dot{\bm{a}}_\alpha$ using the formal definition of the material derivative in terms of the convected coordinates, which yields
	\begin{align} 
		\dot{\bm{a}}_\alpha
		&= \dfrac{\partial}{\partial t} \Big(
		\dfrac{\partial \bm{x} (\theta^\nu, t)}{\partial \theta^\alpha}
		\Big) \Big\rvert_{\xi^\beta}
		~.
		\\
		\intertext{By expressing $\bm{x}(\theta^\alpha, t)$ as $\hat{\bm{x}}(\xi^\mu, t)$ and applying the chain rule, we obtain}
		\begin{split}
			\dot{\bm{a}}_\alpha
			&= \dfrac{\partial}{\partial t} \Big(
			\dfrac{\partial \hat{\bm{x}} (\xi^\lambda, t)}{\partial \xi^\mu}
			\, \, 
			\dfrac{\partial \xi^\mu}{\partial \theta^\alpha}
			\Big) \Big\rvert_{\xi^\beta}
			\\[6pt]
			&= \dfrac{\partial}{\partial t} \Big(
			\dfrac{\partial \hat{\bm{x}} (\xi^\lambda, t)}{\partial \xi^\mu}
			\Big) \Big\rvert_{\xi^\beta}
			\, \,
			\dfrac{\partial \xi^\mu}{\partial \theta^\alpha}
			+ \dfrac{\partial \hat{\bm{x}} (\xi^\lambda, t)}{\partial \xi^\mu}
			\, \,
			\dfrac{\partial}{\partial t} \Big(
			\dfrac{\partial \xi^\mu}{\partial \theta^\alpha}
			\Big) \Big\rvert_{\xi^\beta}
			~.
		\end{split}
		\\
		\intertext{On the right hand side, the temporal and spatial partial derivatives in the first term commute while the time derivative in the second term is zero.
		Accordingly, using equation \eqref{eq:appendix-v-def} we find}
		\dot{\bm{a}}_\alpha
		&= \dfrac{\partial}{\partial \xi^\mu} \Big(
		\dfrac{\partial \hat{\bm{x}}(\xi^\lambda, t)}{\partial t}
		\Big) 
		\, \,
		\dfrac{\partial \xi^\mu}{\partial \theta^\alpha}
		= \dfrac{\partial \bm{v}}{\partial \xi^\mu} \,\,
		\dfrac{\partial \xi^\mu}{\partial \theta^\alpha} \,\,
		= \,\, \bm{v}_{, \alpha}
		~,
	\end{align}
	which is the statement of equation \eqref{eq:a-alpha-dot}.

\end{appendices}
\vspace{45pt}


\phantomsection
\addcontentsline{toc}{section}{References}
\bibliography{refs}
\bibliographystyle{bibStyle}

\end{document}